\documentclass[a4paper,fleqn,usenatbib]{mnras}

\usepackage{newtxtext,newtxmath}

\usepackage[T1]{fontenc}
\usepackage{ae,aecompl}

\usepackage{graphicx}	
\usepackage{amsmath}	
\usepackage{amssymb}	

\title[Short title, max. 45 characters]{MNRAS \LaTeXe\ template -- title goes here}
\title[Interpreting ALMA Observations of the ISM]{Interpreting ALMA Observations of the ISM During the Epoch of Reionization}

\author[H. Katz et al. ]{Harley Katz$^{1}$\thanks{E-mail: hk380@ast.cam.ac.uk}, Taysun Kimm$^1$, Debora Sijacki$^{1}$, and  Martin G. Haehnelt$^{1}$\\
$^1$Institute of Astronomy and Kavli Institute for Cosmology, University of Cambridge, Madingly Road, Cambridge, CB3 0HA\\
} 

\date{Accepted XXX. Received YYY; in original form ZZZ}

\pubyear{2016}

\begin{document}
\label{firstpage}
\pagerange{\pageref{firstpage}--\pageref{lastpage}}
\maketitle

\begin{abstract}

We present cosmological, radiation-hydrodynamics simulations of galaxy formation during the epoch of reionization in an effort towards modelling the interstellar medium (ISM) and interpreting ALMA observations. Simulations with and without stellar radiation are compared at large (Mpc), intermediate (tens of kpc), and small (sub kpc) scales.  At large scales, the dense regions around galaxies reionize first before UV photons penetrate the voids; however, considerable amounts of neutral gas remain present within the haloes. The spatial distribution of neutral gas is highly dynamic and is anti-correlated with the presence of stars older than a few Myrs.  For our specific feedback implementation, most of the metals remain inside the virial radii of haloes and they are proportionally distributed over the ionized and neutral medium by mass. For our most massive galaxy with ${\rm M_h}\sim10^{11}$M$_{\odot}$, the majority of the CII and OI mass are associated with cold neutral clumps. NII is more diffuse and arises in warmer gas while OIII arises in hotter gas with a higher ionization parameter, produced by photo-heating and supernovae. If smaller pockets of high metallicity gas exist in the ISM, the emission from these ions may be observable by ALMA while the low metallicity of the galaxy may cause these systems to fall below the local [CII]-SFR relation. The presence of dust can cause spatial offsets between UV/Ly$\alpha$ and [CII] emission as suggested by the recent observations of Maiolino~et~al.  [OIII] may be spatially offset from both of these components since it arises from a different part of density-temperature phase-space. 

\end{abstract}




\section{Introduction}
In the local Universe, the interstellar medium (ISM) of various different galaxy types has been well studied observationally \citep{Kennicutt2003,Kennicutt2011}.  In low redshift metal-enriched galaxies, fine structure lines are likely to be the dominant coolant of the ISM at $T<10^4$K \citep{Spitzer1978}.  At energies above the ionization potential of neutral hydrogen (13.6eV), forbidden lines such as [NII] and [OIII] trace the ionized medium while at lower energies, other lines such as [CII], [OI] or [CI] can trace the neutral ISM \citep{Carilli2013}.  With the Spitzer Space Telescope, local galaxies can be spatially resolved in the near infrared which has allowed for accurate measurements of various ISM properties such as dust mass, dust-to-gas ratio, dust properties, and stellar properties \citep{Kennicutt2003,Draine2007}.  More recently, the Herschel Space Observatory \citep{Pillbratt2010} has been particularly important for characterising the ISM in local galaxies due to its high spatial resolution in the far infrared, its extended wavelength coverage, and its spectrometer.  This allows many of the primary emission lines (i.e. [OI], [OIII], [NII], [CI], and [CII]), which originate from the different phases of the ISM, to be probed \citep{Kennicutt2011}.  Combining these observations with dust radiative transfer modelling can reveal important insights into dust mass, dust production mechanisms, and star formation \citep[e.g.][]{DeLooze2016}.  

At high-redshift and, in particular, during the epoch of reionization, our understanding of the ISM and internal properties of galaxies is much less advanced.  While hundreds of galaxies have been detected, only the more global properties such as star formation rate, UV luminosity function, and UV continuum slopes have been constrained  \citep{Zheng2012,Coe2013,Ellis2013,McLure2013,Oesch2013,Oesch2014,Bouwens2014,Bouwens2015}.  Various campaigns have targeted [CII] at 158$\mu$m at $z>6$ as this is expected to dominate cooling and to be the strongest emission line \citep{Carilli2013}.  Furthermore, it is well established in low redshift galaxies that [CII] emission correlates with the star formation rate (SFR) \citep[e.g.][]{DeLooze2014}.  The interpretation of this emission is complex as [CII] can be excited in photodissociation regions (PDRs), the warm and cold neutral medium, partially ionized gas, as well as shocked gas \citep{Madden1997,Kaufman1999,Garcia2011,Pineda2014,Cormier2012,Appleton2013,Pineda2014,Velusamy2014}.  Quasar host galaxies have been targeted successfully \citep{Wang2013,Walter2009} and [CII] has even been observed in the most distant spectroscopically confirmed quasar \citep{Venemans2012}.  However, the galaxies which host these bright quasars tend to be the most massive galaxies in the Universe \citep[e.g.][]{Sijacki2009,Costa2014} and do not represent typical star-forming galaxies during the epoch of reionization.  Recent observations with the Atacama Large Millimeter Array (ALMA) are beginning to probe [CII] in galaxies which are more representative of the high-redshift galaxy population and have star formation rates (SFRs) of $\sim10$M$_{\odot}$/yr \citep{Maiolino2015,Capak2015,Willott2015,Knudsen2016,Pentericci2016}.  These observations often find that the high-redshift galaxies fall below the local [CII]-SFR relations which indicates that the internal properties of these high-redshift galaxies are different from what we observe locally.     

Many theoretical studies have attempted to model the far infrared emission which may emanate from star-forming galaxies, both with a semi-analytic approach \citep{Gong2012,Munoz2014,Popping2014} and with numerical simulations \citep{Nagamine2006,Vallini2013,Vallini2015,Pallottini2016}.  \cite{Nagamine2006} post-processed smooth-particle hydrodynamics simulations (SPH) which employed a subgrid model of the ISM and found that the amount of [CII] emission crucially depended on the amount of neutral gas present in the galaxy.  Likewise, \cite{Vallini2013} used an SPH code to model the formation of a single high-redshift ($z=6.6$) galaxy at higher resolution and post-processed the simulation with a Monte Carlo radiative transfer scheme to track the UV photons which are primarily responsible for exciting the [CII] emission line.  However, the underlying simulation included neither star formation processes, radiative cooling, nor supernova feedback which are integral to properly model baryons in galaxies.  Nevertheless, the \cite{Vallini2013} simulations have predicted that the [CII] emission is spatially offset from where the UV continuum of the galaxy is expected to dominate.  This prediction is consistent with a number of observations of high-redshift galaxies which show an offset between the [CII] emission compared to that from the UV and Ly$\alpha$ \citep{Gallerani2012,Capak2015,Willott2015,Maiolino2015}.  This probably means that the [CII] emission likely originates in dense neutral clumps surrounding the galaxy and that high-redshift molecular clouds are rapidly disrupted by strong stellar feedback \citep{Maiolino2015}.  \cite{Vallini2015} have improved on the \cite{Vallini2013} simulations by using a more detailed recipe for the metal distribution and identifying the likely locations of molecular clouds.  They concluded that most of the [CII] emission likely originates in PDRs.  However, the underlying simulations still lacked some of the relevant physical processes (most importantly radiative cooling).  

More recently, \cite{Pallottini2016} used a zoom-in technique to model the formation of a $\sim10^{11}$M$_{\odot}h^{-1}$ halo at 30pc resolution.  These simulations included  a model for the formation of H$_2$ \citep{Krumholz2008,Krumholz2009,McKee2010} which allowed the authors to follow the atomic to molecular transition and make inferences on the correlation between [CII] emission and the location of H$_2$.  They found that 95\% of the [CII] emission originates from the H$_2$ disk, consistent with the \cite{Vallini2015} simulation.  The \cite{Pallottini2016} simulations, however, do not include on-the-fly radiative transfer (RT) and therefore do not properly model the enhancement of the local Lyman-Werner background above the mean metagalactic background, inside of the galaxy.  

Other simulations have been performed which include on-the-fly RT and detailed non-equilibrium chemistry models.  These tend to be full box simulations with small ($<100$ Mpc) volumes or zoom-in simulations of individual haloes or regions \citep[e.g.][]{Wise2012,Kimm2014,Gnedin2014,Xu2016}.  These have generally been performed in the context of reionization.  While it is well established that the Universe becomes fully ionized at some time around $z\sim6-7$ \citep{Fan2006,Ouchi2010,Bolton2011,Choudhury2015,Planck2015}, much remains unknown about reionization.  The sources which provide the majority of UV photons for reionization are still debated \citep{Couchman1986,Haardt1996,Ricotti2002,Ricotti2004,Choudhury2009,Dopita2011,Haardt2012,Chardin2015,Haardt2015}.  Likewise, the time it took to complete reionization is also still uncertain \citep{Bolton2007,Bowman2010,Zahn2012}.  Answering these questions is important for understanding galaxy formation as reionization can have important effects on galaxy properties.  

Unfortunately, RT, non-equilibrium chemistry simulations are computationally very expensive and there must be a compromise between resolution and computational volume.  Here we use cosmological simulations including on-the-fly RT and detailed non-equilibrium chemistry in an attempt towards self-consistently modelling the ISM at high-redshift in order to better interpret current observations of ``normal" star-forming galaxies.  The sacrifice we have to make in order to run a full cosmological box with on-the-fly RT is a moderate spatial resolution of $\sim100$pc.  Our work improves on the previous studies which have attempted to model the far infrared emission in high-redshift galaxies in a number of ways.  Firstly, we explicitly model the relevant star formation and cooling processes which change the underlying structure of the galaxy.  Secondly, we use on-the-fly RT which is coupled to the molecular hydrogen and captures the spatially inhomogeneous flux in the Habing band which is crucial for predicting the far infrared emission.  Thirdly, we relate the internal properties of the galaxy to the global process of reionization.  Finally, we attempt to connect large (few Mpc) and small (100pc) scales and describe how the included physics affects galaxy formation and our understanding of gas properties in the objects which are currently being targeted by ALMA.  

In Section~\ref{sims}, we describe our cosmological, radiation-hydrodynamics simulations and introduce a new variable speed of light approximation which better models photon propagation in both low and high-density regimes.  In Section~\ref{results}, we analyse these simulations on large, intermediate, and small scales with a particular emphasis on the stellar, gas, radiation, and molecular properties.  In Section~\ref{disc}, we compare our simulations to previous work and describe how UV and Ly$\alpha$ emission can be offset from [CII] and [OIII].  Finally, in Section~\ref{conc} we present our conclusions.

\section{Cosmological RT-Hydro Simulations}
\label{sims}

\subsection{Initial Conditions}
We simulate a 10Mpc$/h$ box starting at $z=150$ with a uniform grid of 256$^3$ dark matter particles ($m_{\text{dm}}=6.51\times10^6$M$_{\odot}$) and a similar number of gas cells.  For the initial conditions, we use the software package {\small{MUSIC}} \citep{Hahn2011}.  Lagrangian perturbations and a local Lagrangian approximation are used on the base grid for dark matter particles and gas cells, respectively, so that the initial conditions are accurate to second-order.  Cosmological parameters reported by the Planck Collaboration are assumed ($h=0.6731$, $\Omega_{\text{m}}=0.315$,  $\Omega_{\Lambda}=0.685$, $\Omega_{\Lambda}=0.049$, $\sigma_8=0.829$, and $n_s=0.9655$, \citep{Planck2015}).  {\small{CAMB}} \citep{Lewis2000} was used to generate the transfer function with the relevant cosmology for these initial conditions and the gas is initially assumed to be neutral with 76\% hydrogen and 24\% helium by mass.

\subsection{Numerical implementation of chemistry, and radiation transport}
We use the publicly available adaptive mesh refinement (AMR) code {\small{RAMSES}} \citep{Teyssier2002}, and in particular, the RT version, {\small{RAMSES-RT}} \citep{Rosdahl2013}, to model the gravity, detailed hydrodynamics, non-equilibrium chemistry and RT in the cosmological box.  We have made significant modifications to both the non-equilibrium chemistry and RT packages native to {\small{RAMSES-RT}} in order to better track the H$_2$ abundances as well as the propagation of ionizing radiation in low-density regimes and in particular, the intergalactic medium (IGM).

\subsubsection{H$_2$ chemistry, Radiation, and Cooling}
The standard version of {\small{RAMSES-RT}} is able to accurately follow the non-equilibrium chemistry and cooling for six species: H, H$^+$, e$^-$, He, He$^+$, and He$^{++}$.  There is an additional patch to the code which was used to follow the formation and destruction of a seventh species, H$_2$, in the context of lower redshift galaxies than considered here and post reionization \citep{Tomassetti2015}.  This implementation is not coupled to the inhomogeneous 3D radiation field which requires a complicated treatment of H$_2$ dissociating radiation in multiple bands.  We use this patch as a starting point for our own implementation which follows very closely the methods of radiation-coupled H$_2$ chemistry presented in \cite{Bac15}.  Certain differences in the exact implementation arise due to the fact that \cite{Bac15} use a ray-tracing method to follow the RT while {\small{RAMSES-RT}} is a moment-based scheme which treats the radiation like a fluid. 

We follow radiation in six different energy bins in the range $[5.6$eV$-\infty)$ listed in Table~\ref{ebins}.  The five highest energy bins are used to calculate rates for the following reactions,
\begin{equation}
\text{H}_2+\gamma_{2} \rightarrow \text{H}+\text{H},
\end{equation}
\begin{equation}
\text{H}+\gamma_{3,4,5,6} \rightarrow \text{H}^+ +e^-,
\end{equation}
\begin{equation}
\text{H}_2+\gamma_{4,5,6} \rightarrow \text{H}_2^+ +e^-,
\end{equation}
\begin{equation}
\text{He}+\gamma_{5,6} \rightarrow \text{He}^+ +e^-,
\end{equation}
and
\begin{equation}
\text{He}^++\gamma_{6} \rightarrow \text{He}^{++} +e^-.
\end{equation}
The subscript for each $\gamma$ determines which energy bin listed in Table \ref{ebins} contributes to the reaction.  As in \cite{Bac15}, we have created an extra bin with a lower limit of $15.2$eV which is the ionization energy of H$_2$ and we assume that all H$_2^+$ which is created by this reaction is immediately destroyed by dissociative recombination such that
\begin{equation}
\text{H}_2^++e^-\rightarrow\text{H}+\text{H}.
\end{equation}

\begin{table}
\centering
\begin{tabular}{@{}llll@{}}
\hline
Bin & E$_{\text{min}}$ & E$_{\text{max}}$ & Main Function\\
 & [eV] & [eV] & \\
\hline
1 & 5.60   & 11.20 & CII, OI, OIII, and NII ionization states \\
2 & 11.20 & 13.60 & H$_2$ Photodissociation\\
3 & 13.60 & 15.20 & HI Photoionization\\
4 & 15.20 & 24.59 & HI \& H$_2$ Photoionization\\
5 & 24.59 & 54.42 & HeI, HI \& H$_2$ Photoionization\\
6 & 54.42 & $\infty$ & HeI, HeII, HI \& H$_2$ Photoionization\\
\hline
\end{tabular}
\caption{Energy bins used to track the radiation.  Energy bin 1 represents the Habing band.  It does not affect the temperature or density state of the gas and is propagated so that we can calculated the CII, OI, OIII, and NII ionization states in order to predict their masses.  The photons in these bins can be absorbed in the simulation depending on the metallicity of the cell.}
\label{ebins}
\end{table}

For each reaction listed in Equations 1-5, one must know the average atomic cross-section in order to calculate the photoionization rates.  Furthermore, to self-consistently model the photo-heating from these reactions, one must also store the energy-weighted cross-section (see e.g. \cite{Rosdahl2013}).  Because our photon bins represent averages across multiple frequencies, the cross-sections we use for each of the different species are calculated on-the-fly at each time step and determined by the luminosity weighted mean energy of the  different spectral energy distributions (SEDs) of all of the sources in our box\footnote{Note that this discussion of cross-section is not applicable to H$_2$ photoionization by Lyman-Werner band photons as listed in equation 1.}.  For each individual star particle, photons are injected into the host cell with an SED for a given metallicity and age \citep{Bruzual2003}.  We adopt a Chabrier initial mass function \citep{Chabrier2003}.  The luminosity of the source is scaled to the mass of the star particle.  In order to calculate the average and energy-weighted cross-section of each species in the energy bins greater than 13.6eV, we use Equations B7 and B8 as given in \cite{Rosdahl2013}.  These two equations integrate over the frequency-dependent cross-section for each species.  For H, He, and He$^+$, we use the frequency-dependent cross-sections as given in \cite{Hui1997}.  The frequency-dependence of the H$_2$ cross-section at $E>15.2$eV is consistent with the piecewise fit to the analytical results of \cite{Liu2012} and is listed in Table 1 of \cite{Bac15}.  The average and energy-weighted cross-sections for each species are updated for each of the relevant bins at every coarse time step to account for the formation of new stars.  

While H$_2$ photoionization at $E>15.2$eV is a continuum process, H$_2$ dissociation in the Lyman-Werner band is a line driven process, and therefore, we must treat the cross-section of H$_2$ in this band differently because self-shielding becomes important.  Similar to \cite{Bac15}, we define an effective cross-section,
\begin{equation}
\sigma_{\text{H}_2,\text{LW}}=\sigma_{\text{H}_2,\text{LW},\text{thin}}f_{\text{shd}},
\end{equation}
where $\sigma_{\text{H}_2,\text{LW},\text{thin}}=D/F$, $D=5.18\times10^{-11}$s$^{-1}$ is the photodissociation rate in the optically thin limit \citep{Rollig2007} and $F=2.1\times10^7$s$^{-1}$cm$^{-2}$ is the photon flux in the Lyman-Werner band in the interstellar radiation field \cite{Draine1996}.  For $f_{\text{shd}}$, we follow \cite{Gnedin2009} and assume,
\begin{equation}
f_{\text{shd}}=\frac{1-\omega_{\text{H}_2}}{(1+x)^2}+\frac{\omega_{\text{H}_2}}{\sqrt{1+x}}e^{-0.00085\sqrt{1+x}},
\end{equation}
where $\omega_{\text{H}_2}=0.2$ and $x\equiv N_{\text{H}_2}/(5\times10^{14}\text{cm}^2)$.  $N_{\text{H}_2}$ is the H$_2$ column density and we make the very simple approximation that $N_{\text{H}_2}=n_{\text{H}_2}\Delta x$, where $n_{\text{H}_2}$ is the number density of H$_2$ in the cell and $\Delta x$ is the physical length of the grid cell in the simulation.  

With cross-sections readily available, we can compute the photoionization and photodissociation rates for each of the different species, update their number densities and deplete the number of photons in each bin by the relevant amount, which fully couples the non-equilibrium chemistry to the RT.  

In addition to the individual species as sinks for photons, in principle, there may also be dust present which can absorb photons across a wide energy range.  We do not self-consistently track the formation of dust in the simulation and in order to calculate the dust number density in each cell, we assume that the dust traces the metals.  Similarly to \cite{Gnedin2009}, we assume that the dust-to-gas ratio scales linearly with metallicity and take the dust cross-section to be $\sigma_{\text{dust}}=4.0\times10^{-21}$cm$^{-2}$.  This cross-section is much lower than that of the other species relevant to the more energetic radiation bins in our simulation and the metallicity in our simulations is rather low compared to solar\footnote{At higher resolutions, when the ISM is better resolved}, we are likely to resolve pockets of higher metallicity which would then have a higher dust content and possibly affect the absorption, therefore, photon absorption by dust is a very marginal effect.  The dust number density, however, could become relevant for our lowest energy bin (which is the lower energy range of the Habing band) where none of the species in our simulation can act as absorbers.  The flux in this band is relevant for photoelectric heating as well as for making predictions for [CII] which may be observed by ALMA.  

The total rate equation we use to track the abundance of H$_2$ in our simulation is,
\begin{equation}
\begin{aligned}
\frac{dx_{\text{H}_2}}{dt}={} & -C_{\text{coll}}x_{\text{H}_2}-(k_{\text{UV}}+k_{\text{LW}})x_{\text{H}_2} \\
& +(R_d+R_p)n_{\text{HI}},
\end{aligned}
\end{equation}
where $n_{\rm HI}$ is the number density of neutral hydrogen, $C_{\text{coll}}=\sum_{i=e^-,\text{H},\text{He},\text{H}_2}k_{\text{coll},i}n_i$ and $k_{\text{coll},i}$ are the collisional dissociation rates for each of the four listed species as given in \cite{Glover2008}.  $k_{\text{UV}}$ and $k_{\text{LW}}$ are the UV and Lyman-Werner photoionization and photodissociation rates discussed earlier and are calculated from the cross-section and number density of photons.  Finally, $R_d$ is the formation rate for H$_2$ on dust which is relevant for metal-enriched gas and $R_p$ is the formation rate of H$_2$ via the H$^-$ channel, which becomes relevant in the zero-metallicity limit.  For $R_d$, we use the expression from \cite{Gnedin2009} so that
\begin{equation}
R_d=3.5\times10^{-17}ZC_f{\rm cm}^3{\rm s}^{-1}, 
 \end{equation}
where $Z$ is the metallicity of the gas and $C_f$ is the clumping factor which we set to 10, consistent with \cite{Gnedin2009}.  The expression for the primordial channel of H$_2$ formation is primarily due to the presence of H$^-$ and this reaction is taken from \cite{Glover2010}.  For this we must know the abundance of H$^-$ which we do not explicitly track in the simulation.  In order to calculate this quantity, we assume an equilibrium abundance by solving for the equilibrium rate of the formation of H$^-$ using reactions 1, 2, 5, and 13 in the Appendix of \cite{Glover2010}.  Thus we use the following equations for the primordial channel of H$_2$ formation:
\begin{equation}
k_1n_{\rm HI}n_e=k_2n_{\rm H^-}n_{\rm HI}+k_5n_{\rm H^-}n_{\rm HII}+k_{13}n_{\rm H^-}n_e
\end{equation}
and
\begin{equation}
{\mathcal R}_p\equiv\frac{k_1k_2}{k_2+k_5x_{\rm HII}+k_{13}x_e}.
\end{equation}

In addition to being coupled with the radiation, the H$_2$ in our simulation is also coupled to the thermal state of the gas.  In the low metallicity regime, H$_2$ is the dominant coolant below the atomic cooling threshold and in principle can cool the gas to $\sim100$K.  We use the H$_2$ cooling rates from \cite{Hollenbach1979} which is summed with all of the primordial cooling channels already present in {\small{RAMSES-RT}}, described in \cite{Rosdahl2013}, in order to determine the net cooling rate.  At slightly higher metallicities, metal line cooling becomes the dominant cooling channel below $\sim10^4$K and for these rates, we interpolate tables computed with {\small{CLOUDY}} \citep{Ferland2013} with a \cite{Haardt2012} UV background which were made for the Grackle chemistry and cooling library\footnote{https://grackle.readthedocs.io/en/latest/} \citep{Bryan2014,Kim2014}.  The values for the metal line cooling rates depend on redshift, density, and temperature and we scale the rates with the total metallicity of the cell.  Because we do not resolve the smallest progenitors of $z=6$ galaxies, we assume a metallicity floor with Z$_{\text{min}}=10^{-3.5}$Z$_{\odot}$ at $z=15$ \citep{Wise2012}.

Besides cooling, H$_2$ can also contribute to the volumetric heating rate of the gas and we consider here two processes: heating due to UV photoionization for $E_{\gamma}>15.2$eV as well as heating from photodissociation and UV pumping in the Lyman-Werner band.  The photo-heating rate for H$_2$ due to photoionization from UV photons is treated like the other species where the excess photon energy above the ionization potential contributes to the heating term.  In the Lyman-Werner band, we follow the method of \cite{Bac15} to calculate the volumetric heating rate.  For each photodissociation, an excess energy of $\sim0.4$eV is deposited into the gas as heat \citep{Black1977}.  Not all absorptions of Lyman-Werner photons by H$_2$ lead to photodissociation.  The H$_2$ can instead become vibrationally excited until it either fluoresces back down to the ground state or is collisionally de-excited which can transfer heat to the gas.  We calculate the heating rate due to this UV pumping following Equations 44-48 in \cite{Bac15} which combine the UV pumping rate of H$_2$ from \cite{Draine1996} with the energy released per UV pumping event from \cite{Burton1990}.  Having outlined the thermal coupling of H$_2$ with the gas, we conclude our description of the H$_2$ implementation in {\small{RAMSES-RT}}.  

Thus far, we have neglected the inclusion of our first photon group, Bin 1, which represents the lower energy range of the Habing band.  We follow the radiation in this energy range for two specific reasons: 1) it becomes relevant to calculate the heating rate for photoelectric heating by dust and 2) in order to calculate near-infrared emission from CII, OI, NII, and OIII, one must know the energy density of photons in this band.  The densities we probe in the simulations presented in this paper (especially the density at which we form stars) do not become high enough for photoelectric heating by dust to become relevant and we have therefore neglected it in our simulation.  The exact implementation and usage of photoelectric heating by dust is described in \cite{Kimm2016}.  The latter reason becomes of particular importance when we compare our simulations with ALMA observations.  The origin of the [CII] emission within these high-redshift galaxies is unknown and may originate in low-density neutral or ionized gas as well as from photodissociation regions.  In order to better understand the physical properties of high-redshift galaxies which exhibit strong [CII], [OI], [NII] or [OIII] emission, it is crucial that the Habing band is tracked self-consistently in the simulation.  Note that radiation in this bin does not physically affect the state of the gas in the simulation but the inhomogeneous spatial distribution of this radiation is the quantity which we require to make predictions for ALMA.

\begin{figure}
\centerline{\includegraphics[scale=0.42]{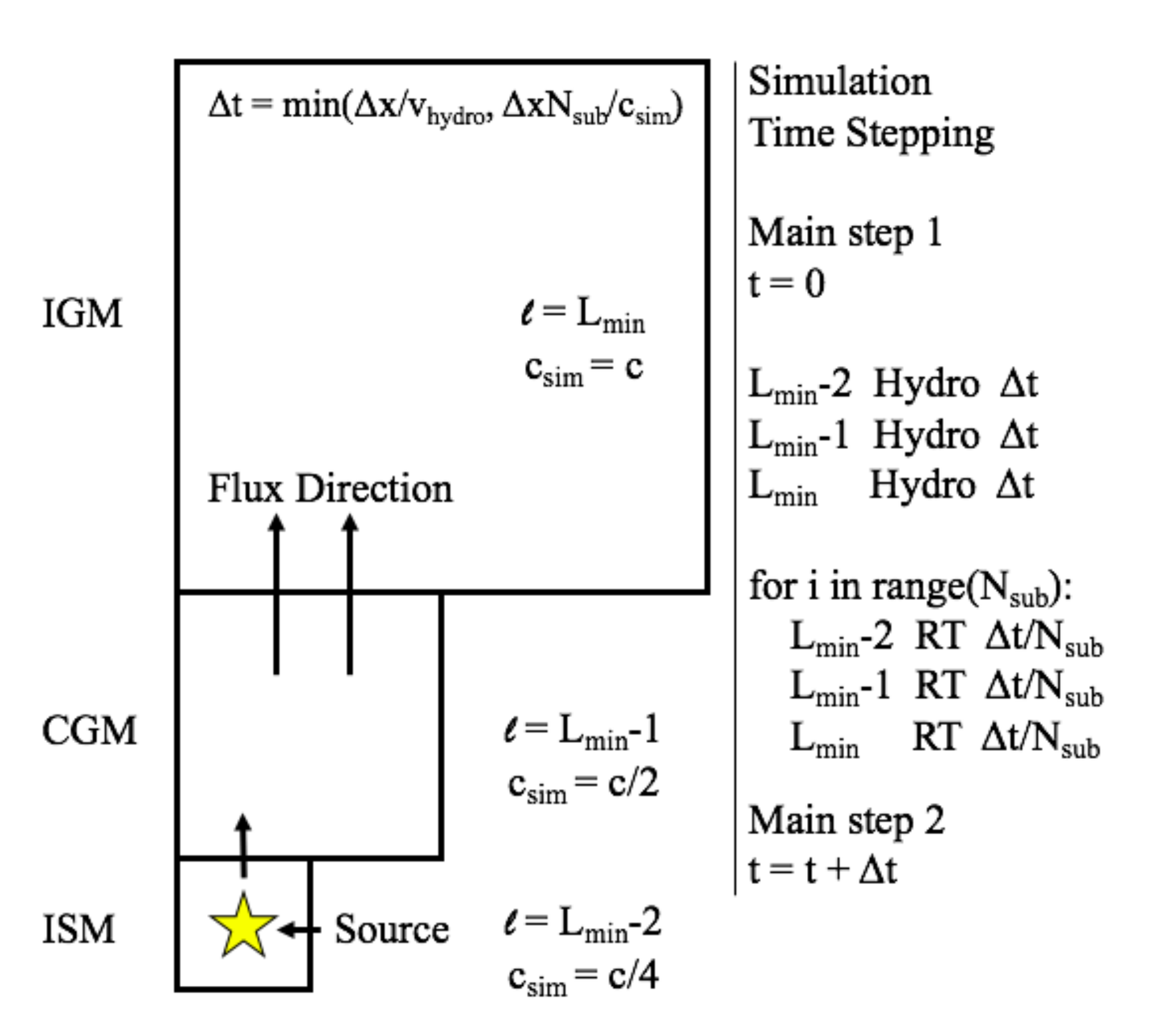}}
\caption{Schematic view of the variable speed of light (VSLA) algorithm.  The speed of light used in the simulation increases by a factor of two at each higher level representing the change in density of the cell.  In this three level example, the most refined level represents the interstellar medium, the middle level represents the circumgalactic medium, and the base grid represents the intergalactic medium.  The source is located in the highest density region and as the photons escape from the ISM, they increase in speed while the total flux is conserved.  On the right hand side, we show pseudocode of how the algorithm is implemented.  The time step is determined and all levels are synched to the smallest time step in the simulation.  First, the hydrodynamics is evolved for all levels starting with the most refined.  Once the hydro time step is completed, the RT calculation is completed for the specified number of subcycles, looping through all levels at each subcycle.  At each RT subcycle, cooling and photon absorption and emission are calculated.}
\label{VSLAscheme}
\end{figure}

\begin{figure*}
\centerline{\includegraphics[]{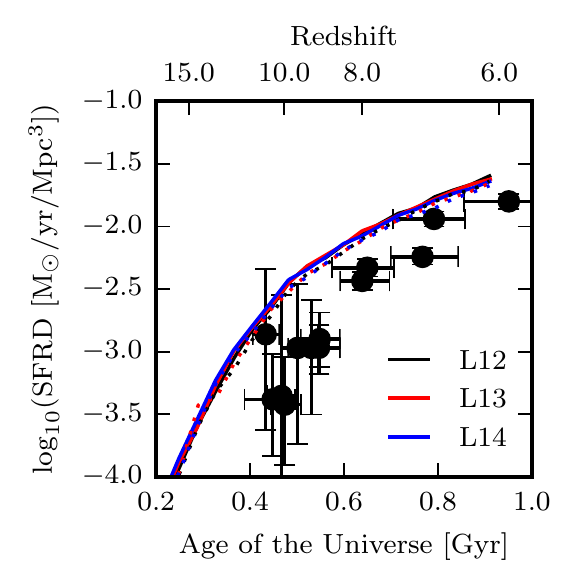}\includegraphics[]{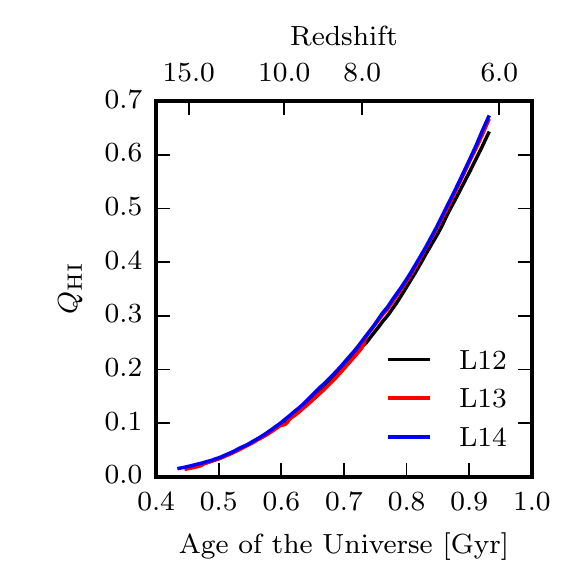}\includegraphics[]{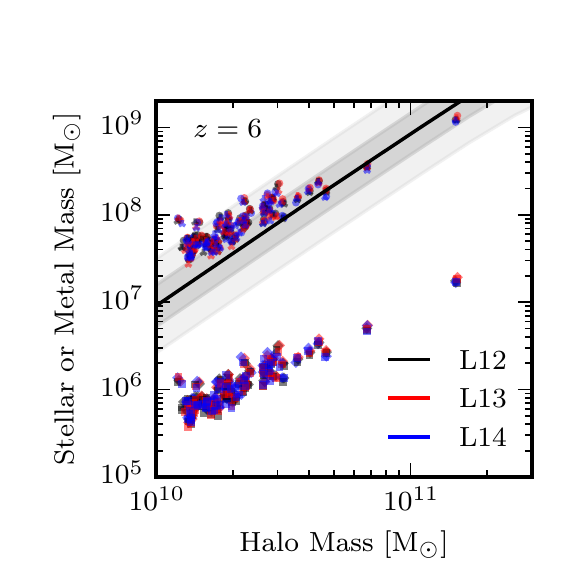}}
\caption{{\it Left}: SFRD of all six simulations compared to observations \protect{\citep{Bouwens2015, McLure2013, Ellis2013, Oesch2013, Oesch2014, Zheng2012, Coe2013, Bouwens2014}}.  Solid and dotted lines represent simulations with and without stellar radiation, respectively, while different colours represent different resolutions as listed in the legend. {\it Centre}: Volume filling factor of ionized hydrogen for the simulations which include stellar radiation at three different resolutions.  {\it Right}: Stellar (circles and triangles, the higher set of points) and metal (diamonds and squares, the lower set of points) masses in our simulations as a function of halo mass at $z=6$ for all haloes with M$_{\rm halo}>10^{10}$M$_{\odot}/h$ at three different resolutions.  Circles and diamonds correspond to simulations without RT, while triangles and squares represent simulations with RT.  The grey band represents the stellar mass-halo mass relation inferred for $z=6$ from \protect\cite{Behroozi2013}.   Note the very good convergence in SFRD, ionization history, dark matter mass, stellar mass, and metal mass.}
\label{SFRD}
\end{figure*}

\subsubsection{The Variable Speed of Light Approximation}  
Our aim is to run a cosmological simulation with on-the-fly RT to model galaxy formation in the high-redshift universe.  {\small{RAMSES-RT}} uses an explicit solver for radiation transport and therefore the RT time step in the simulation is limited by the RT Courant condition such that,
\begin{equation}
\Delta t_{\text{RT}}<\frac{\Delta x}{3c_{\text{sim}}},
\end{equation}
where $c_{\text{sim}}$ is the speed of light used in the simulation.  For high resolution cosmological simulations, using the full speed of light can be prohibitively expensive because it is so much greater than the typical hydrodynamic velocities which govern the time step for non-RT simulations.  One can adopt the reduced speed of light approximation (RSLA, \cite{Gnedin2001}) so that, for example, $c_{\text{sim}}=0.01c$.  This approximation is a good approximation in many regimes (see Sections 4.2 and 4.3 of \cite{Rosdahl2013} for a lengthy discussion), as long as the propagation of the ionization front (I-front) is slower than $c_{\text{sim}}$.  \cite{Rosdahl2013} clearly demonstrate that to track the I-fronts properly in the IGM one must use the full speed of light, while in much higher density regions, such as the ISM, the RSLA is a good approximation.  Furthermore, \cite{Bauer2015} have shown that adopting $c_{\text{sim}}<c$ can lead to delayed reionization for the same source model.

One possible option, which can also be used concurrently with the RSLA, is to subcycle the radiation time step and use optimised boundary conditions (see \cite{Commercon2014}) to make this compatible with adaptive time stepping.  The process of subcycling can be implemented in one of two ways: 1) a fixed value of the speed of light is chosen (i.e. $c_{\text{sim}}=0.01c$) and the number of subcycles is determined by maximising the number of RT time steps which fit into a single hydro time step (potentially up to a maximum number of subcycles, \citep[e.g.][]{Rosdahl2013} or 2) the number of subcycles is kept fixed and the speed of light is changed to perform that number of subcycles \citep[e.g.][]{Gnedin2001}.  In the latter case, the speed of light in the simulation can in principle become very large for short hydro time steps.  The method of subcycling results in a speed-up because it reduces the total number of hydrodynamic time steps, but the number of RT subcycles can still be prohibitively costly if the full speed of light is used.  In this work, we apply this method and for each hydro time step, we perform a maximum of 10 RT subcycles\footnote{Note that in our case, we do not need to adopt any different boundary conditions because our VSLA routine uses the same hydro time step regardless of level of refinement.  Therefore, we can subcycle the RT through all levels at once after every coarse hydro time step.}.  

Alternatively, the problem of short time steps has been circumvented by using uniform grids on GPUs \citep{Aubert2008,Aubert2010}.  With this architecture, using the full speed of light is no longer prohibitively costly; however, the method is limited to a uniform grid to achieve such a large speed-up (although see \cite{Aubert2015} for improvements allowing for AMR grids on the GPU).  In this paper, we present a new approach which we call the variable speed of light approximation (VSLA) where the speed of light in the simulation changes depending on the level of refinement in the simulation which thereby captures the fast and slow moving I-fronts at their proper speed.  A schematic view of this algorithm is shown in Figure~\ref{VSLAscheme} and a full description of VSLA is presented in Appendix A.  For all simulations in this work, we use the full speed of light on the base grid and divide by a factor of two on each subsequently refined level.

\subsection{Resolution and Refinement}
We run a total of six different simulations at three different resolutions with and without on-the-fly RT.  The simulations are listed in Table~\ref{simstab}.  All simulations are run until $z=6$ except for the L14-RT simulation which is run until $z=5.3$.  The dark matter particle resolution is the same for all simulations and we only vary the level the grid can refine to.  The grid is allowed to adaptively refine during the course of the simulation when the density of a cell increases by a factor of eight times that of the previous level in either dark matter or baryons.  We attempt to enforce a constant physical resolution throughout the simulation to keep the physical size of cells at the highest resolution as close to the values listed in Table~\ref{simstab} as possible.  This means introducing further levels of refinement at predefined scale factors such that when $a$ increases by a factor of two, the grid refines a further level.  

Throughout our analysis we will focus mainly on the two runs at the highest resolution with and without stellar RT.  We will return to the lower resolution simulations when discussing the properties of the ISM.

\begin{table}
\centering
\begin{tabular}{@{}lccccc@{}}
\hline
Name & RT & $l_{\text{max}}$ & Box size & M$_{\rm DM}$  & $\Delta x_{\text{min}}$\\
 & & & [cMpc/h] & [M$_{\odot}$] & [pc] \\
\hline
L12       & no   & 12 & 10 & $6.5\times10^6$ & 500\\
L12-RT & yes  & 12 & 10 & $6.5\times10^6$ & 500\\
L13       & no   & 13 & 10 & $6.5\times10^6$  & 250\\
L13-RT & yes  & 13 & 10 & $6.5\times10^6$ & 250\\
L14       & no   & 14 & 10 & $6.5\times10^6$ & 125\\
L14-RT & yes  & 14 & 10 & $6.5\times10^6$ & 125\\
\hline
\end{tabular}
\caption{List of simulations.  The first two columns denote the name of the simulation and whether it includes on-the-fly radiation from star particles.  $l_{\text{max}}$ indicates the maximum level of refinement while $\Delta x_{\text{min}}$ denotes the maximum physical resolution in pc which we maintain throughout the simulation.  The box size and dark matter mass are the same for all simulations.}
\label{simstab}
\end{table}

\begin{figure*}
\centerline{\includegraphics[]{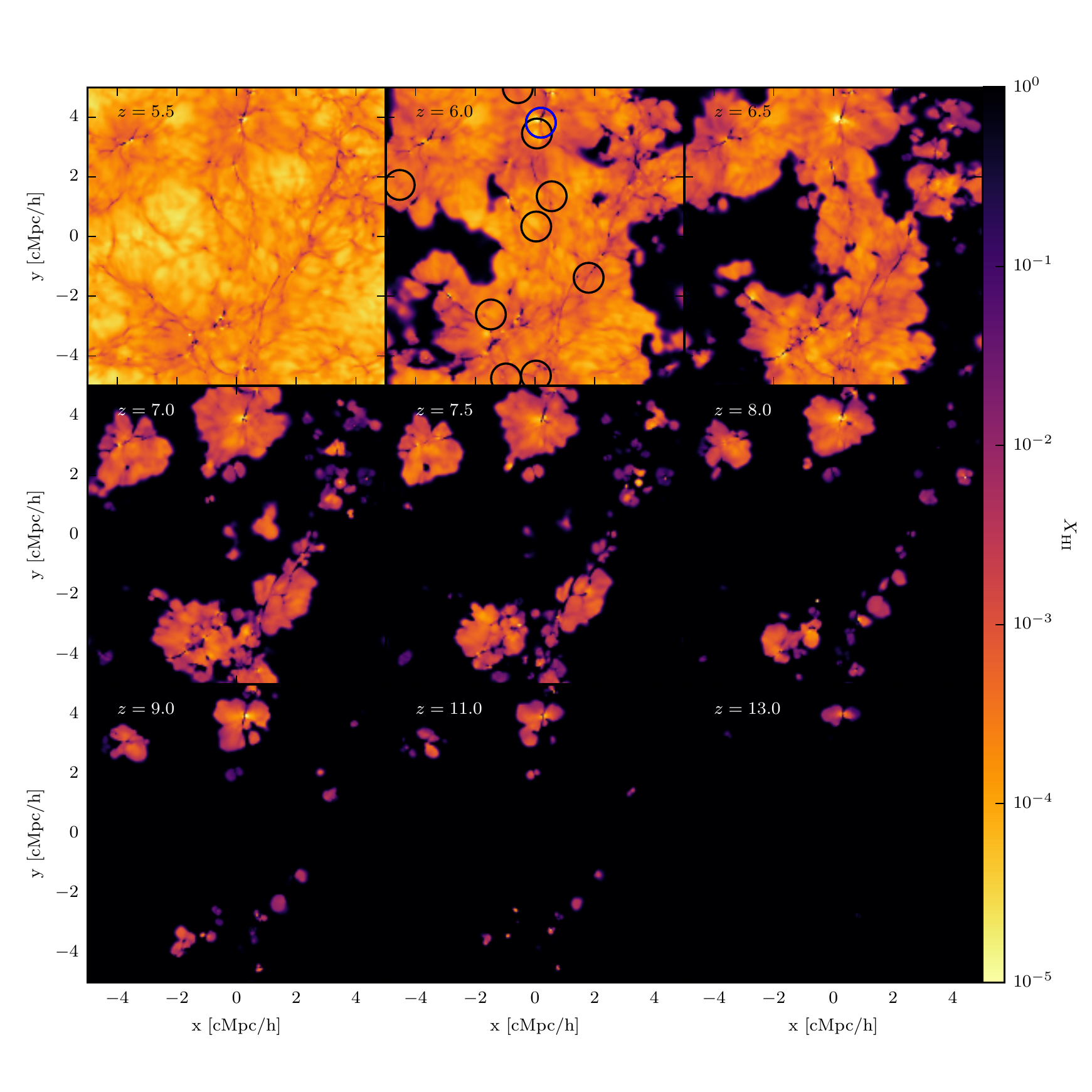}}
\caption{Spacial distribution of the HI fraction of the gas as a function of redshift for the L14-RT simulation.  The z-coordinate of the slice is centred on the most massive halo in the box.  The ten most massive haloes are circled in the top centre panel and the blue circle indicates the position of the most massive halo in the simulation.  Reionization begins around the most massive halo and by $z=5.5$ the Universe is completely ionized.}
\label{reionmaps}
\end{figure*}

\begin{figure*}
\centerline{\includegraphics[]{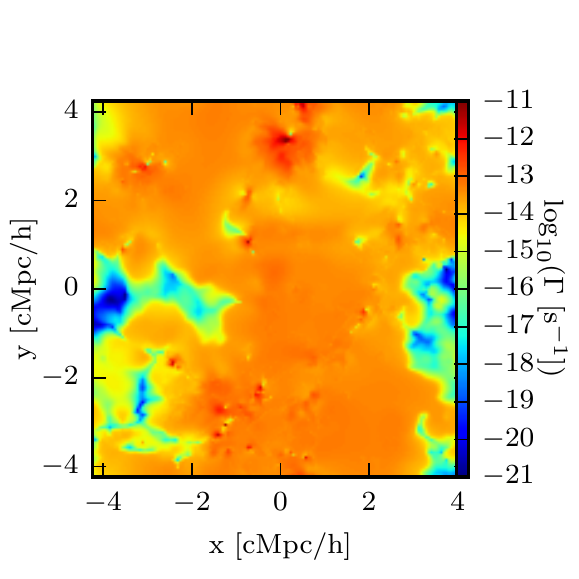}\includegraphics[]{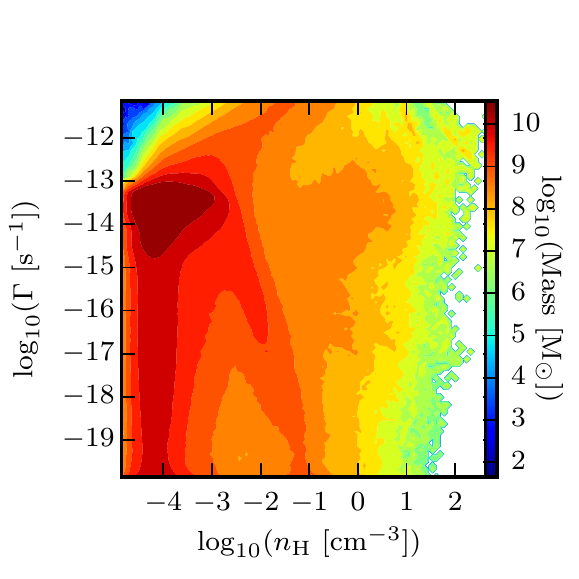}\includegraphics[]{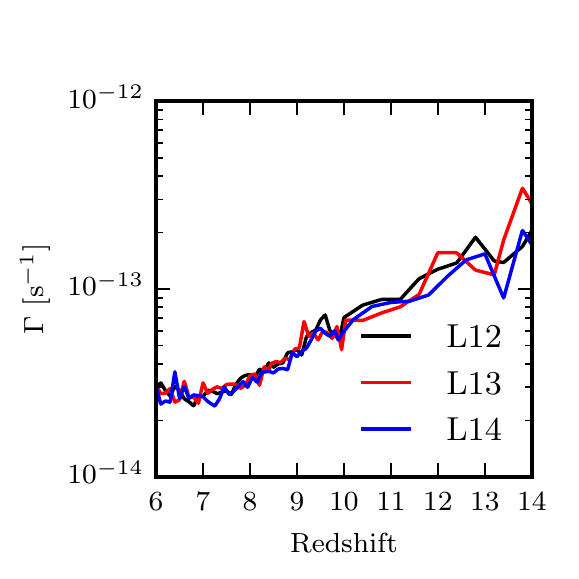}}
\caption{{\it Left}: Spatial distribution of the HI photoionization rate at $z=6$.  The z-coordinate is centred on the most massive halo in the simulation.  The photoionization rate is highest around the most massive halo and is somewhat lower in the voids.  {\it Centre}: Mass-weighted 2D histogram of HI photoionization rate versus number density of the gas. Note that the hydrogen in the simulation is not completely ionized and the photoionization rate extends towards low $\Gamma$ at ${\rm n_H\sim10^{-4}}\ {\rm cm^{-3}}$.  At high densities, we see two competing tracks.  The gas begins to self-shield at densities slightly lower than ${\rm n_H\sim10^{-2}}\ {\rm cm^{-3}}$ where the photoionization rate branches off towards higher values due to the presence of sources within the cells.  {\it Right}: Evolution of the volume-weighted HI photoionization rate in the ionized regions as a function of redshift for the three simulations which include radiation.}
\label{photoionmap}
\end{figure*}

\subsection{Star Formation and Feedback}
Stars are formed in the simulation based on the Schmidt law \citep{Schmidt1959} such that $\dot{\rho}_*=\epsilon_{\text{ff}}\rho_{\text{gas}}/t_{\text{ff}}$, where $\dot{\rho}_*$, $\epsilon_{\text{ff}}$, and $t_{\text{ff}}$ are the star formation rate density, the efficiency of star formation per free-fall time, and the free-fall time of the gas respectively.  We set a star formation density threshold of $n_{\rm H}=0.1$cm$^{-3}$ and require that $\rho_{\text{gas}}>50\bar{\rho}_{b}(z)$ to prevent star formation at very high-redshifts.  Stars are formed with an efficiency of $\epsilon=1\%$ \citep{Kennicutt1998} and the number of star particles formed is drawn from a Poisson distribution, $P(N_*)=(\lambda^{N_*}/N_*!)\exp{(-\lambda)}$, where $\lambda\equiv\epsilon_{\text{ff}}(\rho\Delta x^3/m_{*,\text{min}})(\Delta t/t_{\text{ff}})$.  For all simulations, regardless of resolution, we keep a fixed minimum mass for star particles so that $m_{*,\text{min}}=7.66\times10^4$M$_{\odot}$.  We set a temperature threshold of $2\times10^4$K so that stars cannot form in cells with a temperature higher than this value.  ionization of neutral hydrogen and neutral helium take place at temperatures less than this value and therefore, the temperature increase in gas cells due to ionizing radiation is alone not enough to prevent star formation.  This should allow the simulations which include stellar radiation sources to have similar star formation rate densities (SFRDs) as the simulations without stellar radiation.

Each star particle is assumed to represent a simple stellar population with a Chabrier IMF \citep{Chabrier2005} with a minimum mass of 0.1M$_{\odot}$ and a maximum mass of 150M$_{\odot}$.  After 10Myr, the massive stars are assumed to explode as supernova (SN) and the star particles lose 31\% of their total mass\footnote{This value is calculated assuming a \citep{Chabrier2005} IMF in the mass range $0.1-150$M$_{\odot}$ where stars with $M>6$M$_{\odot}$ undergo supernova, consistent with \cite{Vogelsberger2013,Crain2015}, assuming that stars with $6<M [{\rm M_{\odot}}]<8$ explode as electron capture SNe \citep{Chiosi1992}.} which is recycled back into the gas phase, and we assume that 5\% of the unenriched mass of the ejecta is composed of metals\footnote{We do not include any additional metal enrichment from Type Ia SNe or AGB stars.}.  For each SN, $10^{51}$ergs are injected as thermal energy into the gas.  Since our simulations do not resolve the individual phases of the SN, they are likely to suffer from ``over-cooling" and we therefore employ the ``delayed cooling" model of \cite{Teyssier2013} in order to mitigate this effect.  A delayed cooling parameter is tracked as a passive scalar which decays exponentially as a function of time and cooling is shut off in these cells.  We set the delay timescale to 20Myr.  We have modified the standard feedback routine so that the feedback (i.e. mass, metals, thermal energy, delayed cooling parameter) is spread over the nearest 19 cells (see Appendix A in \cite{Kimm2014} for a visualisation of the geometry).

\section{Results}
\label{results}
\subsection{Calibrating Global Properties}
In order to fairly compare the different simulations, we must ensure that certain global properties are properly calibrated between the simulations.  All six simulations have similar SFRDs as a function of time.  This is achieved by keeping the stellar mass and dark matter resolution constant across all simulations while also using the same density threshold ($n_H\geq0.1$cm$^{-3}$), temperature threshold ($T<2\times10^4$K), and star formation efficiency.  In the left panel of Figure \ref{SFRD}, we compare the SFRDs from each of the six different simulations and see very good convergence.  The SFRD in the simulations slightly over-predict the observed SFRD, although the estimated SFRDs would drop if we consider only haloes with observable star formation rates\footnote{For instance, at $z=6$ for the L14-RT simulation, taking haloes with SFRs~$>0.2$(0.1)M$_{\odot}/yr$ would decrease the SFRD by $\sim0.8(0.73)$dex, which is closer to the observations.  Note that this calculation is based on fewer than 50 haloes due to our small box.  Furthermore, if we only consider those haloes with M$_{\rm halo}>10^{10}$M$_{\odot}/h$ at $z=6$ for the SFRD, consistent with what is used for the stellar mass-halo mass relation in the right panel of Figure~\ref{SFRD}, we see a decrease in the global SFRD by 0.3~dex.}.

\begin{figure*}
\centerline{\includegraphics[]{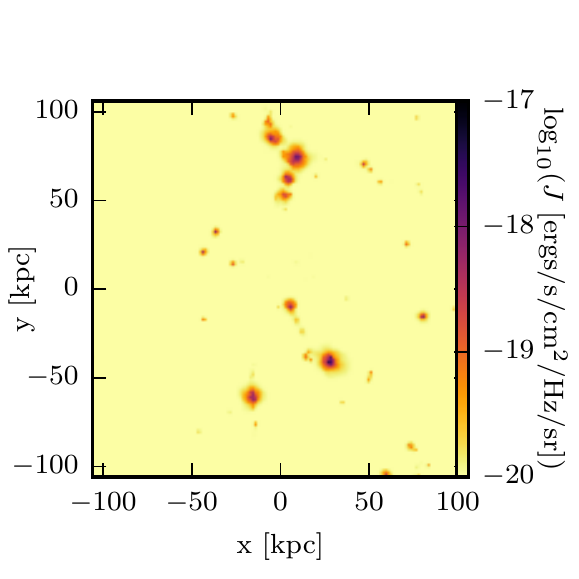}\includegraphics[]{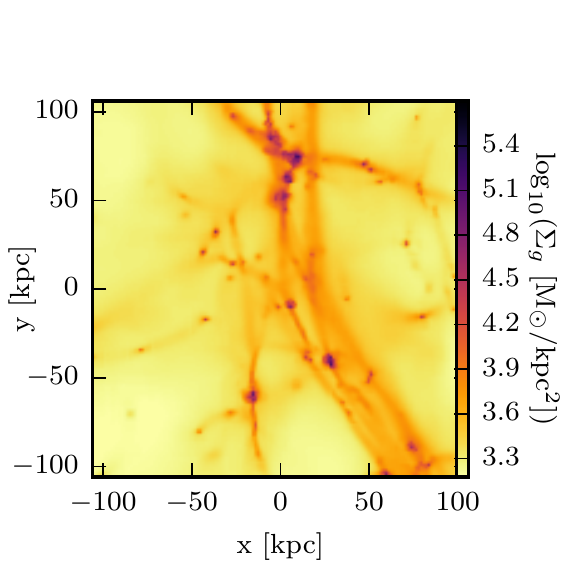}\includegraphics[]{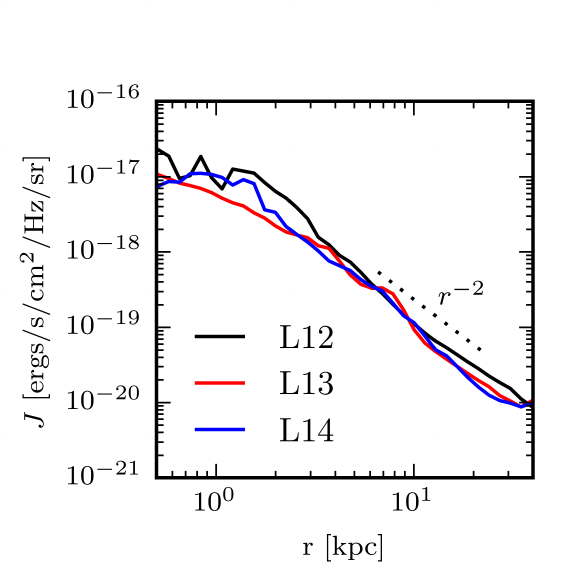}}
\caption{{\it Left}: Map of $J_{\rm LW}$ averaged along the z-axis in a small region, 1/10$^{\rm th}$ of simulation box, for the L14-RT simulation.  In between haloes, $J_{\rm LW}$ is very homogeneous, indicative that the Universe is optically thin to Lyman-Werner radiation.  {\it Centre}: Gas density integrated along the z-axis of the same sub-region for the L14-RT simulation. The gas over-densities correspond well with the sources of Lyman-Werner photons.  {\it Right}: Radial profiles of $J_{\rm LW}$ for the most massive halo in the box at $z=6$ for simulations which include stellar radiation.  $J_{\rm LW}$ falls off as $r^{-2}$ as expected for a source in an optically thin region.  Note the very good convergence in this behaviour between the three simulations.  }
\label{J21}
\end{figure*}

In addition to the SFRDs, we attempt to keep the reionization histories constant across the three simulations which include stellar radiation.  In the centre panel of Figure \ref{SFRD}, we plot the volume filling factor of ionized hydrogen to demonstrate that all simulations agree very well.  Our box is not completely ionized by $z=6$, but note that our 10Mpc$/h$ box is not a representative sample of the Universe and our simulations do not resolve some of the small sources with halo mass of $\sim10^8-10^9$M$_{\odot}$ which are likely very important for reionization \citep{Kimm2016}.  Since SN and UV radiation are two of the dominant mechanisms which govern properties of the ISM, the important point is that we have controlled for them by ensuring that all simulations have similar SFRDs and all RT simulations have similar reionization histories. 

Furthermore, it is crucial that the physical properties of individual haloes are converged in addition to the more global quantities in the simulations.  The most efficient formation channel for H$_2$ is via the surface of dust grains and therefore, it is necessary that the metal masses of the haloes between the different resolutions are converged (as the dust mass is assumed to scale with metallicity).  Note that we achieve the latter for free if we can converge the stellar mass as the total metallicity of a halo can be computed from the integrated star formation rate.  

We use the {\small AHF} halo finder \citep{Gill2004,Knollmann2009} to locate halos in our simulation.  In the right panel of Figure~\ref{SFRD}, we plot the stellar mass and metal mass as a function of halo mass for the 50 most massive haloes in all six simulations at $z=6$ which show very good agreement.  Furthermore, the stellar masses fall within the $2\sigma$ contours of the stellar mass-halo mass relation predicted from abundance matching at $z=6$ from \cite{Behroozi2013} suggesting that our galaxies have produced a reasonable amount of stars.  The stellar mass-halo mass relation in our simulation has a slightly flatter slope than what is derived from abundance matching which could indicate that our lower mass galaxies are forming stars more efficiently than expected from observations.

Having confirmed that the global properties of our simulation are converged over the different resolutions we now proceed to examine the physical properties of our simulations on large, intermediate, and small scales.

\subsection{Large Scales: Global reionization and Thermodynamic Properties of the IGM}
\subsubsection{Ionized Bubble Topology}
Understanding the thermodynamic and temperature state of the IGM on large scales is key for understanding galaxy formation as well as interpreting future measurements of the 21cm signal.  On the largest scales (i.e. Mpc), radiation is the main mechanism which allows galaxies to affect the state of the IGM.  In Figure~\ref{reionmaps}, we show HI maps in thin slices centred around the most massive galaxy as a function of redshift.  This galaxy is one of the first haloes in the simulation to form stars and a HII region has formed around the galaxy as early as $z=13$, which has a diameter across its longest axis of $\sim2$cMpc (the location of this galaxy in the box is circled in blue in the top centre panel of Figure~\ref{reionmaps}). The HII region is rather elongated and flattened and it almost appears as if two separate HII regions are touching at the central location.  This is due to a dense filament of gas which is feeding the galaxy perpendicular to the long axis of the bubble.  The gas in this filament is much denser than the surrounding gas so that it has a much shorter recombination time.  This prevents the ionized bubble from expanding efficiently across the vertical direction of the box leading to this characteristic bipolar shape.  The bipolar shape of the HII region is consistent with other studies which have looked at the escape of photons from a galaxy at a node in the cosmic web \citep{Ciardi2001,Iliev2006a,Abel2007,Wise2008}.

As the simulation evolves in time, more galaxies begin to form and emit ionizing radiation.  A second galaxy forms to the bottom left of the most massive object, as can be seen in the bottom centre panel of Figure~\ref{reionmaps}, and a HII bubble centred around this galaxy appears as early as $z=12$.  By $z=11$, these bubbles have begun to merge and by $z=9$, a single bubble encapsulates both haloes.  Note that the HII bubbles of these haloes are still expanding and have not yet reached the maximum extent of their Stromgren spheres.  To get the bubble merging correct at early times, it is important to use VSLA because it is well established that the RSLA solution will lag behind the true solution until the Stromgren radius is reached \citep{Rosdahl2013}.  For our simulations here, we use a relatively small box size which includes many small sources.  The I-fronts of these sources travel reasonably slowly at later times making RSLA potentially appropriate at later redshifts for hydrogen ionizing radiation bins.  However at early times, using VSLA or the full speed of light is necessary to capture the early evolution of the Stromgren spheres.

\begin{figure}
\centerline{\includegraphics[]{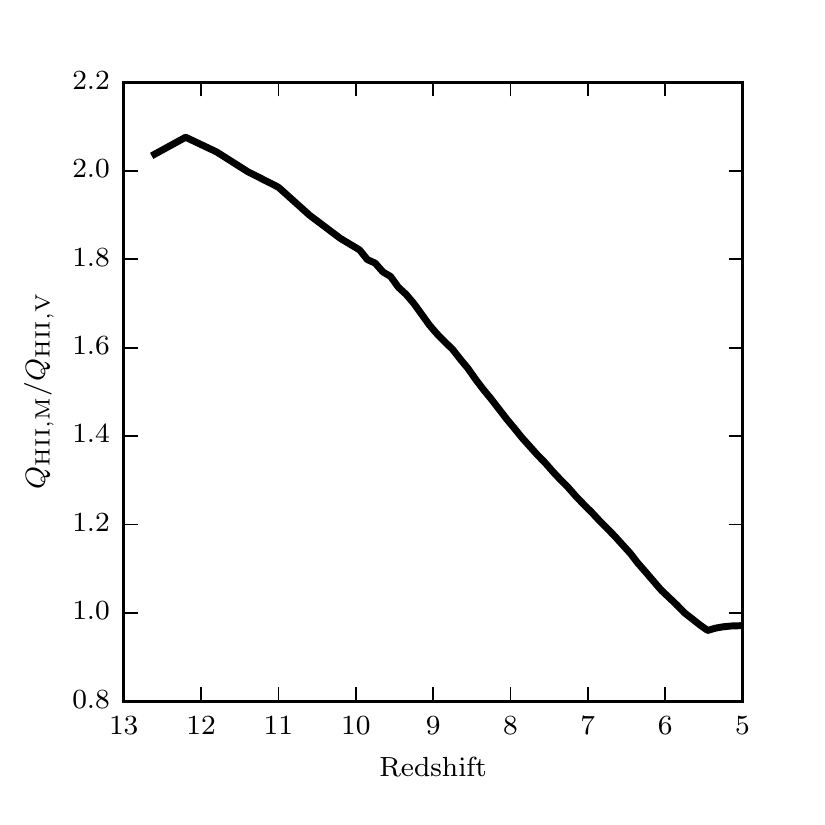}}
\caption{Ratio of the mass-weighted volume filling factor of HII, $Q_{\rm HII,M}$, to the volume-weighted filling factor of HII, $Q_{\rm HII,V}$, as a function of redshift for the L14-RT simulation.  This ratio remains above unity as the highest density regions around the galaxies are ionized first.  The ratio continues to decrease as the voids become ionized and finally drops below one when the only remaining neutral gas resides in the filaments and inside galaxies.}
\label{Qratio}
\end{figure}

Looking towards lower redshifts, by $z=6$, nearly all of the distinct bubbles have merged and the topology of the ionized region is simply connected.  There are still regions in the $z=6$ slice which are not yet ionized and this is not surprising as we know from Figure~\ref{SFRD} that our box is not yet completely ionized by $z=6$.  By comparing the distribution of ionized gas at later redshifts, after the ionized bubbles have begun to merge, with the locations of the massive haloes in Figure~\ref{reionmaps}, we can see that the ionized regions correspond well with the locations of the most massive objects.  The largest portion of residual neutral hydrogen at $z=6$ sits on the left side of the box and has a centre at $y\sim0$.  This corresponds to a void in our simulation.  The highest density regions in our simulation, around the massive haloes, ionise first, before the voids.  In the top left panel of Figure~\ref{reionmaps}, we see that this region is completely ionized by $z=5.5$.  The reionization behaviour in our simulation is consistent with the 'inside-out-middle' scenario \citep{Gnedin2000a,Finlator2009} whereby the remaining high density gas in the post overlap phase (i.e. filaments) are the last regions to become ionized.  To demonstrate this, in Figure~\ref{Qratio} we plot the ratio of the mass-weighted volume filling factor of HII to the volume-weighted filling factor of HII.  At high-redshift, before bubble overlap, this ratio remains greater than one as the high density regions around galaxies are ionized first.  As the UV radiation propagates into the voids, this ratio decreases as more of the volume is ionized.  The ratio drops below one when the voids are completely ionized and slowly approaches unity as the filaments become more ionized.

\subsubsection{HI Photoionization Rate} 
In the left panel of Figure~\ref{photoionmap}, we show the spatial distribution of the photoionization rate for our highest resolution simulation at $z=6$.  Unsurprisingly, the spatial distribution of the photoionization rate corresponds well to that of the distribution of ionized gas.  The gas closest to the most luminous ionizing sources has the highest photoionization rate.  The photoionization rate inside of the bubbles is nevertheless relatively homogenous.  

In the centre panel of Figure~\ref{photoionmap}, we plot a 2D mass-weighted histogram of $\Gamma_{\rm HI}$ versus $n_{\rm H}$.  The majority of the low-density gas $n_{\rm H}<10^{-4}\ {\rm cm^{-3}}$ has $\Gamma_{\rm HI}$ just below $10^{-13}$s$^{-1}$.  This is the isotropic photoionization rate we see permeating throughout the ionized bubbles in the left panel of Figure~\ref{photoionmap}.  There is a plume of gas at these densities with lower $\Gamma_{\rm HI}$ due to the fact the simulation is not completely ionized by $z=6$.  The residual neutral hydrogen absorbs the photons in these regions which strongly reduces $\Gamma_{\rm HI}$.  

At higher densities, there is a bifurcation at $n_{\rm H}\sim10^{-3}-10^{-2}\ {\rm cm^{-3}}$ with one arm pointing to high values and the other towards low values of $\Gamma_{\rm HI}$.  The lower arm at high densities is due to self-shielding in dense regions.  This gas can either be pristine material which is fed into the centres of the haloes along dense filaments, or metal-enriched gas which has been ejected from the central regions of the galaxy from SN feedback and is now cooling and falling back.  For the same temperature, this gas will have much shorter cooling timescales than the gas in the cold flows due to the presence of the metals.  Due to the self-shielding from neutral gas, the radial profiles will not exactly follow $r^{-2}$.  For example, the left panel of Figure~\ref{photoionmap} shows a local minimum in $\Gamma_{\rm HI}$ just above the location of the most massive halo in the box.  This is due to a dense filament of gas which is feeding the galaxy in this direction which leads to large column densities along the line of sight and creates a shadow in the photoionization rate. 

The right panel of Figure~\ref{photoionmap} shows the volume-weighted evolution of $\Gamma_{\rm HI}$ in ionized regions (defined where $x_{\rm HII}>0.5$) as a function of redshift for the three simulations which include stellar radiation.  At high-redshift, the photoionization rate is very large.  This is because the ionized regions in the box are close to the photon sources and thus $\Gamma_{\rm HI}$ remains high (see Figure~\ref{reionmaps} for the locations of the ionized regions).  As the ionized bubbles start to expand, $\Gamma_{\rm HI}$ in the ionized regions begins to decrease as the UV photons reach lower density gas and are spread over a larger volume.  The photoionization rate decreases until $z\sim7$ when the ionized bubbles begin to overlap (see Figure~\ref{reionmaps}).  Once overlap occurs, the mean free path increases and $\Gamma_{\rm HI}$ begins to increase again.  

\begin{figure}
\centerline{\includegraphics[trim={0 1.57cm 0 1.57cm},clip]{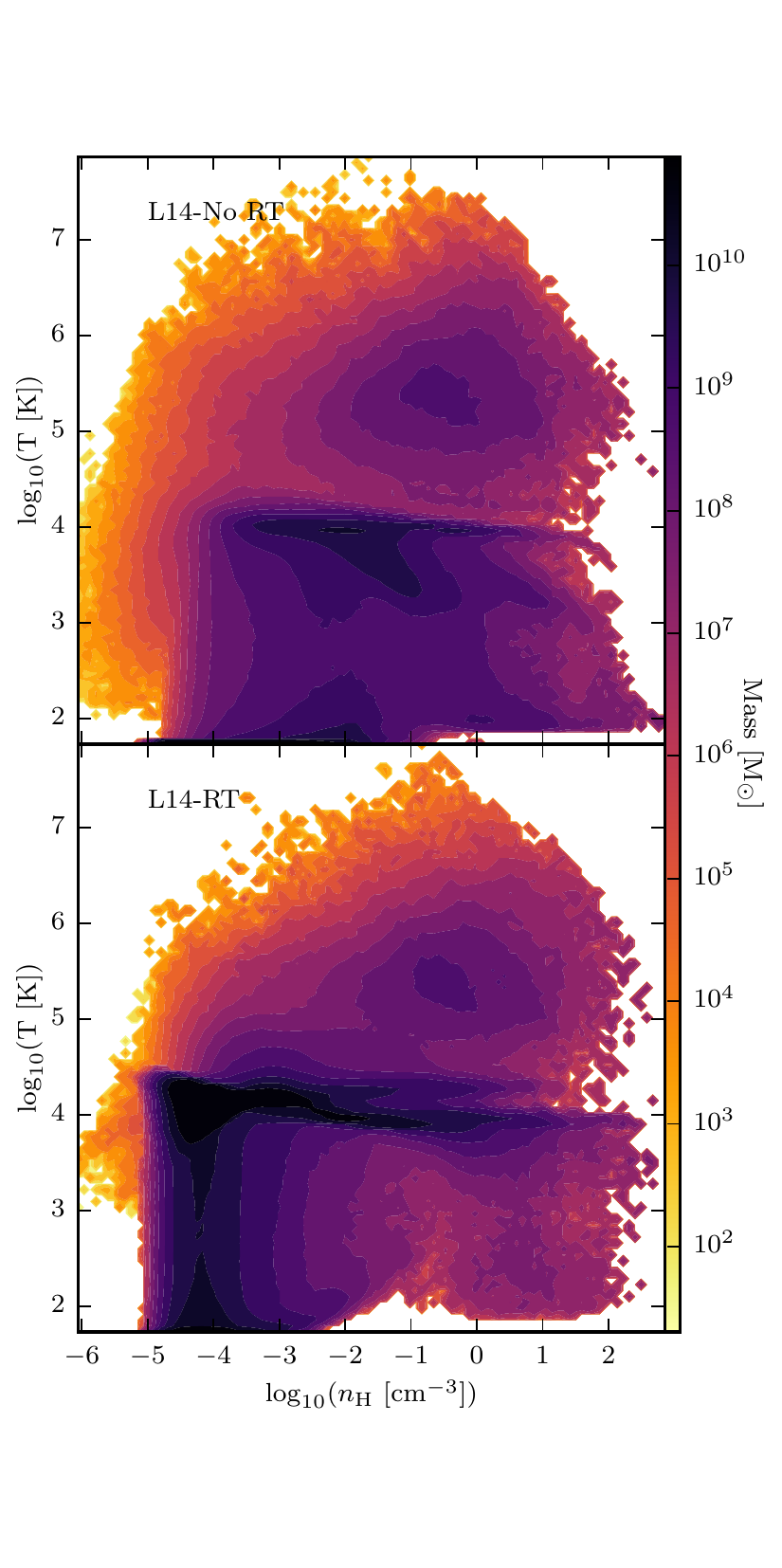}}
\caption{2D histograms of density versus temperature for our two highest resolution simulations at $z=6$.  The upper panel is for the L14-No RT simulation while the lower panel represents the L14-RT simulation.  Very little low-density ($n_{\rm H}<10^{-4}$cm$^{-3}$) gas has temperature $T\geq10^4$K in the L14-No RT simulation because all of the gas in the IGM is neutral.  At higher densities, we see much more similar behaviour between the two simulations with a plume of gas at $10^5<{\rm T\ [K]}<10^6$ due to SN feedback, and gas at $T<10^4$K due to radiative cooling.}
\label{rho_t}
\end{figure}

\begin{figure*}
\centerline{\includegraphics[]{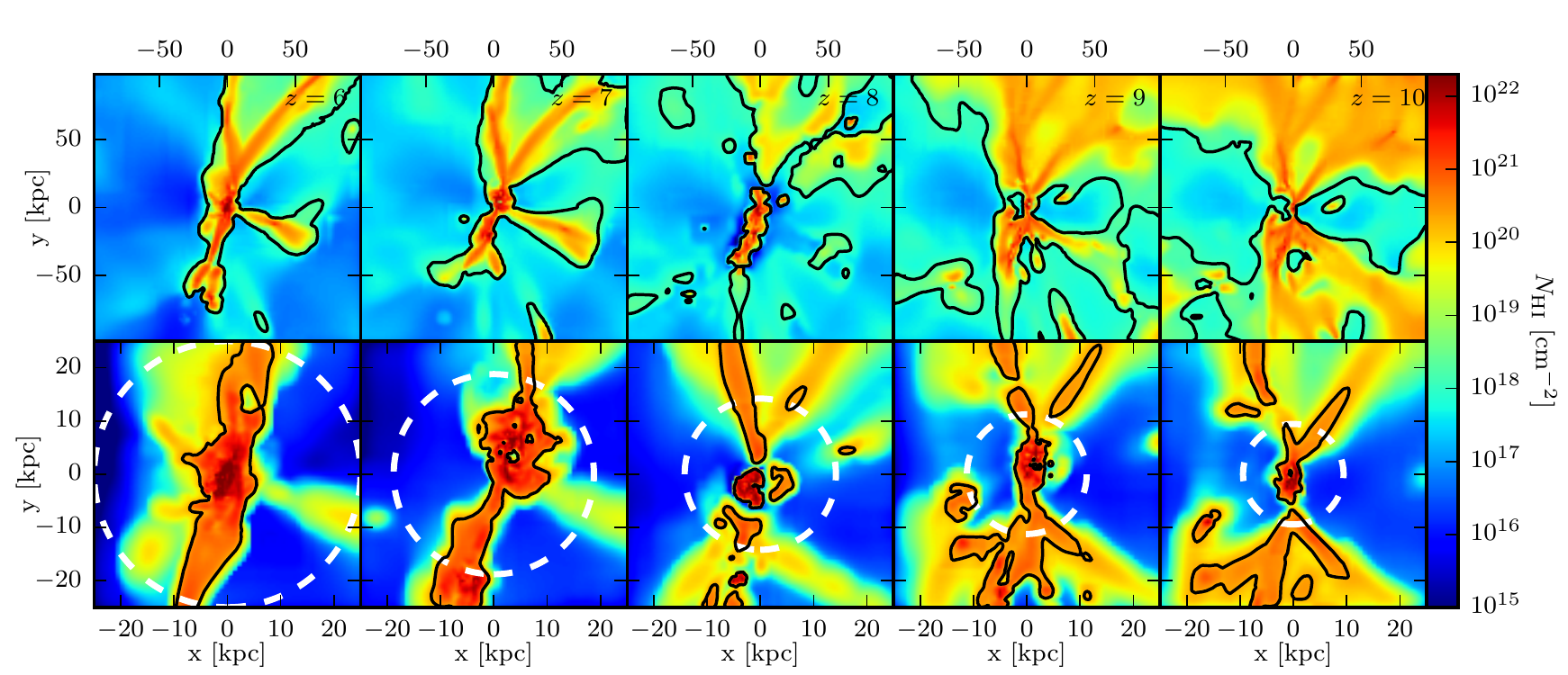}}
\caption{Time series from $z=10$ to $z=6$ of the column density of neutral hydrogen for the most massive halo in the simulation.  Dashed white circles in the bottom panel indicate the extent of the virial radius of the halo.  Black contours in the top panel indicate a column density of $10^{17}\ {\rm cm^{-2}}$ to highlight LLSs.  In the bottom panel, the black contours mark a column density of $2\times10^{20}\ {\rm cm^{-2}}$ to highlight regions causing DLAs.  LLSs tend to map out the filaments between haloes and get wider with increasing redshift while DLAs are only present within the virial radii of haloes.  Both the top and bottom panels are integrated along 50 physical kpc.}
\label{BH_DLA}
\end{figure*}

\subsubsection{Intensity in the Lyman-Werner Band}
In addition to HI ionizing radiation, we study the evolution of the Lyman-Werner background which is important for setting the intergalactic H$_2$ abundance and inhibiting gas cooling in low-metallicity systems.  While the simulation is not yet completely optically thin to hydrogen-ionizing UV radiation at $z=6$, it is completely optically thin to photons in the Lyman-Werner band.  The only two absorbers of Lyman-Werner photons in the box are H$_2$ and dust.  Nevertheless, as we discuss later, our simulations destroy H$_2$ more efficiently than one might expect, possibly due to the moderate resolution of our simulations.  The metallicity in most of the IGM remains at the level $Z=10^{-3.5}Z_{\odot}$ to which it was set at $z=15$.  The left panel of Figure~\ref{J21}, shows the spatial distribution of the intensity of the radiation in the Lyman-Werner band for a $\sim200$kpc region ($1/10^{\rm th}$ the physical box size) in the simulation at $z=6$ while the middle panel shows the surface density of the gas.  The Lyman-Werner intensity ($J_{\rm LW}$) is fairly constant in the IGM and the peaks in $J_{\rm LW}$ are clearly associated with collapsed structures with correspondingly high surface densities.   For gas below the cosmic mean baryonic density, $J_{\rm LW}\sim9.5\times 10^{-21}$erg/s/cm$^{2}$/Hz/sr.  The amplitude slowly decreases with increasing redshift as it follows the integrated star formation rate.  

The right panel of Figure~\ref{J21}, shows the radial profile of $J_{\rm LW}$ for the most massive galaxy in our simulation at $z=6$ for three different resolutions.  In the central regions, $J_{\rm LW}$ is elevated a few orders of magnitude above the background due to the presence of young stars in the galaxy.  For the highest and lowest resolutions, the profile is relatively flat out to a few kpc before decreasing, while for our middle resolution a flat profile is not seen.  The shape of this profile is entirely determined by the distribution of young stars within the galaxy which is reasonably converged between the three different resolutions; however, the middle resolution simulation in this snapshot has a slightly lower SFR compared to the other two simulations.

Since the simulation box is optically thin to Lyman-Werner radiation, any version of RSLA where the photon velocity is decreased significantly will lead to long lags in changes of the Lyman-Werner intensity as the fronts are expanding.  These fronts, in principle, will move at a velocity close to the full speed of light.  A strong burst of star formation in a massive galaxy will raise the Lyman-Werner background locally well above the mean in the IGM.  This, then, should propagate to nearby haloes at the full speed of light and affect the formation of H$_2$.  It is thus important to use the full speed of light or VSLA in order to transport these feedback effects at the correct speed.  

\subsubsection{Thermal State With and Without RT}
Including the radiation in the simulation has significant effects on the temperature of the gas.  In Figure~\ref{rho_t}, we show the density-temperature phase-space diagrams for the two highest resolution simulations at $z=6$.  It is clear that the temperature-density relation is very different with the inclusion of stellar radiation.  Most notably, at low densities ($-5<\log_{10}(n_{\rm H})<-4$) much of the gas in the L14-RT simulation sits at $T\sim10^4$K due to the ionization of hydrogen while for the L14-No RT simulation, the majority of the gas has $T<10^2$K.  Without ionizing radiation, the gas will simply adiabatically cool as it expands and remain at low temperatures.  

Note that the evolution of gas in the density-temperature phase-space in the L14-RT simulation is rather different from a simulation run with a spatially uniform meta-galactic UV background such as those prescribed in \cite{Haardt2012} and \cite{FG2009}.  With a spatially uniform background, the low-density regions will be ionized first and thus, once the UV background turns on, the temperature of the lowest density gas will jump to $\sim10^4$K.  In our L14-RT simulation, the ionized bubbles begin forming in the highest density regions around the haloes while the voids are ionized last.  

\begin{figure*}
\centerline{\includegraphics[]{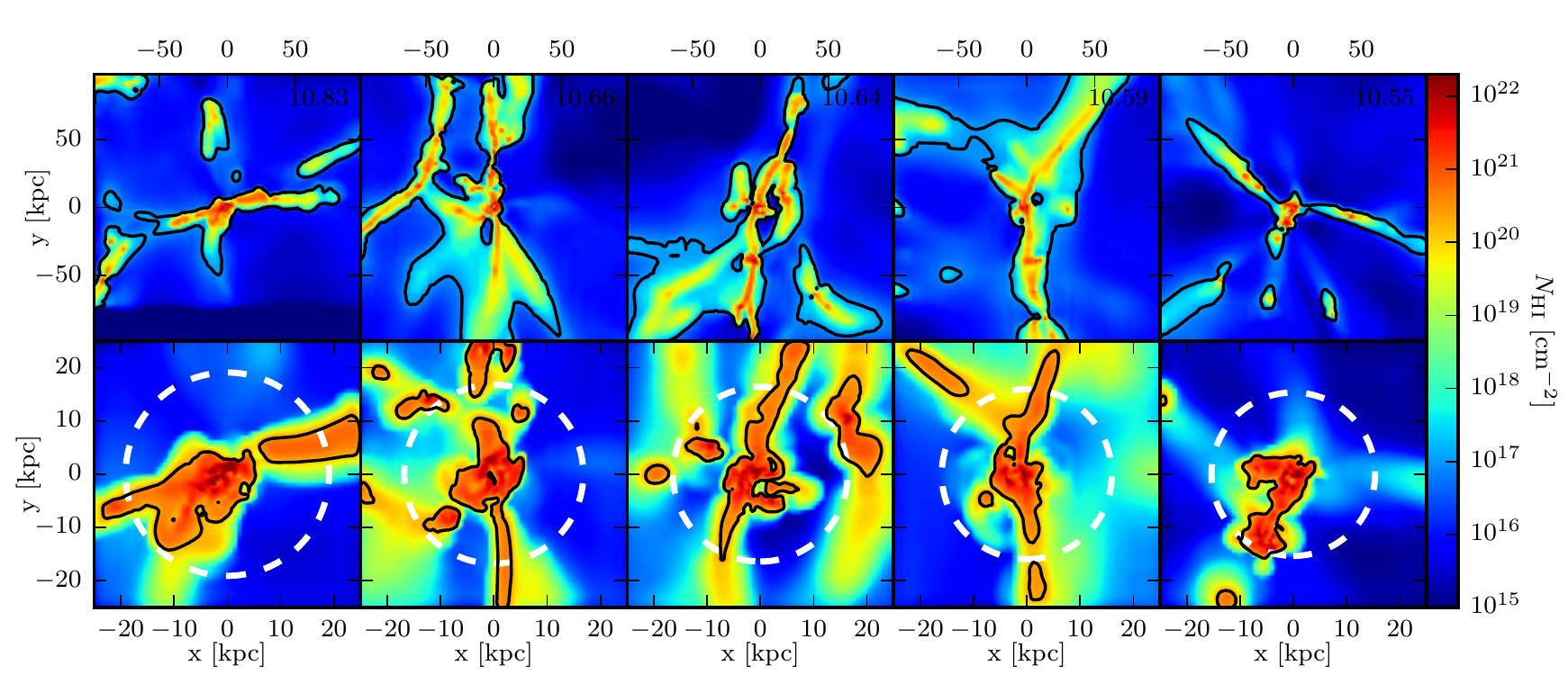}}
\caption{Column density of neutral hydrogen for the next five most massive haloes in the simulation at $z=6$.  Dashed white circles in the bottom panel indicate the extent of the virial radii of the haloes.  Black contours in the top panel indicate a column density of $10^{17}\ {\rm cm^{-2}}$ to highlight LLSs.  In the bottom panel, the black contours mark a column density of $2\times10^{20}\ {\rm cm^{-2}}$ to highlight regions causing DLAs.  The numbers listed in the top row represent $\log_{10}({\rm M_{DM}\ [M_{\odot}]})$ for that object.  The morphology of DLAs and LLSs are very dependent on the system where they are located.  Both the top and bottom panels are integrated along 50 physical kpc.}
\label{AH_DLA}
\end{figure*}

\subsection{Intermediate Scales - Feeding the Circumgalactic Medium from the Cosmic Web}
Characterising the ionization state and metallicity of the circumgalactic gas which is feeding the galaxies will help in interpreting the systems which appear in quasar absorption spectra. The top panel of Figure~\ref{BH_DLA} shows a time series of a 200kpc region of the surface density of neutral hydrogen around the most massive halo from $z=10$ to $z=6$.  At $z=6$, the halo is fed by three dense, mostly neutral filaments which move with the evolving galaxy.  Neutral gas moving along these filaments and penetrates deep into the centre of the halo before merging with the central galaxy.  The filament feeding the galaxy from the bottom is the most dynamic as a major merger is occurring along this direction.  At $z=7$, a large over-density of neutral gas can be seen just outside the virial radius of the central halo.  By $z\sim6.5$, the satellite galaxy has fully entered the virial radius and the merger is complete by $z=6.0$.  

Contours in the top panel of Figure~\ref{BH_DLA} highlight $N_{\rm H}=10^{17}$${\rm cm^{-2}}$ where the gas is expected to self-shield from external ionizing radiation.  Lines of sight passing inside of these contours will result in Lyman-limit absorption systems (LLSs).  This column density threshold encapsulates most of the filamentary structure as well as the central galaxy consistent with \cite{Fumagalli2011} who find that the streams feeding galaxies remain optically thick in hydrodynamics simulations post processed with radiation transfer.  The fraction of the surface area inside the virial radius of the central galaxy which is covered by neutral gas at this surface density changes quite substantially between $z=10$ and $z=6$.  The covering fraction in our simulations at $z=6$ is higher than the 25\% expected from \cite{Fumagalli2011} at $z=3$ and much closer to the expectations for simulated massive haloes at $z=4$ from \citep{Rahmati2015}.  The covering fraction is expected to increase with redshift \citep[e.g.][]{Cen2012,Rahmati2015}.   At $z=6$, more than $50\%$ of the cross-section enclosed by the virial radius is covered by LLSs while at $z=8$, this fraction is significantly smaller.  At this redshift, there is a strong deficit in neutral gas suggesting that a strong burst in star formation has occurred within the past 10Myr.  Figure~\ref{BH_DLA}, shows that the width of the filaments which are above a HI column density of $N_{\rm H}=10^{17}$${\rm cm^{-2}}$ strongly increases towards high-redshift as the photoionization rate in the IGM is decreasing (see Figure~\ref{SFRD}).  By $z=10$, the majority of the gas in a 200 physical kpc cube surrounding this galaxy would results in a LLS. Note however, that at $z\ga8$ gas is still sufficiently neutral that LLS are very extended and not well defined.

Figure~\ref{AH_DLA} shows the HI column density in the 200kpc regions around the next five most massive haloes in the simulation.  Some of the haloes are only being fed by two filaments, while other haloes are fed by three or more.  Some of the filaments are fragmented and others are clear remnants of mergers.  The structure strongly varies between the different haloes.

In the bottom rows of Figures~\ref{BH_DLA} and \ref{AH_DLA}, we zoom in closer to the haloes and highlight the $N_{\rm HI}=2\times10^{20}$${\rm cm^{-2}}$ threshold for damped Lyman-$\alpha$ systems (DLAs).  The regions predicted to cause DLAs can be fragmented and generally have very asymmetric shapes with many of these systems coming in along the filaments and potentially extending out to the virial radius, consistent with \cite{Fumagalli2011,Cen2012}.  There are holes in the distribution of neutral gas as a result of photoionization from young stars and SN blowing out the gas.  Star formation occurs preferentially in the highest density regions at the centres of haloes and this energetic feedback is very efficient at disrupting the HI.  

\begin{figure*}
\centerline{\includegraphics[]{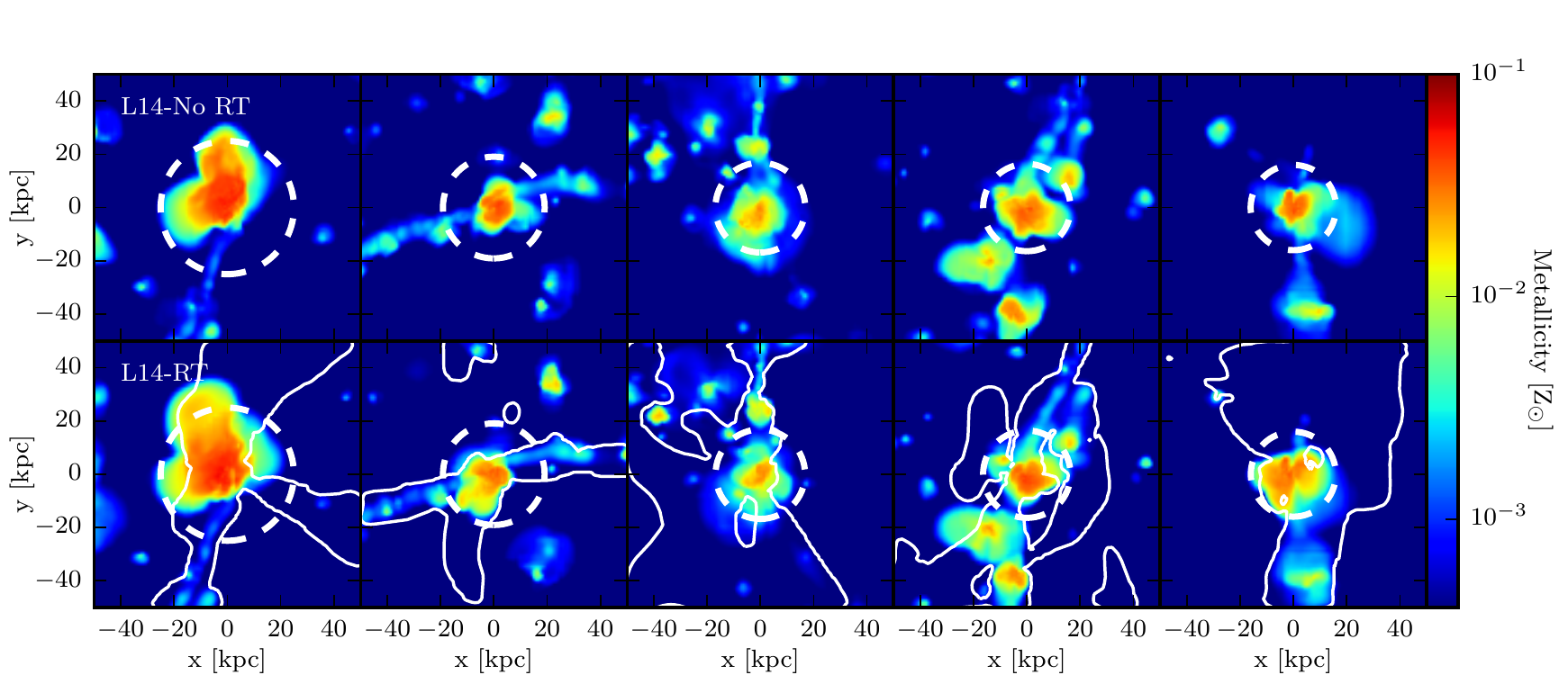}}
\caption{Spatial distribution of metallicity for the 100kpc region centred on the five most massive haloes in the L14-No RT (top) and L14-RT (bottom) simulations at $z=6$.  The integration along the line-of-sight (50 physical kpc) has been weighted by density.  The most massive halo is shown in the left panels and the halo mass decreases in each subsequent panel.  The virial radius of each galaxy is indicated as a white circle.  The white contours in the bottom row outline neutral gas with a column density $>10^{17}$cm$^{-2}$.  Most of the metals tend to remain inside the virial radius of the haloes.  The locations where the filaments have been enriched above the metallicity floor generally correspond to a smaller galaxy which has formed along the filament.  However, some filaments remain unenriched.}
\label{metalmaps}
\end{figure*}

We turn now to the spatial distribution of metals in our simulations which can be probed by associated metal absorption lines in QSO spectra during the epoch of reionization at $z>6$ \citep{Ryan2009,Becker2011,Simcoe2011,Dodorico2013}.  In Figure~\ref{metalmaps}, we show the density-weighted spatial distribution of the metallicity surrounding the five most massive haloes in the L14-RT and L14-No RT simulations at $z=6$.  The metallicity inside the enriched bubbles is relatively uniform in the inner regions and remains rather low compared to solar.  For the most massive halo, the metallicity reaches a maximum of a few~$\times10^{-1}\ {\rm Z_{\odot}}$, and is somewhat smaller for the less massive haloes.  Interestingly, very little of the metals actually penetrate out of the virial radius and into the IGM and this is similar to the results of \cite{Ma2016} at $z=0$ where their haloes of this mass in their simulations have retained nearly 100\% of their metals.  However, much of the circumgalactic medium CGM (in cross-section) is enriched to values slightly above the metallicity floor.  The gas feeding the galaxies along the filaments appears to be close to primordial metallicity.  Where the metallicity of the filament is raised above the metallicity floor, this is due to smaller haloes which have formed stars and are accreting along the filament.  Note again that we do not resolve the small mini-haloes which may pre-enrich the IGM at earlier epochs \citep{Wise2012} so the exact metallicity of the filaments is not reliably predicted by our simulations.  In the simulations with and without stellar radiation, the distribution of metals and stellar mass are similar however which is consistent with the expectations of \cite{Okamoto2008} where it was shown that the galaxy filtering mass is likely smaller than the resolution limit of DM haloes in our simulations.

\subsection{Galactic Scales - From the Circumgalactic Medium to Stars and the ISM}
\label{sec:ss}
We will now focus our attention on the most massive galaxy in the simulation, which is the best resolved object, and discuss the properties of its ISM.

\subsubsection{Spatial Distribution of Young, Middle, and Old Aged Stars}
\begin{figure*}
\centerline{\includegraphics[]{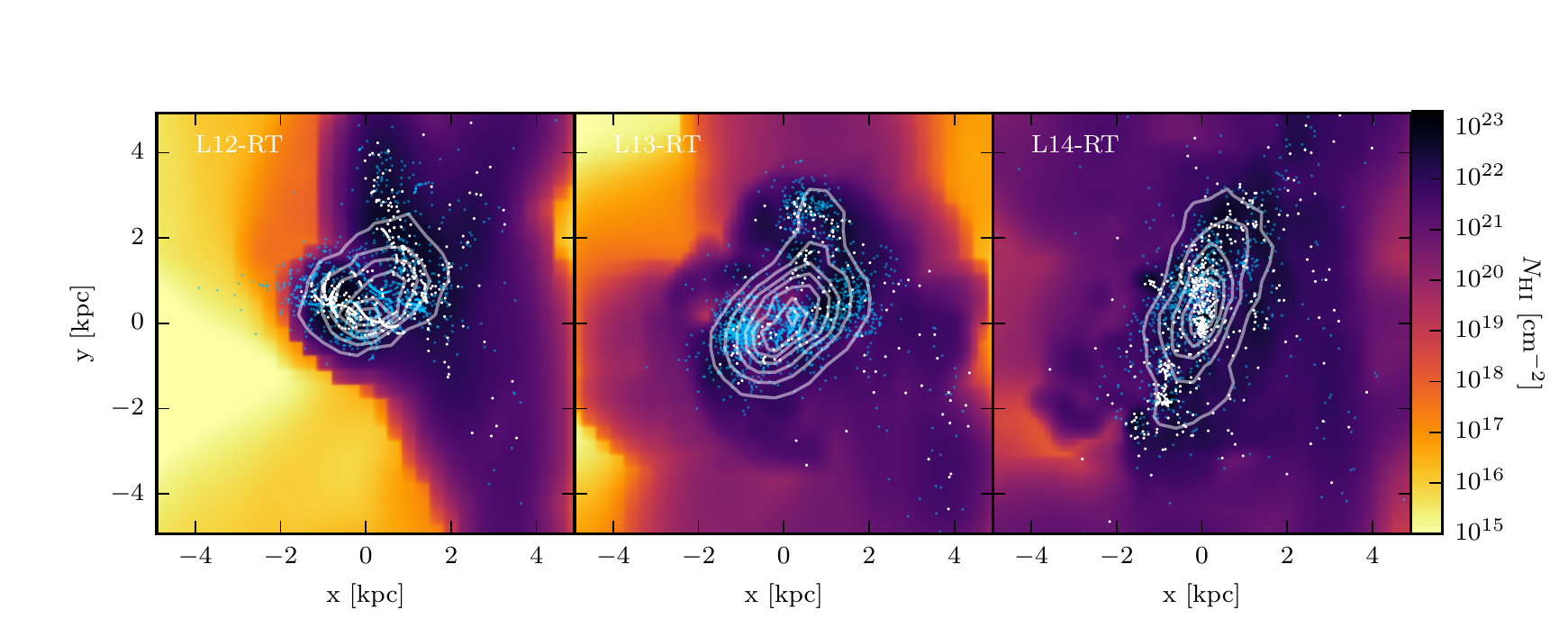}}
\caption{Distribution of young (white), middle age (blue) and old (white contours) stars for the three different simulations which include stellar radiation.  The underlying coloured map shows the neutral hydrogen column density.  The young stars tend to form in clumps while middle aged stars have spread out more.  The old stellar population forms a smooth distribution coincident with the dense gas.}
\label{spat_star}
\end{figure*}

At high redshifts, galaxies are predominantly identified through their UV emission \citep[e.g.][]{Bouwens2015}, and here we study the different stellar populations which give rise to the emission at different wavelengths.  We identify three specific groups of stars: young stars (${\rm age}<10$Myr), middle aged stars ($10{\rm Myr}<{\rm age}<30$Myr) and old stars (${\rm age}>30$Myr).  The first group represents the stars that have yet to go SN, the second group represents the stars which have gone SN recently and the remnants of the explosions are keeping the gas at a high temperature since we delay cooling for $20$Myr, and the final group represents the passive stellar population which are remnants of previous episodes of star formation.  In Figure~\ref{spat_star}, we show the surface density of these three populations at $z=6$ for the most massive halo for the three different resolution simulations which include stellar radiation.  While all three populations are centred around the same location, the young stellar population exhibits a clumpy distribution, the middle aged population is slightly more spread out, while the spatial distribution of the old population is very smooth.

Once the stars have gone SN, the density of the host gas cells decreases as matter is expelled out of the central regions of the galaxy.  This decreases the local potential well allowing the middle aged stars to spread out and respond to the overall gravitational potential well of the galaxy.   

\subsubsection{Time Evolution of Star Formation}
\begin{figure*}
\centerline{\includegraphics[]{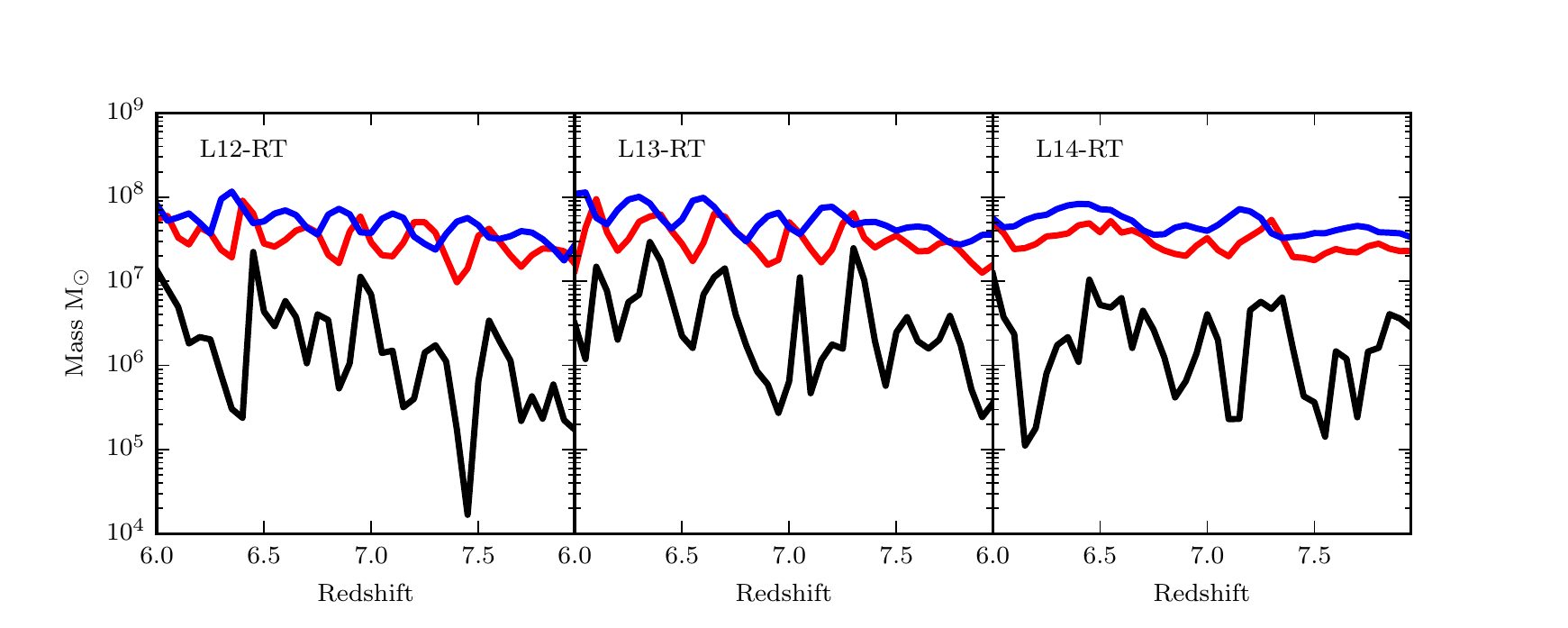}}
\caption{H$_2$ mass (black), the mass of young stars (red), and the mass of middle aged stars (blue) as a function of redshift in the three different simulations which include stellar RT.  The formation of H$_2$ is synched with the formation of young stars as both tend to form in dense neutral gas.  Because of strong radiation feedback, the fluctuations in H$_2$ mass are anti-correlated with those in the mass of middle aged stars.}
\label{H2ofz}
\end{figure*}

Observations of high-redshift galaxies only provide an instantaneous snapshot of the current state of a galaxy.  Understanding these observations therefore requires understanding duty cycle of star formation in these systems which puts limits on a galaxy's observability at high-redshift.  In Figure~\ref{H2ofz}, we show how the mass in young and middle aged stellar populations changes as a function of redshift between $z=8$ and $z=6$ for the three simulations with different resolutions which include stellar radiation.  For the two lowest resolutions, there is a clear sinusoidal pattern in the mass of both young and middle aged stars where the peaks have a separation of $\Delta z\approx0.25$, corresponding to a timescale of $\sim30-50$Myr in this redshift range.  For the young stars, the timescale between first star formation and when the last SN remnants allow the gas to cool again is $\sim40$Myr\footnote{10Myr for the timescale between first stars forming plus 20Myr for delayed cooling plus an additional 10Myr because young stars will form during the 10Myr before the first stars have undergone SN and thus the final SN explode 10Myr after the first.} which corresponds well with the time scale of this sinusoidal pattern.  The peaks in middle aged stellar mass are somewhat offset from the peaks in young stellar mass.  This is simply because the middle aged stellar mass increases as the young stars age into this category.  SN feedback from the young stars will prevent more stars from forming and thus decrease the young stellar mass while the middle aged stellar mass increases.    

In the highest resolution simulation, much of the prominent sinusoidal pattern in the young and middle aged stellar mass is not evident.  This is likely due to the fact that as the resolution increases, the gas can reach higher densities and fragment more.  In the two lower resolution simulations, the distribution of the gas is dominated by only a few large clumps.  Once SN explode in these clumps, it is as if the entire galaxy becomes regulated by these episodic star formation events.  As more and more clumps form at higher resolution, disrupting them all simultaneously becomes difficult and thus the pattern becomes washed out.  The right panel of Figure~\ref{H2ofz} shows that the similar features we see at the two lower resolutions are present at the highest resolution as well.  In particular, there is a clear peak in young stellar mass at $z\sim7.3$.  As the young stellar mass decreases, the middle aged stellar mass increases as expected.  There are obvious other features such as the wide peak in young stellar mass spanning $z=6.8-6.2$ which is similarly followed by a delay in the middle aged stellar mass.

These patterns can be recognised in the spatial distribution in Figure~\ref{props_time}, where we show surface density maps of the young stellar mass, middle aged stellar mass, and old stellar mass as a function of redshift from $z=6.8-6.0$.  A prime example of this is the large hole in the distribution of young stars at $z=6.8$ due to an earlier burst of star formation.  A trough in the young stellar mass can be seen at this redshift in the right panel of Figure~\ref{H2ofz}.  By $z=6.6$, this hole is once again filled with young stars and at this redshift, there is a peak in the young stellar mass in the right panel of Figure~\ref{H2ofz}.  It is evident from this figure that both the mass and distribution of young stars in the galaxy, which are the primary sources of UV and Lyman-$\alpha$ radiation, are both dynamic in spatial distribution and in mass over short time scales.  Looking at the old stars, we do not see much of a change in the distribution as a function of redshift.  In this redshift range, the total stellar mass of the galaxy increases from $4.3\times10^8$M$_{\odot}$ to $8\times10^8$M$_{\odot}$.  Given the large dynamic range of the colour bar in Figure~\ref{props_time} a small change will not be visible in the distribution and therefore the old stellar mass looks relatively constant. 

\subsubsection{Spatial Variations in the Intensity of Radiation}
\label{radss}
Star formation will leave a unique signature on the radiation field within the galaxy as young stars are the primary emitters of ionizing photons, both at $E>13.6$eV and in the Lyman-Werner band. In Figure~\ref{props_time}, we characterize the spatially inhomogeneous radiation field which is key for predicting the ionization states of metals whose emission can be observed by ALMA.  We show $\Gamma_{\rm HI}$ and the intensity in the Lyman-Werner band ($J_{\rm LW}$) as a function of redshift between $z=6.8-6.0$.  Comparing the two quantities, one can see a stark contrast.  At $z=6$, $\Gamma_{\rm HI}$ has a strong peak in the centre of the halo corresponding to the highest density in young stars.  This quickly fades by a few orders of magnitude even at very small radii although small secondary peaks in $\Gamma_{\rm HI}$ can be seen throughout the image.  These are due to secondary star formation sites which have less mass in young stars compared to the central regions.

By contrast the intensity of the radiation in the Lyman-Werner band is extremely smooth at $z=6$.  There is a clear central peak in the radiation field and this intensity falls off as $r^{-2}$ at larger radii (see the right panel of Figure~\ref{J21}).  This difference arises because of the nature of the absorbers in each band.  Photons with $E>13.6$eV are readily absorbed by neutral hydrogen.  The mass- (volume-) weighted mean free path for hydrogen ionizing photons is $\sim500$kpc ($\sim11$Mpc) inside the virial radius of the halo while for the Lyman-Werner band, we find a significantly longer mass- (volume-) weighted mean free path of $\sim700$Mpc ($\sim1.2$Gpc).

\subsubsection{Time Evolution of the Thermal and ionization State of the ISM}
\begin{figure*}
\centerline{\includegraphics[trim={0 0 0.2cm 0},clip]{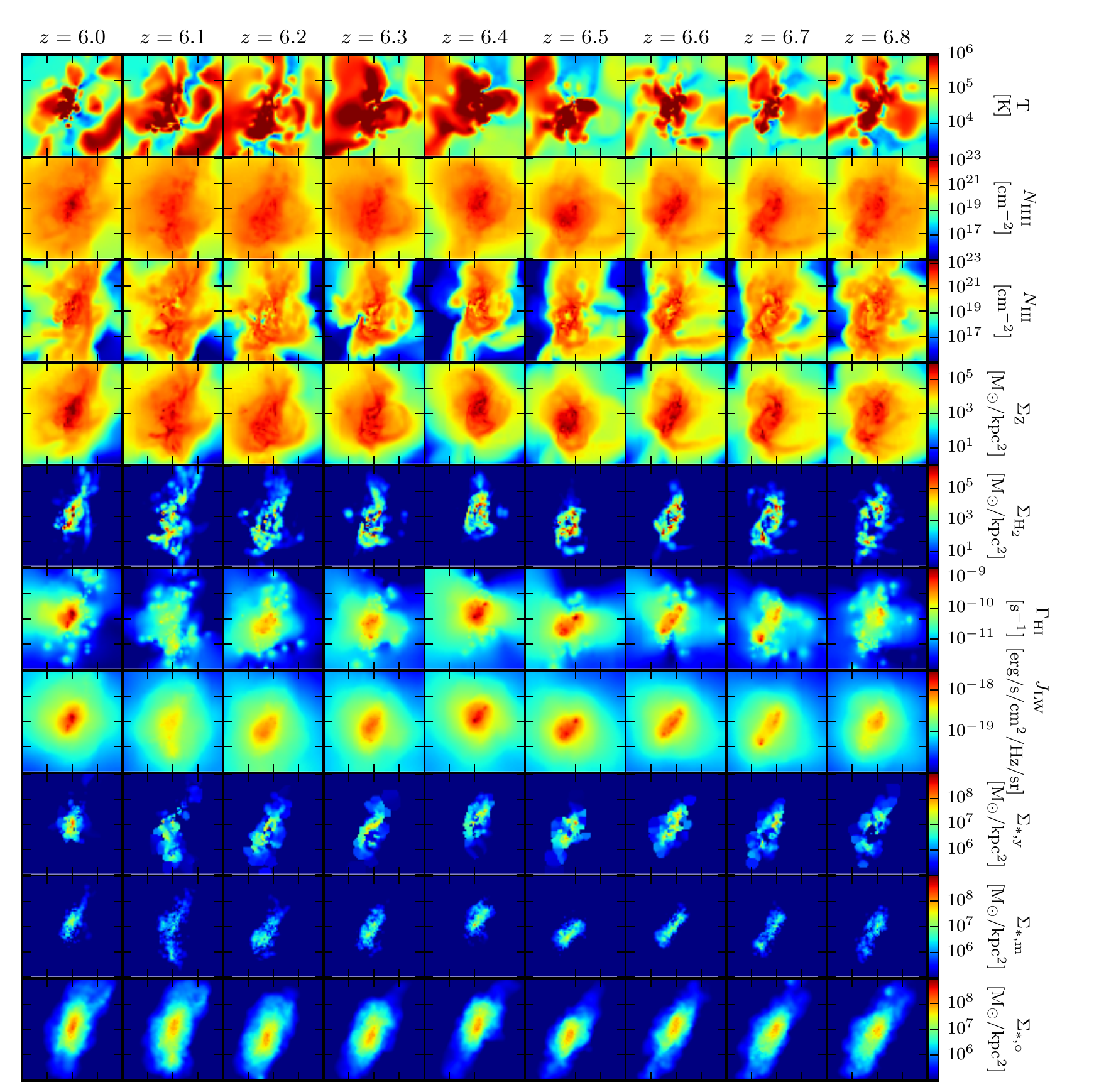}}
\caption{Time series of various physical properties of the most massive galaxy in the L14-RT simulation shown from $z=6.0$ to 6.8 in intervals of $\Delta z=0.1$  starting from the left.  Each image is 20 proper kpc in width.  From  top to bottom, the quantities are temperature (slice), HII column density, HI column density, metal surface mass density, H$_2$ surface density, $\Gamma_{\rm HI}$ (slice), $J_{\rm LW}$ (slice), young stellar mass surface density, middle age stellar mass surface density, old stellar mass surface density.  Properties shown as slices represent a small section of the simulation box centred on the middle of the halo.}
\label{props_time}
\end{figure*}

The radiative and SN feedback significantly affects the ionization state and temperature of the gas and here we show how these properties correlate with star formation.  The top panel of Figure~\ref{props_time}, shows the time evolution of temperature.  A large ionization bubble exists around this galaxy (see Figure~\ref{reionmaps}) so it is unsurprising to see gas at $T\sim10^4$K surrounding the galaxy.  The temperature changes from $10^3$K to $10^6$K and the peaks in temperature correspond to peaks in middle aged stellar mass (i.e. compare middle age star surface mass density to temperature in the $z=6.3$ and $z=6.7$ columns of Figure~\ref{props_time}).  

The second and third rows of Figure~\ref{props_time} show the surface density of ionized and neutral gas as a function of redshift.  Towards the outskirts of the galaxy, there is significantly more ionized gas compared to neutral gas.  This gas is diffuse and the recombination timescale is large and thus the gas is maintained in an ionized state (hence why the temperature is $\sim10^4$K in these regions).  A minimum in the surface density of ionized gas can be seen between $z=6.1$ and $z=6.3$ which is associated with the decline in young stellar mass (see Figure~\ref{H2ofz}).  The intensity of ionizing radiation is sensitive to the presence of young stars (see Section~\ref{radss}) and in particular, stars with ${\rm age}\lesssim3$Myr as the emission of ionizing photons drops quickly for stars past this age \citep{Bruzual2003}. 

In the neutral gas, we see a markedly different structure compared to that in the ionized gas.  The dynamic range is significantly higher and many more strong spatial features are present.  The neutral gas distribution changes rapidly.  We see many examples where holes appear in the distribution due to star formation and SN feedback.

In Table~\ref{halostats}, we list the physical properties for the ten most massive haloes in the L14-RT simulation at $z=6$.  In 9 of 10 haloes, the HII mass is dominant over the HI mass indicating that these haloes are sufficiently ionizing their surroundings.  The metal mass is spread fairly proportionally over the neutral and ionized gas consistent with the mass of these two quantities (i.e. for the most massive halo 32\% of the total metal mass is associated with neutral gas while 34\% of the total gas is in the neutral phase).  Most haloes have retained their cosmic baryon fraction of gas, consistent with the expectations from \cite{Okamoto2008,Gnedin2014b} for the halo masses considered here.  The feedback from radiation and photoionization does not seem to be strongly affecting the accretion of gas onto these massive haloes.  Most of the gas is being fed through cold, neutral, dense filaments (see Figure~\ref{AH_DLA}) and the outflows are not preventing fresh gas from penetrating down to the centres of the haloes \citep{Kimm2015}.  

\begin{table*}
\centering
\begin{tabular}{@{}lccccccccc@{}}
\hline
 Halo & ${\rm M_{DM}}$ & ${\rm M_{*}}$ & ${\rm M_{gas}}$ & ${\rm M_{Z,HI}}$ & ${\rm M_{Z,HII}}$ & ${\rm M_{HI}}$ & ${\rm M_{HII}}$ &  ${\rm M_{H_2}}$ & SFR \\
 & $\log_{10}({\rm M_{\odot}})$ & $\log_{10}({\rm M_{\odot}})$ & $\log_{10}({\rm M_{\odot}})$ & $\log_{10}({\rm M_{\odot}})$ & $\log_{10}({\rm M_{\odot}})$ & $\log_{10}({\rm M_{\odot}})$ & $\log_{10}({\rm M_{\odot}})$ &  $\log_{10}({\rm M_{\odot}})$ & ${\rm M_{\odot}\ yr^{-1}}$\\
\hline
1 & 11.17 & 9.08 & 10.43 & 6.73 & 7.06 & 9.85 & 10.13 & 6.78 & 4.91\\
2 & 10.83 & 8.51 & 10.08 & 6.25 & 6.45 & 9.56 & 9.74 & 5.76 & 2.12 \\
3 & 10.66 & 8.20 & 9.91 & 5.90 & 6.20 & 9.34 & 9.61 & 3.38 & 0.72 \\
4 & 10.64 & 8.37 & 9.92 & 5.97 & 6.42 & 9.29 & 9.64 & 3.54 & 0.69 \\
5 & 10.59 & 8.26 & 9.86 & 5.78 & 6.34 & 9.20 & 9.59 & 3.98 & 0.47 \\
6 & 10.55 & 8.19 & 9.81 & 5.73 & 6.24 & 9.14 & 9.55 & 4.31 & 0.68 \\
7 & 10.51 & 7.96 & 9.76 & 5.64 & 5.96 & 9.17 & 9.46 & 3.40 & 0.53 \\
8 & 10.48 & 8.24 & 9.72 & 5.71 & 6.26 & 8.96 & 9.49 & 5.28 & 0.93 \\
9 & 10.48 & 8.11 & 9.71 & 5.70 & 6.10 & 9.10 & 9.42 & 3.69 & 0.63 \\
10 & 10.41 & 8.18 & 9.73 & 6.05 & 6.06 & 9.32 & 9.29 & 4.53 & 0.43 \\
\hline
\end{tabular}
\caption{Properties of the ten most massive haloes in the L14-RT simulation at $z=6$.  ${\rm M_{DM}}$, ${\rm M_{*}}$, and ${\rm M_{gas}}$ represent the dark matter mass, stellar mass, and gas mass in the haloes, respectively.  ${\rm M_{Z,HI}}$ and ${\rm M_{Z,HII}}$ are the metal masses associated with neutral and ionized gas.  ${\rm M_{HI}}$, ${\rm M_{HII}}$,  ${\rm M_{H_2}}$ are the masses in neutral hydrogen, ionized hydrogen, and molecular hydrogen.  SFR is the current star formation rate of the halo.}
\label{halostats}
\end{table*}

\begin{figure}
\centerline{\includegraphics[]{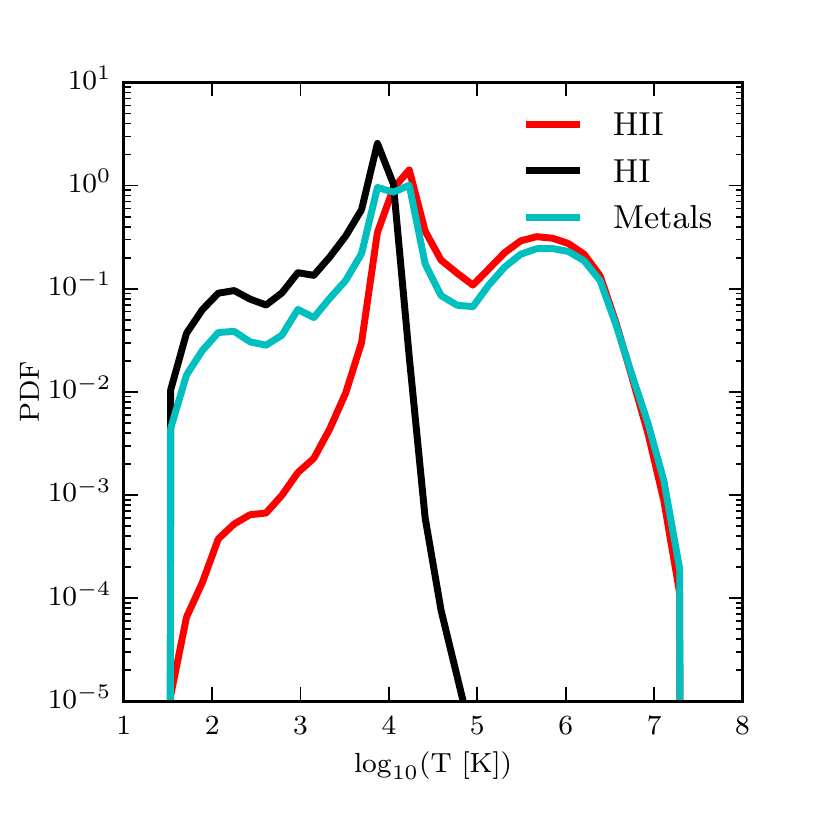}}
\caption{PDFs of the gas temperature of HI, HII and metals inside the virial radius of the most massive halo in the L14-RT simulation at $z=6$.  The majority of high temperature gas ($T>2\times10^4$K) is ionized while most of the gas at $T<2\times10^4$K is neutral.  The metals are proportionately spread between the two components by mass inside of the haloes.  Approximately $2/3$ of the metal mass is found in the ionized component while the other $1/3$ remains neutral.}
\label{gasPDF}
\end{figure}

For observations, it is imperative to know the temperature states of each of the different gas components inside the halo.  In Figure~\ref{gasPDF}, we show the probability distribution function (PDF) of the temperatures for HI, HII, and metals inside the virial radius of the most massive halo in the L14-RT simulation at $z=6$.  For HII, we see a double peaked profile: the lower temperature peak is due to photo-heating while the higher temperature peak is from SN feedback.  Hydrogen only efficiently recombines at $T<10^4$K and thus we only see HI at temperatures lower than this value.  The metals are very evenly spread across the galaxy and for this reason, we see that the temperature PDF of this component is a combination of both the HI and HII distributions.

\subsubsection{The Evolution of H$_2$ with and without RT}
\label{h2sec}
Of particular interest is H$_2$ because it is well known to correlate with star formation \citep{Kennicutt1998,Bigiel2008}.  Furthermore, in low metallicity environments, as is the case in many regions of our simulation box, H$_2$ becomes the primary coolant below $10^4$K and thus alters the thermodynamic state of the gas.

\begin{figure}
\centerline{\includegraphics[]{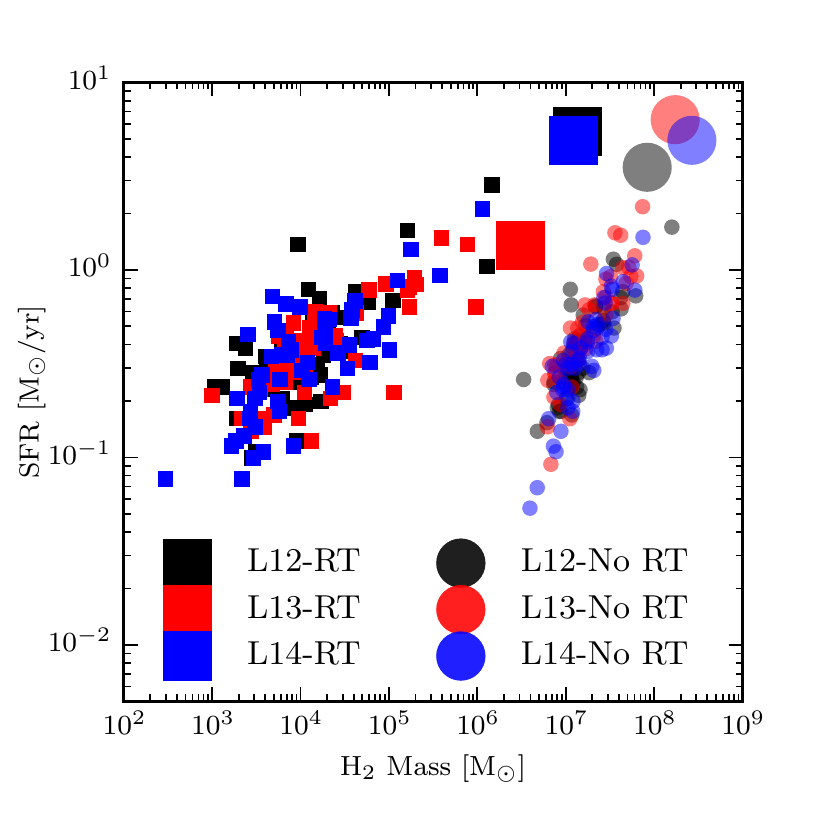}}
\caption{H$_2$ masses versus SFR for all haloes with M$_{\rm halo}>10^{10}$M$_{\odot}/h$ at $z=6$.  Simulations with and without RT are shown as squares and circles respectively.   Black, red, and blue points represent simulations refined to L12, L13, and L14.  The enlarged points represent the most massive halo in the box.  The scatter in the positive correlation between H$_2$ mass and SFR increases significantly when stellar radiation is included.}
\label{gasH2}
\end{figure}

In Figure \ref{gasH2}, for each of the six simulations, we plot the total mass of H$_2$ contained within the virial radius at $z=6$ against the star formation rate (SFR, measured by averaging over the past 10Myr) for all haloes with M$_{\rm halo}>10^{10}$M$_{\odot}/h$.  Without RT, the H$_2$ masses form a very tight, positive and steep correlation with the SFRs of the haloes.  For the best resolved halo (the three circles at the top right of Figure \ref{gasH2}), we see that the instantaneous SFR and H$_2$ masses are converged within a factor of $\sim2-3$ at $z=6$.  When RT is introduced into the simulations, the total mass of H$_2$ decreases drastically because of the presence of UV radiation and the scatter in the H$_2$-SFR relation significantly increases to about $\sim1$dex.  The range of SFRs exhibited by the haloes in the runs with RT is similar to those in the runs without \citep[e.g.][]{Rosdahl2015}.  

\begin{figure*}
\centerline{\includegraphics[]{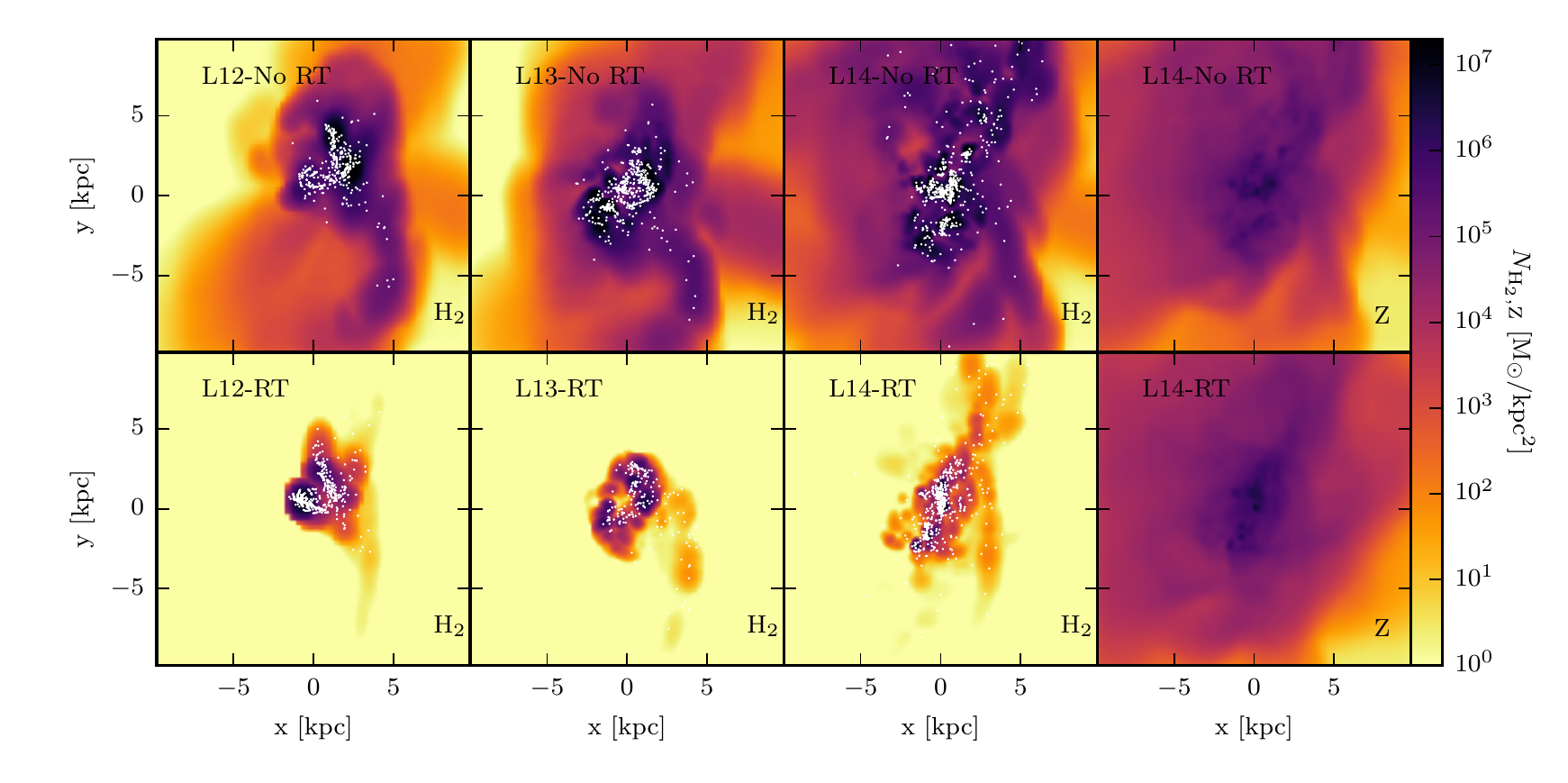}}
\caption{{\it First three columns}: Column density maps of H$_2$ in 20kpc cubes surrounding the most massive galaxy in our simulations at $z=6$.  The locations of the young stars ($t<10$Myr) are indicated as white points.  The young stars are the most likely sources of the UV and Ly$\alpha$ flux while the H$_2$ regions may be the dominant source of [CII] emission. {\it Fourth column}: Surface mass density of metals.  In both simulations, the locations of the young stars are coincident with the locations where H$_2$ has formed.  However, there is significantly less H$_2$ in the simulations which include stellar radiation due to strong Lyman-Werner feedback.}
\label{H2SD}
\end{figure*}

In Figure \ref{H2SD}, we show surface density maps of H$_2$ in the most massive halo for each of the six different simulations at $z=6$.  Moving from lower to higher resolution increases the amount of structure in the H$_2$ regions regardless of whether or not we include RT.  The differences between the H$_2$ surface density maps for a given halo are likely due to a number of reasons.  The H$_2$ abundance is sensitive to density, temperature, metallicity, and the local radiation field.  At higher resolutions, we resolve higher densities so, in principle, H$_2$ should be able to form slightly more efficiently in the most resolved cells of the highest resolution simulations.  Furthermore, the star formation algorithm is based on random Poisson sampling and for these low resolution simulations, we cannot expect perfect agreement as a function of resolution because of the different densities sampled.  We also cannot expect a perfect agreement inside a specific halo for the temperature which is once again dependent on the density of the gas and where the stars are located.

\begin{figure}
\centerline{\includegraphics[trim=0 1.57cm 0 1.57cm,clip]{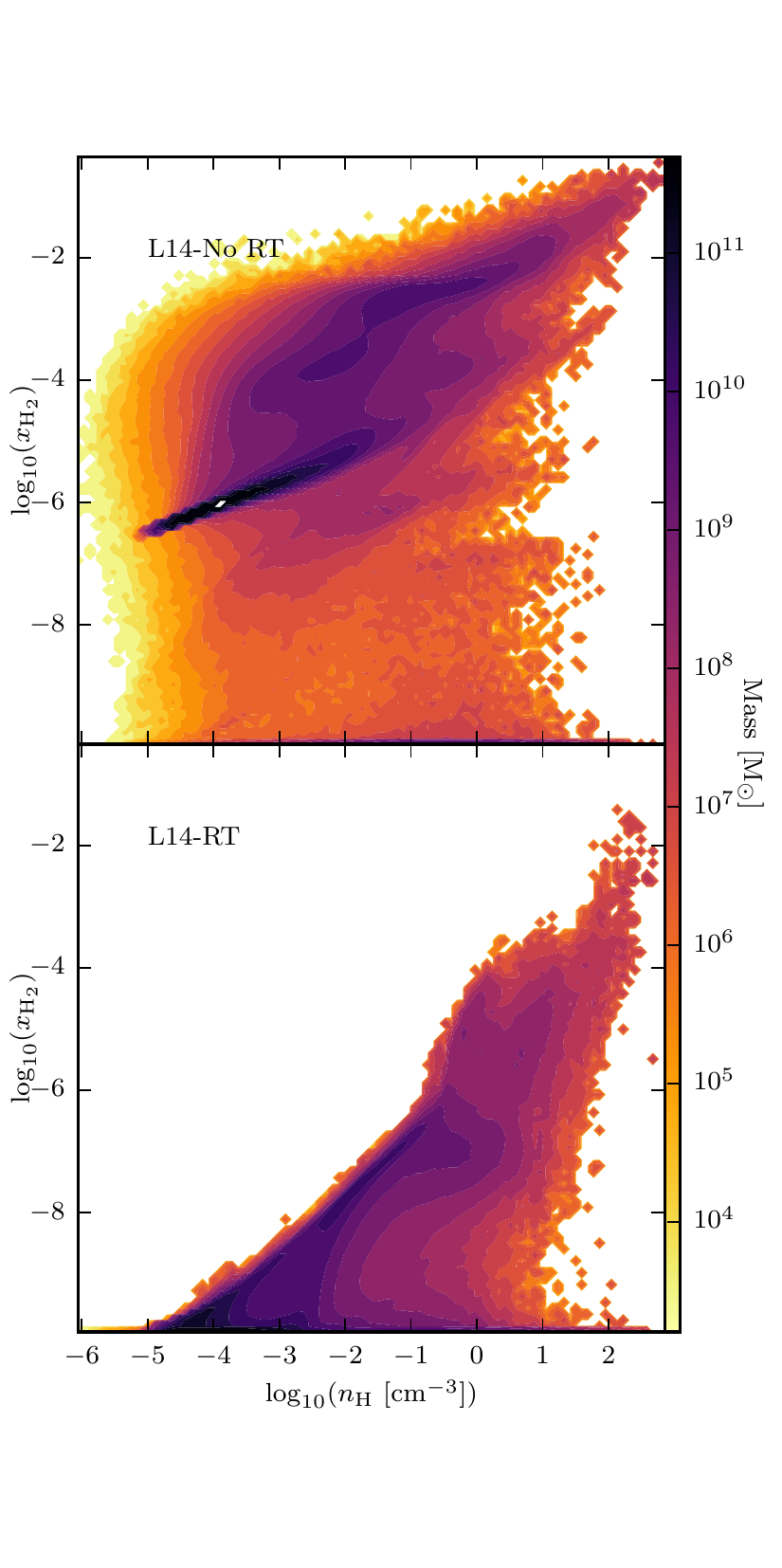}}
\caption{Mass-weighted phase-space diagrams of the H$_2$ fraction as a function of gas mass-density at $z=6$ for the L14-No RT simulation and the L14-RT simulation.  When stellar radiation is included, any residual H$_2$ at low densities is destroyed by the Lyman-Werner background.  Overall, significantly less H$_2$ forms in the simulations which include stellar radiation.}
\label{H2ps}
\end{figure}

There is a much more extended halo of H$_2$ in the simulations without RT.  This is because the haloes in our simulation are optically thin to Lyman-Werner radiation and all low-density H$_2$ is easily destroyed.  This is better seen in Figure~\ref{H2ps} where we plot a 2D histogram of the H$_2$ fraction versus density for all cells in the two highest resolution simulations at $z=6$.  In the simulations without RT, the H$_2$ fraction of the IGM is $\sim10^{-6}$ but as it approaches the more metal-enriched and slightly denser CGM, it increases to a value of $\sim10^{-3}$ as its formation is not stopped since there is no Lyman-Werner background.  As the gas reaches higher and higher densities, the H$_2$ fraction continues to increase.  From the bottom panel of Figure~\ref{H2ps}, we see a markedly different dependence of the H$_2$ fraction as a function of density in the simulations which include stellar RT.  At volume densities below $n_{\rm H}\sim10^{-2}$cm$^{-3}$, the Lyman-Werner background is extremely efficient at destroying any residual H$_2$ and the molecular fractions remains at $\sim10^{-10}$.  It is only at these densities where even marginal self-shielding can begin to take place in the metal-enriched gas.  However, only at the highest densities probed by our simulation does the H$_2$ fraction even approach $\sim10^{-2}$.  Effectively, we find that the amount of H$_2$ in our simulations with stellar radiation is very small. The consumption time of this gas due to star formation is very short.  Our simulation is either too low resolution to model H$_2$ self-shielding well or we have not increased the subgrid clumping factor enough in order to form a sufficient quantity of H$_2$.  

In Figure~\ref{H2ofz}, we also plot the H$_2$ mass as a function of redshift for the most massive halo in the three simulations which include stellar radiation.  We see that the H$_2$ mass for a single galaxy can fluctuate by more than $\sim2$dex within a few Myr.  The peaks in the H$_2$ mass correspond well with the peaks in the young stellar mass.  Likewise, the peaks in the H$_2$ mass are associated with the troughs in the middle aged stellar mass.  Even for the highest resolution run where the sinusoidal pattern is smoothed out, there is still a clear offset between the middle aged stars and the young stars and H$_2$.  This offset occurs in time as well as spatially.  In Figure~\ref{H2SD}, one can see that the location of the young stars agrees well with the highest surface density of H$_2$ because the two will both form in cold dense gas clouds.  The middle aged stars are much more spread out (see Figure~\ref{spat_star}) and are not associated with these cold dense clumps of gas and SN feedback has already effectively destroyed the birth clouds.  

The mass and spatial distribution of H$_2$ in the simulated galaxy is very dynamic and changes on short time scales.  H$_2$ is very sensitive to the temperature, density, and ionization state of the gas so it is not surprising that we find an anti-correlation between H$_2$ and middle aged stars.  One major shortcoming of our simulations is the density threshold at which star formation occurs.  We have set this density threshold to $n_{\rm H}=0.1$cm$^{-3}$ which happens to coincide with the density above which H$_2$ will begin to form efficiently.  Realistically, star formation occurs at much higher densities and thus if we increase the density threshold for star formation, there will be a time offset between when H$_2$ begins to form and when the stars begin emitting high energy, UV photons.  During this time delay, enough H$_2$ may form where it can self-shield better and thus our estimate for the H$_2$ mass will change.  

\subsubsection{The Spatial Distribution of CII, NII, OI, and OIII}
Four spectral lines of particular interest for ALMA observations at high-redshift are [CII] at $158\mu$m, [NII] at $122\mu$m, [OI] at $63\mu$m, and [OIII] at $88\mu$m.  Because our RT simulations follow the temperature, metallicity, and radiation at all frequencies including and shorter than the Habing band, we can make predictions for the spatial distribution and total mass of CII, NII, OI, and OIII in our galaxy.  We post-process the central region ($r<20$ckpc/h) of the most massive galaxy in the L14-RT simulation at $z=6$ using {\small CLOUDY} \citep{Ferland2013} to calculate the ionization states of each of these four ions in our simulation.  We use the total hydrogen density, temperature, and metallicity of each of the $\sim20,000$ cells in the central region of the galaxy as inputs.  For the radiation field, we assume a spectral shape within each radiation bin which is similar to the spectrum in the Milky Way ISM \citep{Black1987}.  We renormalise this spectrum in each bin such that the intensity within the energy bins is consistent with the intensity measured for each individual cell in our simulation.  For energies lower than the Habing band, we normalise the spectrum by the same scale factor as we measure for the Habing band in the simulation so that the shape is continuous at low energies.  We add an additional radiation field to model the contribution of the CMB at $z=6$. 

\begin{figure*}
\centerline{\includegraphics[]{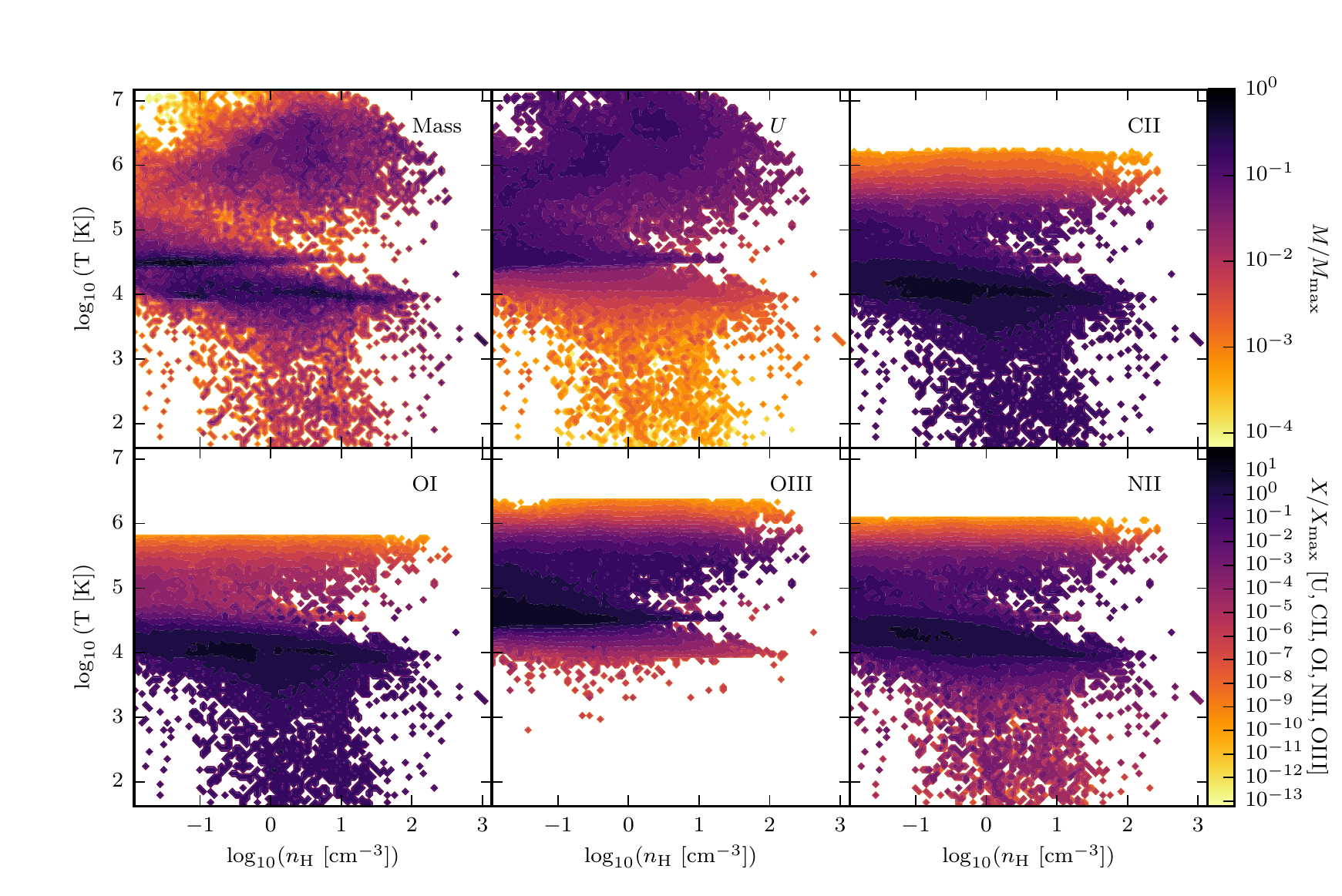}}
\caption{Phase-space diagrams of density versus temperature for all cells within viral radius of the most massive halo in the L14-RT simulation at $z=6$.  The cells in the top left panel are mass-weighted while the cells in the other panels are weighted by ionization parameter ($U$), CII, NII, OI, and OIII fractions, respectively.  The top colour bar shows the colour scale of the mass-weighted plot while the bottom colour bar represents the colour scale for the other five plots. $U$ remains enhanced in the high temperature gas where we see the highest fractions of OIII.  CII and OI occupy a very similar part of the phase-space diagram at low temperatures corresponding to the neutral gas.}
\label{ps_spec}
\end{figure*}

In Figure~\ref{ps_spec}, we plot the phase-space diagrams of temperature versus total gas density for all cells in the central region of the most massive galaxy in the L14-RT simulation at $z=6$.  The top left panel of this figure shows the mass-weighted phase-space diagram.  We see that most of the gas mass has $T\geq10^4$K while a few cells scatter to lower temperatures.  The additional five panels show the same phase-space diagram which has been weighted by the ionization parameter ($U=n_{\gamma}(E>13.6\ eV)/n_{\rm H}$), CII, NII, OI, and OIII fractions, respectively.  While most of the gas has $T\geq10^4$K, the cells which have high CII fractions have $T\leq10^4$K.  There still exists some CII at warmer temperatures up to $T\sim10^5$K.  By contrast, almost all of the NII resides in gas with $T\sim10^4$K and not much NII exists at lower temperatures.  OI looks similar to CII in that it is most prevalent at $T\leq10^4$K.  However, we see a much stronger decline in OI at warmer temperatures than we see for CII.  OIII exists at somewhat higher temperatures ($10^4<{\rm T\ [K]}<10^5$), and coincides with ionized hydrogen.  We can also see that OIII exists where the ionization parameter is high.

\begin{figure*}
\centerline{\includegraphics[]{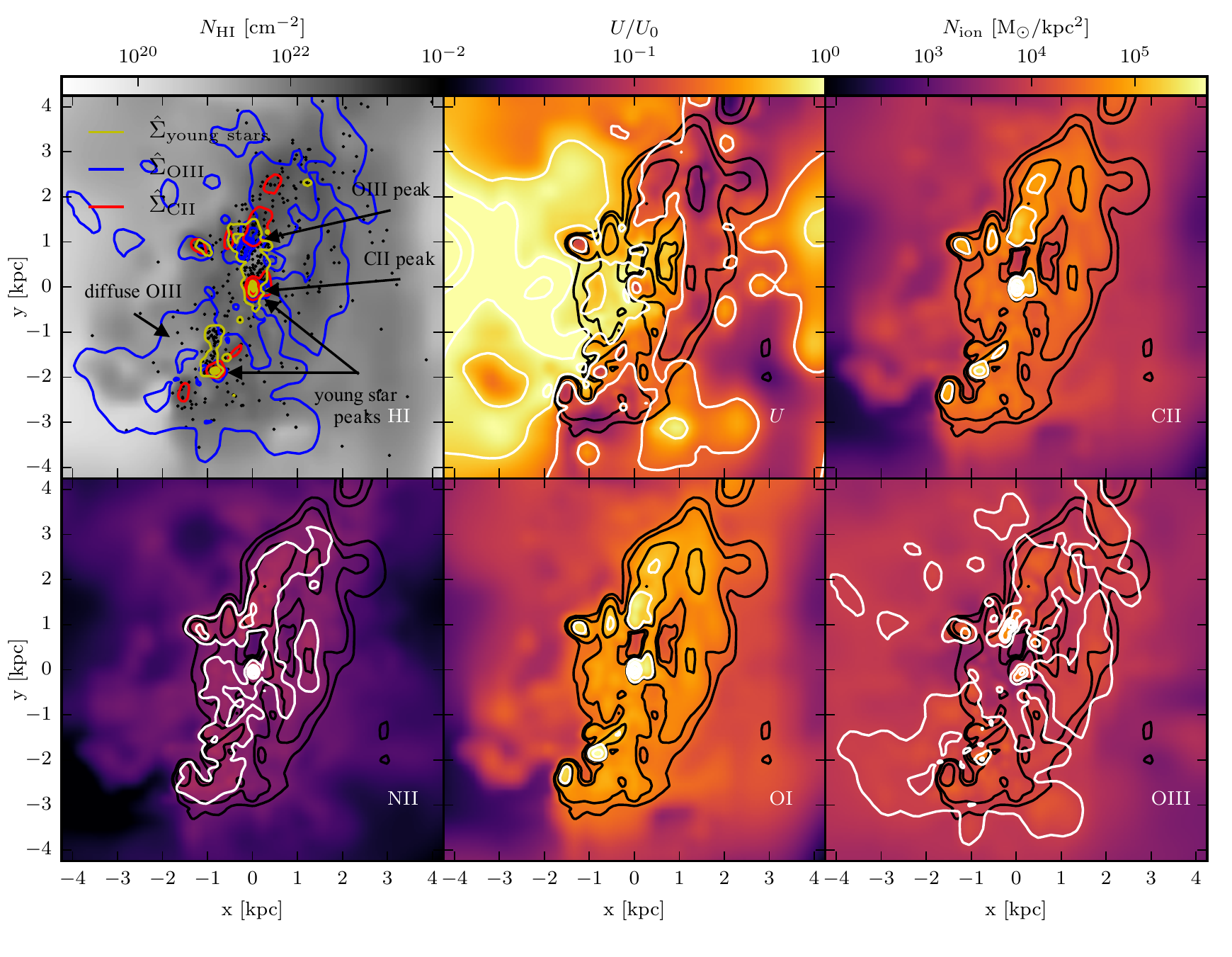}}
\caption{Surface density plots of HI (top left), $U$ (top centre), CII (top right), NII (bottom left), OI (bottom centre), and OIII (bottom right) around the most massive halo in the L14-RT simulation at $z=6$.  In the top left panel, the column density contours of OIII, CII, and young stars are shown in blue, red, and yellow respectively.  The contours have been normalised to the peak column density in order to best show the locations of the peak surface densities of each ion.  The projected location of young stars are shown as black points.  The OIII peak is spatially offset from the CII peak and young star peaks while the CII peak is co-located with one of the young star peaks.  In the other five panels, the black contours represent HI column densities of $2\times10^{20}$cm$^{-2}$, $3\times10^{20}$cm$^{-2}$, and $7\times10^{20}$cm$^{-2}$.  The white contours show the normalised surface densities representing 10\%, 30\%, 50\%, 80\%, and 90\% of the maximum value of the quantity plotted in that image.  The maximum surface densities of CII, NII, OI, and OIII are $7.0\times10^5$M$_{\odot}$/kpc$^2$, $2.5\times10^4$M$_{\odot}$/kpc$^2$, $1.6\times10^6$M$_{\odot}$/kpc$^2$, and $9.3\times10^4$M$_{\odot}$/kpc$^2$, respectively.  The colour scale for $U$ has been normalised to have a maximum of 1 while the top right colour bar applies to the CII, OI, NII, and OIII maps.}
\label{abund}
\end{figure*}

In Figure~\ref{abund} we plot the surface density of each of the four ions and the features we identified in the phase-space diagrams are reflected spatially.  The surface density of CII is clumpy and these clumps are associated with the cold neutral gas.  The features in NII are not as strong as in CII because the neutral gas at $T\sim10^4$K has a more extended distribution than the much colder, neutral gas (see Figure~\ref{props_time}).  Unsurprisingly, OI looks very similar to CII, but is not quite as extended as CII because it cannot exist at slightly warmer temperatures.  In general, the surface density of OI is higher than for CII mainly because the total abundance of oxygen is greater than that of carbon by a factor of $\sim2$ \citep{Asplund2009,Grevesse2010}.  The spatial distribution of OIII is different from that of OI and CII because it is the dominant ionization state of oxygen in a much different part of phase-space.  Inside the regions shown, we find that the total mass of CII, NII, OI, and OIII are $6.3\times10^5$M$_{\odot}$, $5.3\times10^4$M$_{\odot}$, $1.3\times10^6$M$_{\odot}$, and $2.6\times10^5$M$_{\odot}$, respectively, compared to a total metal mass in the same region of $1.7\times10^7$M$_{\odot}$.  We also show in Figure~\ref{abund} the spatial distribution of the ionization parameter along the line-of-sight, weighted by density and we can see how this anti-correlates with the location of CII and OI.  Finally, in the top left panel of Figure~\ref{abund}, we plot the neutral hydrogen column density which correlates well with CII and OI as expected and anti-correlates with the ionization parameter.  The red and blue contours show the normalised distributions of CII and OIII column density, respectively.  We can see that OIII is more extended and there is an offset between the peaks.  Likewise, the yellow contours show the normalised distribution of the column density of young stars.  We see two distinct peaks of young stars, one of which is associated with the peak in CII and the other occurs to the bottom left, off centre from the densest region of the galaxy.

Ideally, we would convert the CII mass into a luminosity and discuss its observability for ALMA.  Various works have attempted to do this \citep[e.g.][]{Vallini2013,Vallini2015,Vallini2016}; however modelling this emission is complicated.  The fine structure line is likely excited by collisions with neutral hydrogen, free electrons, and protons.  Our modelling allows us to make predictions for these quantities, but they are likely to fail in the highest density, metal-enriched regions, because we do not take into account the free electrons which come from dust grains, photoelectric-heating, and the ionization states of the metals \citep{Draine1978,Helou2001}.  The structure of the ISM at these scales is likely important due to self-shielding across the face of the molecular clouds.  For these reasons, we refrain here from estimating the [CII] and [OIII] emission of our galaxy.

\section{Discussion}
\label{disc}

\subsection{Comparison with Previous Simulations}
Various previous simulations have attempted to model the IR emission coming from galaxies during the epoch of reionization \citep{Nagamine2006,Vallini2013,Vallini2015,Pallottini2016}.  Our work builds on these previous results and in particular, the inclusion of on-the-fly, multi-frequency, RT and radiation-coupled H$_2$ chemistry, allows for an improved modelling of star-forming galaxies during the epoch of reionization.  In this section, we compare some of our results to these previous simulations.

\cite{Vallini2013} and \cite{Vallini2015} have run detailed simulations attempting to identify the origin of [CII] emission in a ``normal" high-redshift star-forming galaxy.  \cite{Vallini2013} use cosmological SPH simulations to model the formation of a $10^{11}$M$_{\odot}$ halo at $z=6.6$ which has very similar mass to the most massive galaxy in our work.  These simulations were then post-processed with UV radiation transfer using a Monte-Carlo ray-tracing code.  They find that the warm neutral medium resides in over-dense clumps which are displaced from the central star-forming regions and that cold gas resides in very dense clumps.  Furthermore, they show that at solar metallicity, 95\% of [CII] emission originates from cold gas.

We find a large difference in the morphology of our galaxy compared to \cite{Vallini2013} which forms one large dense clump of stars (see their Figure~1).  This likely arises because the simulations used in \cite{Vallini2013} include neither radiative cooling nor SN so a complex morphology cannot arise and radiation post-processing will not affect the structure of the gas.  Because the stars are concentrated in one central clump we find a different temperature distribution within our galaxy which can significantly affect the ionization state of carbon.

The work of \cite{Vallini2015} significantly improves on \cite{Vallini2013} by using the same underlying SPH simulation but including a recipe for the inhomogeneous distribution of metals, the formation of molecular clouds, and the effect of the cosmic microwave background on [CII] emission.  Although our simulations likely under-predict the formation of H$_2$ compared to \cite{Vallini2015}, we find that the molecular clouds form in the central and most dense regions of the galaxy, consistent with \cite{Vallini2015}.  They find that most of the [CII] emission arises from PDRs rather than the cold neutral medium indicating that the [CII] is likely coming from illuminated molecular clouds.  If this is in fact the case, we can use our simulation to potentially understand the morphology of the [CII] emission as well as the UV and Ly$\alpha$ to better understand the observations of \cite{Maiolino2015} (see Section~\ref{maiolino}).

\cite{Pallottini2016} (hereafter P16) performed a similar exercise using a zoom-in simulation and attempted to model the [CII] emission from a single high-redshift $z=6$ galaxy of similar mass to our work and \cite{Vallini2013,Vallini2015}.  Contrary to \cite{Vallini2013,Vallini2015}, this simulation models radiative cooling, feedback and metal enrichment and forms a disk galaxy with a star formation rate of $\sim15$M$_{\odot}$/yr.  This allows the authors to explicitly track the location of metals.  They find that 30\% of the CII mass has been ejected from the central, radiation pressure-supported disk due to outflows, but 95\% of the [CII] luminosity comes from the disk.  Furthermore, they show that this galaxy is under-luminous compared to the local [CII]-SFR relation. 

Comparing our work to P16, the total stellar mass of our most massive galaxy is significantly lower (by an order of magnitude).  The different feedback implementations are likely to be the main reason for this difference.  The delayed cooling stellar feedback model used in our work efficiently regulates star formation in the galaxy which allows our stellar mass estimate to be consistent with \cite{Behroozi2013}.  To support this conclusion, P16 find that the metallicity of the galaxy drops to the artificial metallicity floor at 12kpc whereas we find metals at nearly double this radius, indicative that our feedback model is driving stronger outflows.  Because the metals in our galaxy are much more spread out, the average metallicity of the gas is lower which can have drastic effects on the observability of [CII] \citep{Vallini2013}.  This would mean that our galaxy would fall even lower on the [CII]-SFR relation compared to P16.  

Interestingly, the H$_2$ mass measured in P16 is comparable to what we find in our simulations which do not include stellar radiation.  P16 uses a different model for H$_2$ formation which is not explicitly coupled to the radiation field and the equations are independent of the flux in the Lyman-Werner band (see their Equation 2d).  At our resolution, the amount of H$_2$ which forms in the galaxy is very sensitive to the local enhancement of the Lyman-Werner flux above the background (see Figure~\ref{H2ofz}).  Our simulations likely under-predict the H$_2$ mass in these haloes due to low resolution which hampers self-shielding.  Because the H$_2$ mass in our simulation corresponds well with the high-density peaks, we predict that [CII] luminosity will also derive from the H$_2$ regions, consistent with P16 and \cite{Vallini2015}.  

\subsection{Interpreting ALMA Observations}
\label{maiolino}

Various observations targeting the [CII] emission line during the epoch of reionization have been successful \citep{Maiolino2015,Capak2015,Willott2015,Knudsen2016,Pentericci2016}.  These works have revealed a number of interesting properties of star-forming high-redshift galaxies.  In particular, they find that the far-IR emission of these galaxies is weaker than similar galaxies at lower redshift.  This could indicate a decrease in dust mass at higher redshifts and/or low metallicities.  The galaxies tend to fall low on the [CII]-SFR relation derived from local galaxies.    

\cite{Maiolino2015} observed three spectroscopically confirmed Lyman break galaxies at $6.8<z\leq7.1$ with ALMA that have SFRs of $5-15$M$_{\odot}$/yr.  Interestingly, \cite{Maiolino2015} detect no [CII] emission at the location of the {\it Y}-band counterpart which probes the rest-frame Ly$\alpha$ and UV emission and likely the location of young stars.  Where [CII] is detected, it is spatially offset by a few~kpc from the central core of the galaxy.

Our cosmological RT simulations support the interpretation of the offset between the UV/Ly$\alpha$ and [CII] emission proposed by \cite{Maiolino2015}.  The UV and Ly$\alpha$ emission comes from young stars which are predicted to form in tight clumps from our simulation.  The [CII] emission is likely to originate in cold, neutral gas, or in PDRs close to young stars.  Our simulation includes modelling of the inhomogeneous radiation field of the galaxy which leads to a complex morphology in the HI distribution and temperature.  We probably under-predict the total mass in H$_2$, although the locations where H$_2$ forms are likely to be spatially consistent with the cold neutral medium.  The final two pieces of the puzzle are the metal and dust distributions.  Our simulations do not capture small scale inhomogeneities in the metal distribution and we do not explicitly follow dust so the distribution of both quantities are uncertain.  Dust is particularly important because of its ability to obscure star-forming regions.  

\begin{figure}
\centerline{\includegraphics[]{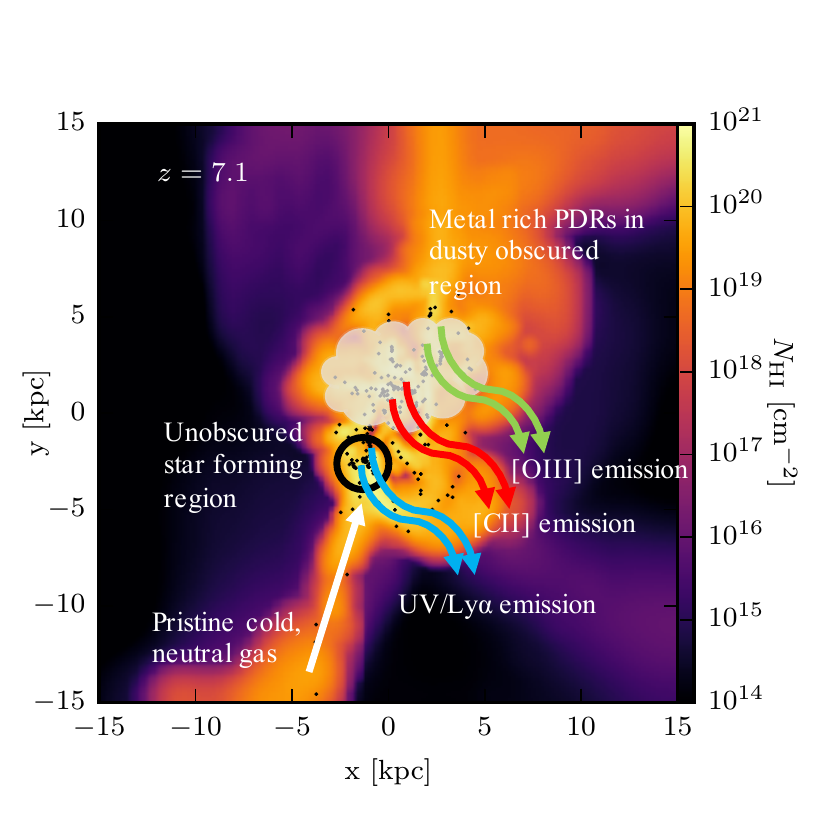}}
\caption{Schematic view for how an offset can arise between UV/Ly$\alpha$, [CII], and [OIII] emission.  The underlying image shows the surface density of neutral gas for the most massive halo in the L14-RT simulation at $z=7.1$, the redshift of the \protect\cite{Maiolino2015} observation.  The location of young stars which emit in the UV/Ly$\alpha$ are shown as black points.  The white arrow shows the direction by which cold, neutral, low metallicity gas is fed onto the galaxy.  The intersection point of the filament and the galaxy creates a star-forming region circled in black which we assume is relatively unobscured.  Towards the central regions of the galaxy, we have placed a dust cloud where we expect the metallicity and dust content to be slightly higher.  We propose that here, the UV/Ly$\alpha$ emission are obscured but the [CII] emission coming from PDRs and [OIII] emission from higher temperature, ionized gas can escape.  The expected UV/Ly$\alpha$ emission is shown as the blue arrows while the [CII] and [OIII] emission is shown as the red and green arrows respectively.}
\label{schematic}
\end{figure}

In Figure~\ref{schematic}, we show how spatial offsets between UV/Ly$\alpha$ and [CII] emission can arise depending on where the dust is located.  We show how this may work by annotating a neutral hydrogen surface density map of the most massive galaxy in our simulation at $z=7.1$, the redshift of the \cite{Maiolino2015} observation.  In the bottom of the image, a cold neutral filament is feeding pristine or low metallicity gas onto the central regions of the galaxy.  At the intersection of the galaxy and the filament, a large star-forming clump has formed which we envision to have reasonably low metallicity and thus little [CII] emission.  Because of the low metallicity and dust content, the UV and Ly$\alpha$ emission escape without being reprocessed and are visible.  Towards the central regions of the galaxy, one may expect that the metallicity and dust content are higher and thus star-forming clouds in this region are much more obscured and no UV or Ly$\alpha$ emission escapes this region.  However, in these dense clouds, PDRs produce [CII] emission which can pass through the dust; hence an offset is created between the UV/Ly$\alpha$ and [CII] emission.  To calculate the dust attenuation, what is required is the dust optical depth, $\tau_d$. Our simulations do  not self-consistently trace dust. If we make again the simplistic assumption that the dust optical depth scales linearly with neutral hydrogen column density and metallicity \citep{Gnedin2009}, we find that the maximum $\tau_d$ is a factor of about three times higher in the central star forming regions compared to the outer regions of our simulated galaxies due to metallicity and density gradients in the galaxy. The difference in surface mass density of young stars between the inner and outer regions is a factor of $\sim1.5$. Note that the  attenuation scales as $e^{-\tau_d}$ and that we expect more attenuation in the central regions. These values should, however, be considered with some caution as a more sophisticated model for dust physics will be needed to get more accurate estimates of the attenuation.  Furthermore, resolution effects in our simulations limit the maximum densities in the simulations and therefore put an upper limit on $\tau_d$.

If [NII] and [OI] could also be observed, we predict that these will be spatially coincident with the [CII], although the [NII] might be slightly more extended.  By contrast, [OIII] arises from higher temperature gas at a higher ionization parameter and thus would be offset from [CII].  Therefore, we also might expect an offset between [OIII] and [CII].

\section{Caveats}
As with all numerical simulations, there are a number of caveats, many of which we have already discussed in the text.  At large scales, reionization is sensitive to the resolution and star particle mass of the simulation \citep{Aubert2010} and the galaxies resolved. Here we do not resolve the atomic cooling threshold haloes which may help drive reionization \citep{Kimm2016}.  Most of the metals in our simulations remain inside the virial radius of the haloes and do not efficiently enrich the IGM.  This is rather sensitive to the feedback method implemented in the simulation \citep{Keating2016} and thus our conclusions may be subject to our chosen feedback model.  

The density threshold of star formation in our simulations is much lower (0.1cm$^{-3}$) than the typical densities observed in local molecular clouds ($>100$cm$^{-3}$).  Because, the galaxies in our simulations have low metallicity, H$_2$ only forms very efficiently at densities higher than the density threshold where we assume stars form.  For this reason, the Lyman-Werner radiation is extremely efficient at destroying the H$_2$ as the resolution of our simulations is too low to model self-shielding properly.  The formation of H$_2$ depends on the clumping factor chosen.  We have experimented with increasing the clumping factor and unsurprisingly, significantly more H$_2$ forms.  Higher resolution simulations will be needed to properly model the formation of H$_2$.  Nevertheless, the conclusion that H$_2$ is located in dense, metal-enriched gas will continue to hold.  

Likewise, because of the still rather moderate resolution of our simulations, the metals efficiently spread throughout the central regions of the haloes.  A higher resolution simulation might resolve a more patchy metal distribution in the ISM.  If small patches of gas have significantly higher metallicity, this will affect where the [CII] emission originates from and may change the conclusion with regard to its observability \citep{Vallini2013,Vallini2015}.     

Finally, our simulations make no attempt to model the spatial distribution of dust which plays a role in  various astrophysical processes such as the cooling and fragmentation of gas, the formation of H$_2$, absorption of UV photons, and the obscuration of UV/Ly$\alpha$ emission.  

\section{Conclusions}
\label{conc}
We have presented six cosmological simulations, three of which include on-the-fly multi-frequency RT and three without, at various resolutions which demonstrate how inhomogeneous stellar radiation affects galaxy formation at large (Mpc), intermediate (tens of kpc), and small (sub kpc) scales.  The simulations employ a new variable speed of light approximation which allows ionization fronts to travel at the appropriate speed in both low and high-density regions before Stromgren spheres have reached their maximum extent.  The simulations include a non-equilibrium chemistry model for the formation and destruction of H$_2$ which is coupled to the spatially inhomogeneous radiation from star particles.  Our findings can be summarised as follows:

\begin{itemize}
  \item At large scales, reionization proceeds inside-out.  The highest density regions surrounding haloes are ionized first.  However, a significant amount of neutral gas remains inside the haloes with a typical ratio of masses of ionized to neutral gas of 2:1.  
   
  \item The simulation becomes optically thin to radiation in the Lyman-Werner band at a significantly higher redshift than it does for photons which ionise hydrogen and helium.  Inside of haloes the flux in the Lyman-Werner band follows a characteristic $r^{-2}$ profile while hydrogen ionizing radiation is dominated by the individual star-forming clumps as it is readily absorbed internally by cold neutral gas.
  
  \item At intermediate scales, the filaments remain mostly self-shielding with HI column densities above the Lyman limit threshold ($N_{\rm H}>10^{17}$cm$^{-2}$).  The spatial distribution of the DLAs is very dynamic and strongly anti-correlated with the presence of young stars.  
    
  \item The distribution of young stars inside of haloes is strongly clustered and the locations of these stars are closely associated with their birth clouds.  As stars age, they redistribute throughout the galaxy which creates a smooth, old stellar component.  The UV/Ly$\alpha$ radiation will be dominated by these clumps and its observability will be subject to the presence of dust.    
  
   \item We find rapid changes in the spatial distribution of ionized and neutral gas inside of the haloes due to young stars ionizing their surroundings and SN feedback blowing holes in the neutral medium.
  
  \item The metal mass in our simulations is proportionally distributed between the ionized and neutral gas inside of the haloes.  The ten most massive haloes in our simulation with masses $10^{10.4}<{\rm M_{DM}}<10^{11.2}$ all exhibit the expected cosmic baryon fraction within their virial radii.
  
   \item For our specific feedback implementation, most of the metals are confined inside the virial radius of haloes.  Many small haloes form along the filaments which increase the metallicity of this gas slightly above the metallicity floor of our simulations; however, much of the gas at these densities has a metallicity which remains close to primordial. 
  
  \item H$_2$ masses in the haloes are time variable and extremely sensitive to the presence of young stars.  Without stellar radiation, a strong correlation arises between the SFR and the H$_2$ mass while with RT, this correlation weakens and the scatter significantly increases.  Our simulations, however, likely under-predict the total mass of H$_2$ in the haloes because they have not reached H$_2$ densities where it can effectively self-shield.
   
 \item The CII in the galaxy is associated with the locations of the cold neutral clumps of gas.  OI follows a similar trend; however has a less extended distribution compared to CII.  By contrast, the distribution of NII is more diffuse and associated with much warmer gas and OIII exists in hotter, ionized gas.
 
 \item Depending on the location of dust in the galaxy, spatial offsets are expected between UV/Ly$\alpha$ and [CII] emission which would explain the recent observations of \cite{Maiolino2015}.  A spatial offset also naturally arises between [CII] and [OIII] because of the different part of phase-space where each of the ionization states are dominant.  It should thus be expected to regularly find spatial offsets between UV/Ly$\alpha$, [CII], and [OIII] emission in the same object.
  
\end{itemize} 

With our simulations, we have shown that the inclusion of local radiation from young stars is crucial to realistically model the ISM in high-redshift galaxies and in particular, the spatial distribution of the different ionization states of oxygen and carbon which can be observed by ALMA.  To make meaningful comparisons with observations, future simulations will require a further improved modelling of dust in order to better model the formation of H$_2$ as well as better understand the obscuration which might cause a spatial offset between [CII] and UV emission. Higher resolution is also required to better model the distribution of metals in the haloes and to better resolve the physics governing the [OIII] and [CII] emission in a multiphase ISM.

\section*{Acknowledgements}
We thank the referee for their review of this manuscript and very constructive comments.  We thank Roberto Maiolino, Stefano Carniani, and Joki Rosdahl for comments on the manuscript.  HK is grateful to Laura Keating and Fernanda Ostrovski for useful conversations and revision of the manuscript.  Furthermore, we thank the MIAPP for the reionization workshop in Garching where some of this work was completed.

This work made considerable use of the open source analysis software
{\small PYNBODY} \citep{pynbody}. HK's thanks Foundation Boustany, the Cambridge Overseas Trust, and an the Isaac
Newton Studentship. Support by ERC Advanced Grant 320596
``The Emergence of Structure during the Epoch of reionization" is
gratefully acknowledged.  DS acknowledges support by STFC and ERC
Starting Grant 638707 ``Black holes and their host galaxies: coevolution across cosmic time''.

This work was performed using the DiRAC/Darwin Supercomputer
hosted by the University of Cambridge High Performance
Computing Service (http://www.hpc.cam.ac.uk/), provided by Dell
Inc. using Strategic Research Infrastructure Funding from the
Higher Education Funding Council for England and funding from
the Science and Technology Facilities Council. 

This work used the DiRAC Complexity system, operated by the University of Leicester IT Services, which forms part of the STFC DiRAC HPC Facility (www.dirac.ac.uk ). This equipment is funded by BIS National E-Infrastructure capital grant ST/K000373/1 and  STFC DiRAC Operations grant ST/K0003259/1. DiRAC is part of the National E-Infrastructure

Furthermore, this work used the DiRAC Data Centric system at Durham University, operated by the Institute for Computational Cosmology on behalf of the STFR DiRAC HPC Facility (www.dirac.ac.uk).  This equipment was funded by the BIS National E-infrastructure capital grant ST/K00042X/1, STFC capital grant ST/K00087X/1, DiRAC operations grant ST/K003267/1 and Durham University.  Dirac is part of the National E-Infrastructure.




\bibliographystyle{mnras}
\bibliography{./HkTkDsMh_2016.bib} 



\appendix
\section{The Variable Speed of Light Approximation}
The motivation for using a variable speed of light is that in high-density regimes (ISM) and for low luminosity sources, a reduced speed of light generally gives correct results.  However, in low-density regimes (IGM) and for highly luminous sources, it is necessary to use the full speed of light to properly model the propagation of I-fronts.  Most AMR simulations use a density criterion to initiate refinement.  The VSLA we introduce here interpolates between low values of the speed of light and high values of the speed of light depending on AMR refinement level.  This can make the simulation far less expensive than using the full speed of light everywhere while properly modelling I-front propagation in low-density regions.  

\subsection{Implementation}
To implement the VSLA, we first identify the grid resolution and density at which we wish to use the full speed of light.  In general, we set this as the level of the base grid in the simulation (in our case Level 8).  Choosing a higher level would result in a slower, albeit more conservative approach.  For each higher refinement level, we successively divide the speed of light by a factor of two up to the highest resolved level, or until we no longer wish to further reduce the speed of light.  We choose to divide the speed of light by a factor of two at each higher level in order to keep the global time step constant across all levels such that,
\begin{equation}
c_{\text{sim}}(l)=\frac{c}{2^{l-l_{\text{min}}}},
\end{equation}
where $l$ is the level of interest and $l_{\text{min}}$ is the level where we set $c_{\text{sim}}=c$.
 
With oct-based refinement, as is used in {\small{RAMSES}}, upon refinement, the cell length is divided by a factor of two.  Dividing $c_{\text{sim}}$ by a factor of two therefore also keeps $t_{\text{RT}}$ constant.  Dividing $c$ by a larger factor would mean lower, less resolved levels act on a shorter time step than the more refined levels.  Dividing by a factor less than two would correspond to a more conservative approach but would not fully exploit the speed-up possible with VSLA and we, therefore, choose to use factors of two in this work.  

Consider the specific case where our base grid is resolved at Level~$=l_{\text{base}}$ which is where $c_{\text{sim}}(l_{\text{base}})=c$ and we impose 7 additional levels of refinement so that at Level~$=l_{\text{base}}+7$, $c_{\text{sim}}(l_{\text{base}}+7)=c/2^7=7.8\times10^{-3}c$.  Using the RSLA, this simulation would in principle be $\sim2^7=128$ times faster than using the full speed of light at all refinement levels; however, this would not model the I-fronts properly in the IGM.  With the VSLA, our maximum time step for our most refined levels is the same as in the RSLA with a similar gain, in principle, in terms of speed-up.  What prohibits the VSLA from achieving the same speed-up as the RSLA is that by definition, we abandon multi-stepping in time (i.e. all levels are run on the same time step which is generally dictated by the most refined cells).  Depending on what fraction of all cells are refined, abandoning the multi-stepping advantage of AMR can correspond to different costs.  If the computational load is completely dominated by the highest resolution cells (i.e. a zoom-in simulation) then VSLA is unlikely to become much more expensive than RSLA.  In the case where very few cells are at the highest level of refinement, abandoning multi-stepping can become more costly at the expense of more accurate RT and the refinement criteria must be optimised\footnote{Note that in the newest version of VSLA (not used in this work), it is no longer required that multi-stepping be abandoned for the AMR.  Rather, hydro and radiative time steps can be adaptive and we instead update the number of subcycles in the radiation sub-steps depending on the speed of light in the cells on that level.}.  

We can improve the speed-up even further by introducing RT subcycling in the simulation where the radiative time step is semi-decoupled from the hydro time step.  Even with RSLA, the RT time step is often much shorter than the corresponding hydro time step and therefore a speed-up can be achieved by subcycling multiple times over the RT calculation for every single hydro time step.  This has been implemented in the main version of {\small{RAMSES}}, however, because the hydro time steps are adaptive, one has to adopt suitable boundary conditions at the interfaces between coarse and refined cells because multiple fine steps are executed without the coarse cells being fully updated (see \cite{Commercon2014}).  Nevertheless, even with these special boundary conditions, we have confirmed that the method is accurate to a few per cent.  By contrast, using the VSLA forces all hydro time steps to be of the same length and we are no longer forced to subcycle the RT calculation individually at each level.  Instead, we have restructured {\small{RAMSES}} so that the RT calculation is performed after every coarse time step.  This allows us to subcycle through all AMR levels at once.  The only downside to performing the calculation this way is that for large cosmological simulations, accessing the same array multiple consecutive times is faster than cycling through all of the different levels one after another because one must continually access slower areas of memory to retrieve the arrays which hold information for other levels.  We have attempted to optimise our routines to achieve the speed-up one obtains with the original version; however there is a moderate computational price for increased accuracy\footnote{Note again that in the newest version of VSLA (not used in this work), we have made our algorithm fully consistent with the boundary conditions adopted in \cite{Commercon2014}, which allows for a greater increase in speed.  Thus the slow down of VSLA compared to RSLA reported in this work (especially for the cosmological test) are overestimates of what we have seen with the newest version of the algorithm.}. 

\subsection{Conservation of Flux and Photon Number Density}
To maintain the conservation properties of the code, we demonstrate a simple example of how photon flux and number density are advected across cell boundaries.  

\begin{figure*}
\centerline{\includegraphics[scale=0.5]{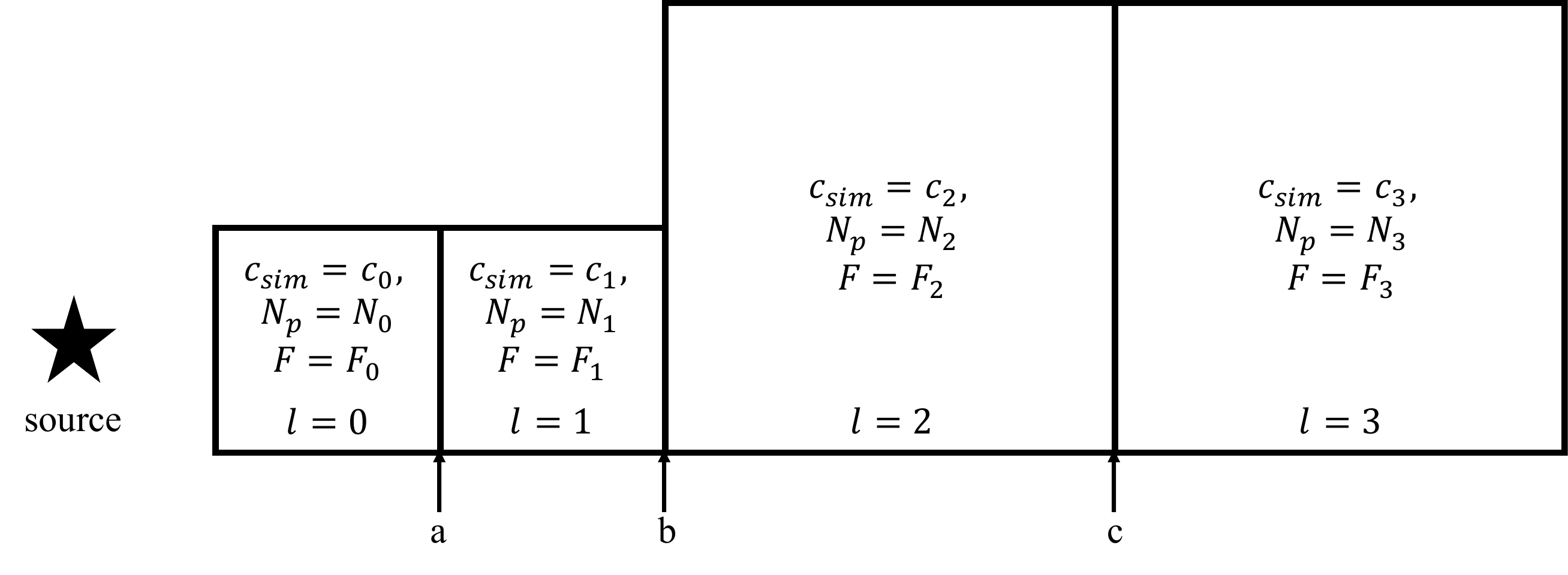}}
\caption{Simple set up of three cells to show how photon properties are advected across the refinement boundary.}
\label{fluxdiag}
\end{figure*}

To calculate the intercell fluxes, we use the global Lax-Fredrich (GLF) flux function where
\begin{equation}
\label{GLF}
\mathcal{F}_{\rm GLF}=\frac{\mathcal{F}_l+\mathcal{F}_{l+1}}{2}-\frac{c_{\rm sim}(l)}{2}(\mathcal{U}_{l+1}-\mathcal{U}_{l}).
\end{equation}
Here $\mathcal{U}_l=[N_l,{\bf F}_l]$ and $\mathcal{F}_l=[{\bf F}_l,c_l^2\mathbb{P}_l]$ where $N_l$ is the number density of photons in the cell, ${\bf F}_l$ is the flux vector of the cell, and $\mathbb{P}_l$ is the pressure tensor of the cell.  To calculate the pressure tensor, we have
\begin{equation}
\mathbb{P}_l=\mathbb{D}_lN_l.
\end{equation}
We compute this for each of the different photon groups. $\mathbb{D}_l$ is the Eddington tensor of the cell, and for this, we use the M1 closure such that
\begin{equation}
\mathbb{D}_l=\frac{1-\chi_l}{2}{\bf I}+\frac{3\chi_l-1}{2}{\bf n_l}\otimes{\bf n_l},
\label{eddten}
\end{equation}
where
\begin{equation}
{\bf n}_l=\frac{{\bf F}_l}{|{\bf F}_l|},\ \chi_l=\frac{3+4f^2_l}{5+2\sqrt{4-3f^2_l}},\ f_l=\frac{|{\bf F}_l|}{c_lN_l}.
\end{equation}
If we consider the 1D case, which we will do for the rest of this derivation, $\mathbb{P}_l\rightarrow N_l$ and therefore $\mathcal{F}_l=[{\bf F}_l,c_l^2N_l]$.

In Figure~\ref{fluxdiag}, we show a setup of four cells with indices $l=[0:3]$.  Each cell has a speed of light $c_l$, a number density of photons $N_l$, and a photon flux ${\bf F}_l$.  We begin by calculating the intercell fluxes between the central two cells and we can use Equation~\ref{GLF} to calculate these values.  In order to account for the differences in speed of light between the different cells we make a slight modification to $\mathcal{U}$ and $\mathcal{F}$ for the cells to the left and right of the cell of interest so that 
\begin{equation}
\mathcal{U}_{l\pm1}(N_{l\pm1},{\bf F}_{l\pm1}) \rightarrow\mathcal{U}_{l\pm1}(N^{'}_{l\pm1},{\bf F}_{l\pm1})
\end{equation}
and 
\begin{equation}
\mathcal{F}_{l\pm1}({\bf F}_{l\pm1},c_{l\pm1}^2N_{l\pm1})\rightarrow\mathcal{F}_{l\pm1}({\bf F}_{l\pm1},c_l^2N^{'}_{l\pm1})
\end{equation}
where
\begin{equation}
N_{l\pm1}^{'}=\frac{c_{l\pm1}}{c_l}N_{l\pm1}.
\end{equation}

Starting with the photon number densities, the intercell flux at {\bf b} as seen by $l=1$ is
\begin{equation}
\label{nphotdn}
\begin{split}
\mathcal{F}_{\rm GLF,{\bf b},l=1}^{N,l=1} & =\frac{ {\bf F}_{1}+{\bf F}_{2}}{2}-\frac{c_{1}}{2}(N_{2}^{'}-N_{1})\\
& =\frac{ {\bf F}_{1}+{\bf F}_{2}}{2}-\frac{c_{2}}{2}N_{2}+\frac{c_{1}}{2}N_{1}.
\end{split}
\end{equation}
Likewise, we can calculate the same value at {\bf b} from the perspective of $l=2$ as
\begin{equation}
\label{nphotup}
\begin{split}
\mathcal{F}_{\rm GLF,{\bf b}}^{N,l=2}& =\frac{ {\bf F}_{2}+{\bf F}_{1}}{2}-\frac{c_{2}}{2}(N_{2}-N^{'}_{1})\\
& =\frac{ {\bf F}_{2}+{\bf F}_{1}}{2}-\frac{c_{2}}{2}N_{2}+\frac{c_{1}}{2}N_{1}.
\end{split}
\end{equation}

Clearly the fluxes calculated in Equations~\ref{nphotdn}~and~\ref{nphotup} are identical and thus it is trivial to show that the total number of photons is conserved regardless of the speed of light in each of the cells as long as we substitute the appropriate value of N into Equation~\ref{GLF}.  This will be detailed further below.

We can perform a similar exercise with the fluxes and the intercell flux at {\bf b} as computed by $l=1$ is  
\begin{equation}
\begin{split}
\mathcal{F}_{\rm GLF,{\bf b}}^{F,l=1} & =\frac{c_1^2N_2^{'}+c_1^2N_1}{2}-\frac{c_1}{2}({\bf F}_2-{\bf F}_1)\\
& =\frac{c_1c_2N_2+c_1^2N_1}{2}-\frac{c_1}{2}({\bf F}_2-{\bf F}_1).
\end{split}
\end{equation}

On the contrary, if we compute the intercell flux at $l=2$, we have 
\begin{equation}
\label{fluxup}
\begin{split}
\mathcal{F}_{\rm GLF,{\bf b}}^{F,l=2}& =\frac{c_2^2N_2+c_2^2N_1^{'}}{2}-\frac{c_2}{2}({\bf F}_2-{\bf F}_1)\\
& =\frac{c_2^2N_2+c_1c_2N_1}{2}-\frac{c_2}{2}({\bf F}_2-{\bf F}_1).
\end{split}
\end{equation}
Thus there is a asymmetry between the intercell flux of the flux as computed by $l=1$ and $l=2$ such that
\begin{equation}
\mathcal{F}_{\rm GLF,{\bf b}}^{F,l=2} = \frac{c_2}{c_1}\mathcal{F}_{\rm GLF,{\bf b}}^{F,l=1}.
\end{equation}

In order to assure that no spurious flux is created or destroyed at the intercell boundary, one must update the cells at the boundary with a different flux.  For the cell $l=1$, we use $\mathcal{F}_{\rm GLF,{\bf b}}^{F,l=1}$ and for $l=2$, we use $\frac{c_2}{c_1}\mathcal{F}_{\rm GLF,{\bf b}}^{F,l=1}$.

This becomes clearer when we plug in steady-state values for this setup where $c_3=c_2=c$ and $c_1=c_0=\frac{c}{2}$ as we would use in the cosmological simulation.  In this case {\bf F}$_0=$~{\bf F}$_1=$~{\bf F}$_2=$~{\bf F}$_3=$~{\bf F} and $N_1=N_0=N$ and $N_3=N_2=\frac{N}{2}$\footnote{The change in $N$ is so that the photoionization rate remains constant which is what is expected for steady state.}.  The updated state of a cell is computed as
\begin{equation}
\label{update}
\mathcal{U}_l^{n+1}=\mathcal{U}_l^{n}+\frac{\Delta t}{\Delta x}(\mathcal{F}_{{\rm GLF},l-1/2}^n-\mathcal{F}_{{\rm GLF},l+1/2}^n).
\end{equation}
If we consider the photon density update for cell $l=2$, we have computed $\mathcal{F}_{{\rm GLF},l-1/2}^n$ in Equation~\ref{nphotup} and $\mathcal{F}_{{\rm GLF},l+1/2}^n$ is easily computed as
\begin{equation}
\mathcal{F}_{\rm GLF,{\bf c}}^{N,l=2}=\frac{ {\bf F}_{2}+{\bf F}_{3}}{2}-\frac{c_{2}}{2}(N^{'}_{3}-N_{2}).
\end{equation}
Plugging in the steady-state values in the equations for  $\mathcal{F}_{{\rm GLF},l-1/2}^n$ and $\mathcal{F}_{{\rm GLF},l+1/2}^n$, we find the updated number density in cell $l=2$ to be
\begin{equation}
\begin{split}
\mathcal{U}_{l,N}^{n+1} & =\mathcal{U}_{l,N}^{n} + \frac{\Delta t}{\Delta x}\Biggl(\frac{{\bf F}_1+{\bf F}_2}{2}-\frac{c_2}{2}N_2+\frac{c_1}{2}N_1\\
                                 & -\frac{{\bf F}_3+{\bf F}_2}{2}+\frac{c_2}{2}N^{'}_3-\frac{c_2}{2}N_2 \Biggr) \\
                                 &= \mathcal{U}_{l,N}^{n} + \frac{\Delta t}{\Delta x}\left({\bf F}-\frac{c}{2}\frac{N}{2}+\frac{c}{4}N-{\bf F}+\frac{c}{2}\frac{N}{2}-\frac{c}{2}\frac{N}{2}\right)\\
                                 &= \mathcal{U}_{l,N}^{n}.
\end{split}
\end{equation}
Similarly for the flux, we have computed  $\mathcal{F}_{{\rm GLF},l-1/2}^n$ in Equation~\ref{fluxup} and $\mathcal{F}_{{\rm GLF},l+1/2}^n$ is computed as
\begin{equation}
\mathcal{F}_{\rm GLF,{\bf b}}^{F,l=2}=\frac{c_2^2N_2+c_2^2N_1^{'}}{3}-\frac{c_2}{2}({\bf F}_3-{\bf F}_2).
\end{equation}
Plugging into Equation~\ref{update}, we find the updated flux in cell $l=2$ as
\begin{equation}
\begin{split}
\mathcal{U}_{l,{\bf F}}^{n+1} & =\mathcal{U}_l^{n} + \frac{\Delta t}{\Delta x}\Biggl(\frac{c_2^2N_2+c_2c_1N_1}{2}\\
                                 &-\frac{c_2}{2}({\bf F}_2-{\bf F}_1)-\frac{c_2^2N_2+c_2^2N^{'}_3}{2}+\frac{c}{2}({\bf F}_3-{\bf F}_2)\Biggr)\\       
                                 &= \mathcal{U}_{l,{\bf F}}^{n}+ \frac{\Delta t}{\Delta x}\Biggl(\frac{c^2\frac{N}{2}+\frac{c^2}{2}N}{2}-\frac{c}{2}{\bf F}+\frac{c}{2}{\bf F}\\
                                 &-\frac{c^2\frac{N}{2}+c^2\frac{N}{2}}{2}-\frac{c}{2}{\bf F}+\frac{c}{2}{\bf F}\Biggr)\\
                                 &= \mathcal{U}_{l,{\bf F}}^{n}.
\end{split}
\end{equation}
Clearly, the steady state is maintained which demonstrates the validity of the VSLA algorithm.  A similar exercise can easily be computed for the cell $l=1$.  Recall that when updating the flux, this is done asymmetrically depending one which side of the boundary one is on which is key to maintaining the steady state.

\begin{figure*}
\centerline{\includegraphics[]{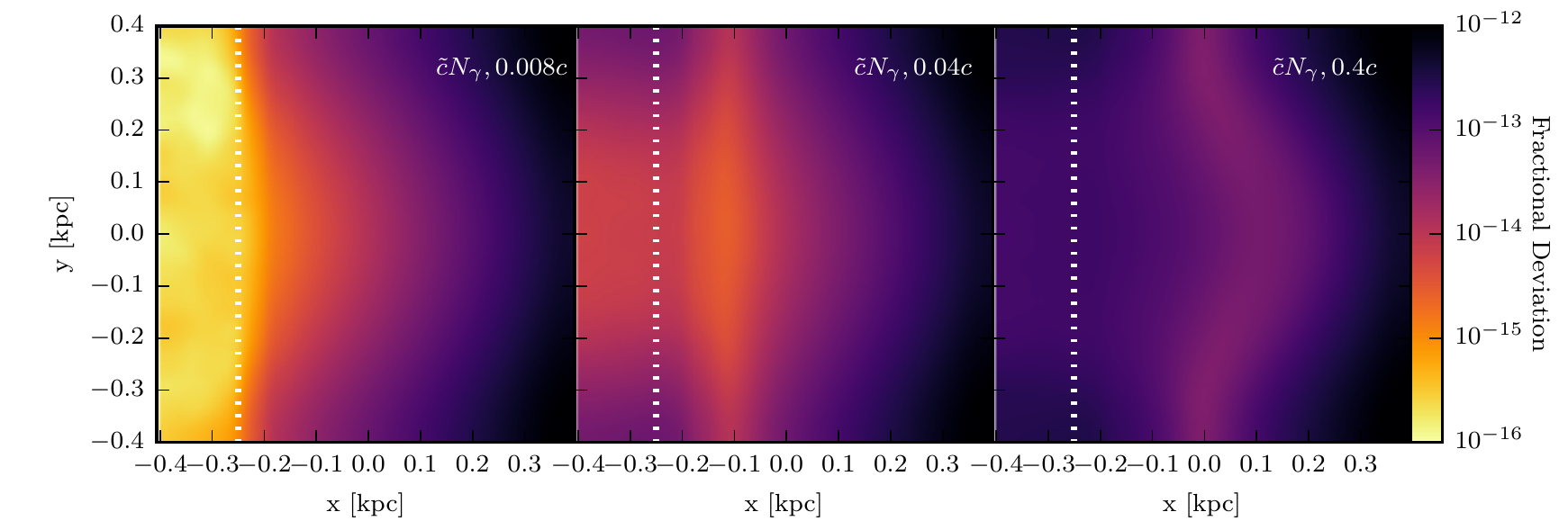}}
\caption{Fractional deviations of the VSLA run with a 2 (top), 10 (middle), and 100 (bottom) times faster speed of light in the coarse cells compared to the reference RSLA run.  The snapshots are takes at 4Myr, long after the crossing time of the box and at a point where the simulation should be in a steady state.  The total photon flux is well conserved between all of the runs.  The vertical dashed line shows the location where the refinement level changes.  The coordinates of the box have been re-centred on (0.4,0.4,0.4)kpc for these plots.}
\label{twosame}
\end{figure*}

We present here a series of idealised tests that verify that our new VSLA implementation exhibits the properties we have described.

\subsubsection{Test 1: Point Source in an Optically Thin Box}
In this test, we set up a low resolution square box with sides of length 0.8 kpc and put a point source which emits $10^{50}$photons/s at position (0, 0.4, 0.4)kpc such that it is emitting from the left wall of the box.  All walls have outflow boundary conditions so that any radiation which reaches the edges of the box is advected out of the computational domain.  Cells with $x<0.15$kpc are refined to Level 5 ($32^3$) all other cells are refined to Level 4 ($16^3$) and this refinement map is kept static throughout the test.  The source luminosity is split into three bins with mean energies of 18.85, 35.05, and 65.67eV, respectively, and the number of photons emitted in each bin is consistent with that of a $10^5$K blackbody spectrum.  The box is entirely filled with hydrogen gas with a uniform density of $n_{\rm H}=0.1$cm$^{-3}$ and a temperature of 100K.  In this test, all hydrogen atom cross-sections are set to zero so that the photons travel unimpeded across the box and thus it as if the radiation is propagated in a vacuum.  

\begin{figure*}
\centerline{\includegraphics[]{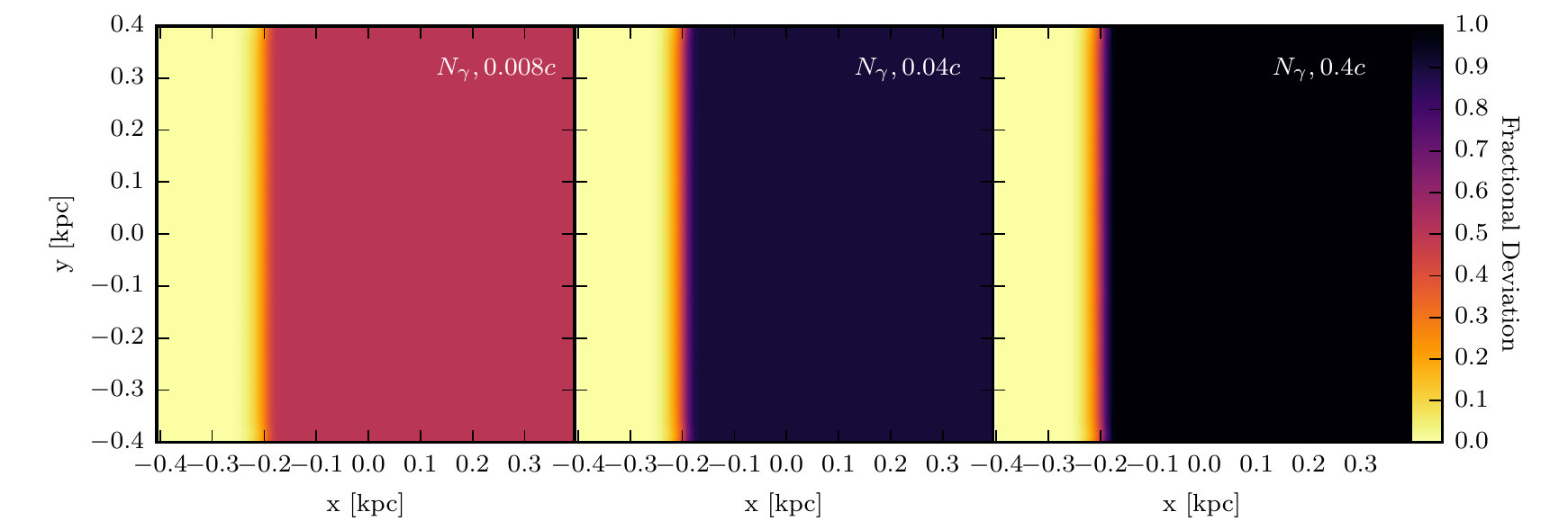}}
\caption{Fractional deviations of the VSLA run with a 2 (left), 10 (centre), and 100 (right) times faster speed of light in the coarse cells compared to the reference RSLA run.  The snapshots are taken at 4Myr, long after the crossing time of the box and at a point where the simulation should be in a steady state.  The number density of photons changes as expected in order to maintain a constant photoionization rate.  The coordinates of the box have been re-centred on (0.4,0.4,0.4)kpc for these plots.}
\label{twong}
\end{figure*}

We begin by running this test four times.  In our first experiment, we set $c_{\text{sim}}=0.004c$ everywhere in the box, regardless of the level of refinement, which represents the control run.  With this setup, the effective crossing time of the box is $\sim650,000$yrs.  The simulation is run for 4Myr, much longer than the crossing time, so that the simulation has reached a steady state.  In the next two experiments, we increase $c_{\text{sim}}$ in the coarse cells (i.e. at $x\geq0.15$kpc) by a factor of 2 (our default VSLA implementation), by a factor of 10, and by a factor of 100 in order to demonstrate conservation of flux and photon number density in all cases.

Figure~\ref{twosame} shows the fractional deviation of the total photon flux ($\tilde cN_{\gamma}$) in the second three experiments compared with the reference control run ($|$Control - Experiment$|$ / Control).  The total photon flux is well conserved regardless of the speed of light we use in the less refined region.  In all cases, the luminosity of the source is the same and therefore the flux, which depends only on luminosity and distance should be the same regardless of the value of the speed of light we choose in the simulation (assuming the simulation has reached a steady state). We have confirmed that we conserve photon flux in each individual direction to a precision of one part in $\sim10^{-13}$ when a flux limiter is not imposed.  When a second-order flux limiter\footnote{This determines how the cell properties are interpolated from coarse to fine cells and likewise in the reverse.} is used, the conservation changes to less than one part in $10^{-2}$ and is slightly different for the x direction compared to y and z.  This is almost certainly due to the geometry of the problem.  The photon source is located in a position such that all photons inherit a positive flux in the x direction while this is not the case for both the y and z directions and the refinement boundary is placed at a fixed x value.  In the case of VSLA, it may be useful to adopt a lower-order slope limiter for the radiation while keeping it on for the hydro as we expect the radiation to be a much smoother field.  

We also show in Figure~\ref{twong}, how the photon number density changes across the boundary such that the photoionization rate remains constant.  Here we see the fractional deviations expected for changing the speed of light by a factor of 2, 10, and 100, respectively.

Because the total photon flux is conserved across the boundary, the number density of photons in the cells is different by the same factor as the change in $\tilde c$.  However, because the absorption cross-section is scaled with the speed of light, for a constant density medium, the same number of absorptions will occur even with a change in photon number density across the refinement regions.  This is further shown in Appendix~\ref{psld}.

\begin{figure*}
\centerline{\includegraphics[scale=0.5]{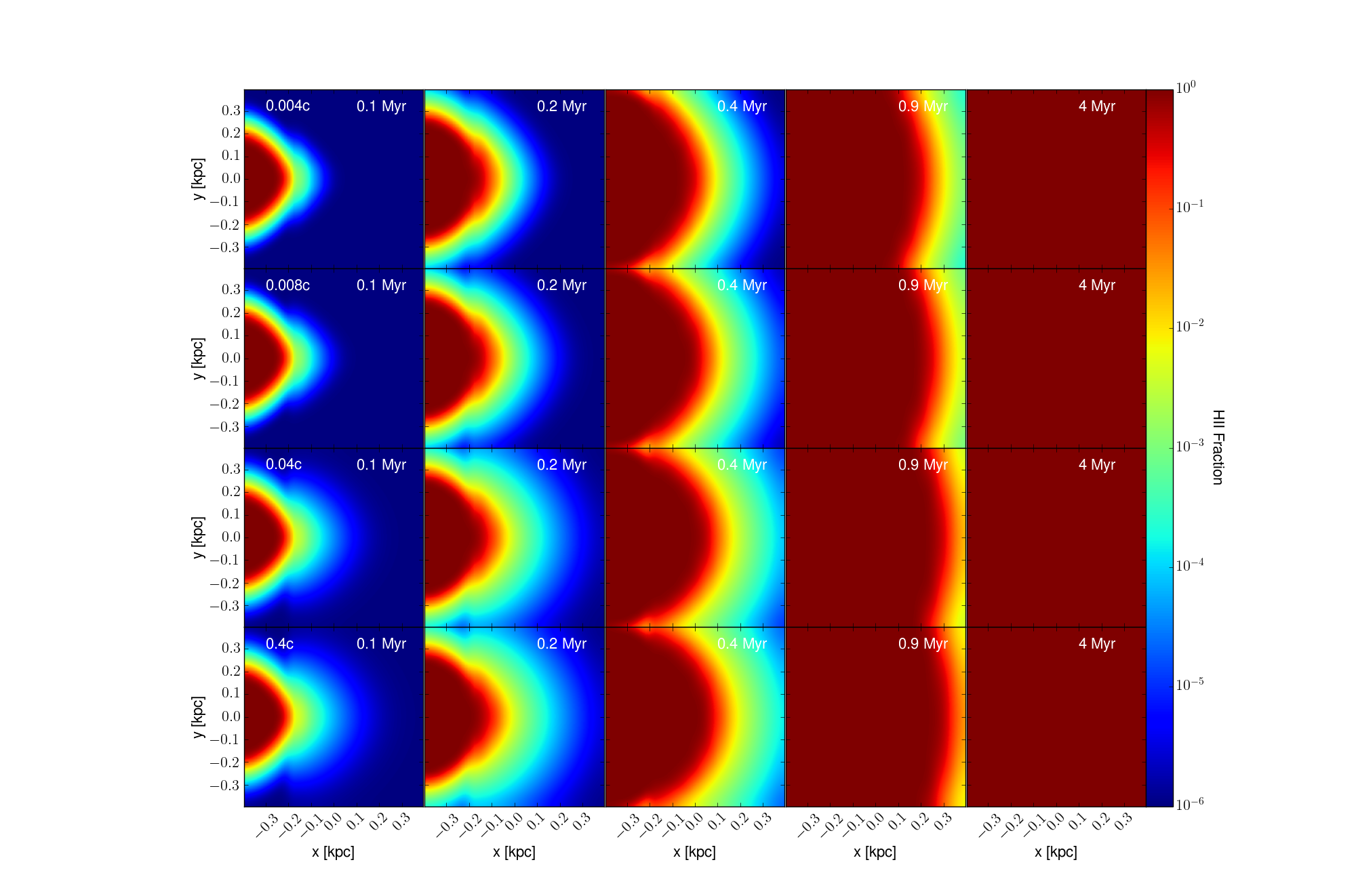}}
\caption{Evolution of the ionization front at 0.1, 0.2, 0.4, 0.9, and 4Myr for the control run (top) and VSLA runs (bottom three panels).  The refinement boundary occurs at $x=-0.25$ where to the left of the boundary, the speed of light is set to be that of the control run while to the right of the boundary, we vary the speed of light as indicated on the plot.  Hence, the I-fronts travel faster when the speed of light is increased.}
\label{HIIfraction}
\end{figure*}

We have also checked to make sure that our implementation is robust to changes of the location of the source as well as the direction of the refinement boundary.  We have run the same test with the exact reverse setup by putting the source at position (0.8, 0.4, 0.4)kpc  and reversing the orientation of the refinement boundary and obtained very similar results.  Furthermore, we have run the test both in the y, and z directions and once again find similar results.

These tests have all been run on four cores and we provide the timing results in Table~\ref{timetest1}.  All tests have exactly the same domain decomposition and number of outputs.  We have run an additional control where adaptive time stepping is turned off (i.e. all levels are run on the fine time step and thus there is no hydro subcycling) and we see that the timing is the same for the run with the twice enhanced speed of light.  Because this test is at a such low resolution and we use more cores than necessary (this was done to test the robustness of our algorithm to MPI calls), the timing between the control with and without adaptive time stepping is fairly comparable.  In the run with a ten times enhanced speed of light, the time step is 5 times shorter than the control run and we see that it takes 4.6 times as long to run.  Likewise, the run with the 100 times enhanced speed of light has a time step that is 50 times shorter than the control run and this takes 55 times as long to run.  Note that some of the time is spent on I/O and the amount of time needed to print an output is independent of the value of the speed of light used.  For the shorter runs, this becomes more important in timing than for longer runs.  

\begin{table}
\centering
\begin{tabular}{@{}lc@{}}
\hline
$c_{\rm coarse}$ & wall clock time (s) \\
\hline
0.004$c$     &    27.45    \\
0.004$c^a$ &    33.77     \\
0.008$c$     &    32.80     \\
0.04$c$       &    155.30   \\
0.4$c$         &    1686.92 \\
\hline
\end{tabular}
\caption{Wall clock time for test 1. $^a$ We have run a second control run where we turn off adaptive time stepping (i.e. nsubcycle is set to 1 for all levels). }
\label{timetest1}
\end{table}

\subsubsection{Test 2: Point Source in a Low-Density Medium}
\label{psld}
The setup for this test is exactly the same as in the previous test except that the HI cross-sections are no longer set to zero, rather the energy-weighted and photoionization cross-sections are set as appropriate for a $10^5$K blackbody spectrum.  We have set the RAMSES-RT parameter ``static=.false.'' so that the hydro quantities can be updated.  Furthermore, we decrease the density in the box to $n_{\rm H}=10^{-3}$cm$^{-3}$ so that the Stromgren radius is much larger than the box size.  In addition to the control where $c_{\text{sim}}=0.004c$, we perform a VSLA run where we enhance $c_{\text{sim}}$ by a factor of two at the coarse level.  After 4Myr, the box is completely ionized and optically thin to radiation.  Flux in all directions is conserved to a few parts in $10^{-6}$ and the maximal deviation in ionized fraction is $2\times10^{-6}$ which occurs right at the edge of the far side of the box and away from the source.  Timings for this test can be found in Table~\ref{timetest2}.

\begin{table}
\centering
\begin{tabular}{@{}lc@{}}
\hline
$c_{\rm coarse}$ & wall clock time (s) \\
\hline
0.004$c$     &    32.43    \\
0.008$c$     &    31.42    \\
0.04$c$ & 162.08\\
0.4$c$ & 1423.24 \\
\hline
\end{tabular}
\caption{Wall clock time for test 2.  The control experiment has adaptive time stepping turned off.}
\label{timetest2}
\end{table}

\begin{figure*}
\centerline{\includegraphics[scale=0.4]{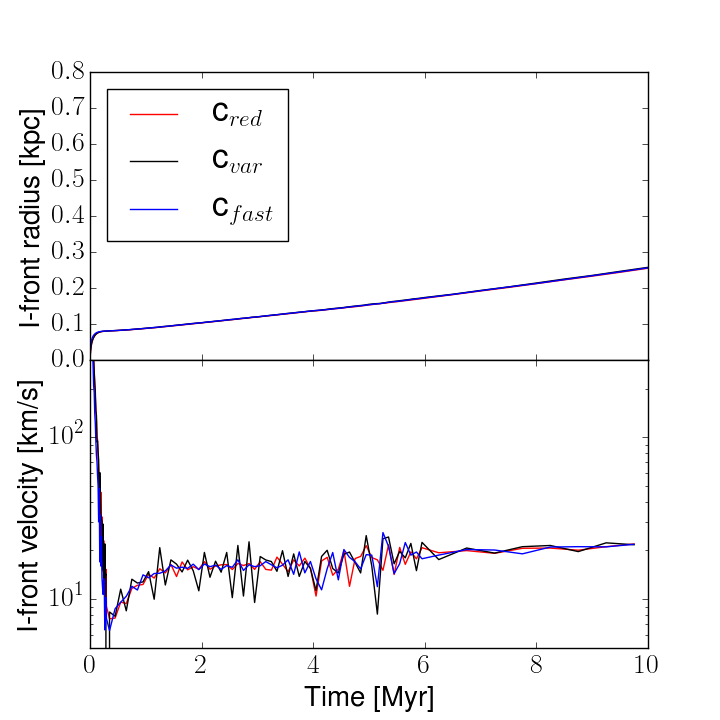}\includegraphics[scale=0.4]{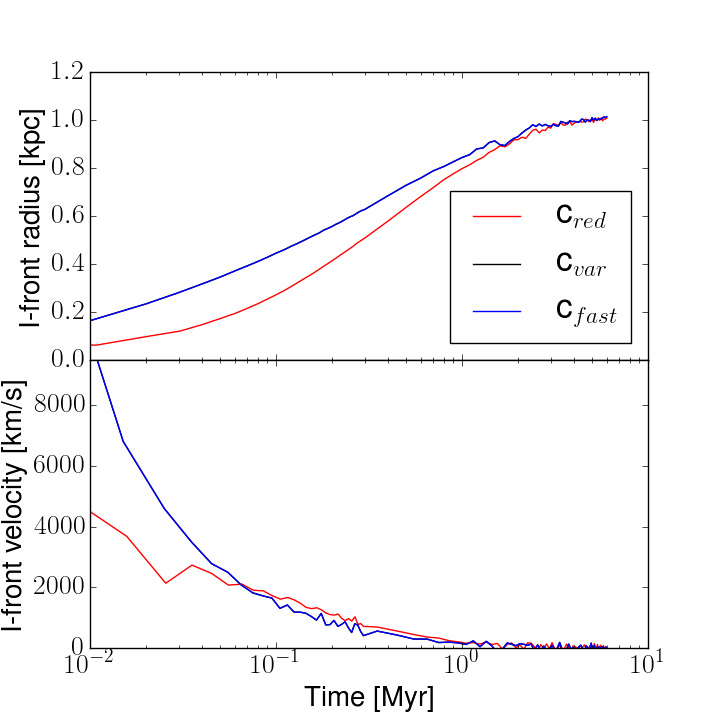}}
\caption{{\it Left.} Position (Top) and Velocity (Bottom) of the ionization front as a function of time for each of the three different simulations for Iliev test 6. {\it Right.} Position (Top) and Velocity (Bottom) of the ionization front as a function of time for each of the three different simulations for Iliev test 6 reversed.  The fluctuations in velocity are due to sampling.}
\label{i6}
\end{figure*}

In Figure~\ref{HIIfraction}, we show the propagation of the ionization front at 0.1, 0.2, 0.4, 0.9, and 4Myr.  The position of the ionization fronts is very similar in the left panels of Figure~\ref{HIIfraction} since the I-front is only just beginning to cross into the coarse cells where the speed of light differs.  In the left centre panel of the figure, we see a small mushroom effect where it appears that the radiation has expanded a bit further in the y direction after the refinement boundary.  This is due to the low resolution of the cells and the rendering that comes with averaging the quantities along the z-axis.  The mushroom effect is enhanced in the lower left panels because the speed of light is faster so the radiation should be diffusing quicker than in the control run.  We can see in the right four panels of Figure~\ref{HIIfraction} that the I-front of the control run is lagging behind the VSLA runs as expected.  This is further tested in Section~\ref{I6}.  After 4Myr, the I-front is far beyond the regions of the box and the simulation is in a steady state as described above.

\subsection{Additional Tests}
In addition to the previous two tests, we have checked that our algorithm is robust to adaptive refinement, both refining and deferring cells on-the-fly.  Furthermore, imposing multiple levels of refinement also poses no issues.

\subsection{VSLA Comparison Tests}
Having demonstrated that VSLA successfully conserves both number of photons and flux across level boundaries, we now test our new VSLA implementation in a series of controlled experiments which demonstrate the advantage of VSLA over both RSLA and using the full speed of light.  We present three tests which demonstrate that the physical properties of the I-front are conserved in regimes where the velocity of the I-front (the rate at which the radius HII region grows) is much slower than the speed of light and that VSLA better reproduces the results of the full speed of light in low-density regimes at a much reduced computational cost.  

\subsubsection{Iliev 2009 Test 6}
\label{I6}
\cite{Iliev2009} test 6 is designed to represent a radiative source emitting at the centre of a dense cloud of gas.  A luminous source ($\dot{N}_{\gamma}=5\times10^{50}$photons/s summed over three radiation bins), mimicking a $10^5$K blackbody is placed at the corner of a box of length 0.8kpc and centred on a dense cloud of gas with the following density profile,
\begin{equation}
n_{\text{H}}(r)=
\begin{cases}
\; n_0 & \mathrm{if} \, r\le r_0
\\
\; n_0(r_0/r)^2 & \mathrm{if} \, r\ge r_0,
\end{cases}
\end{equation}
where $n_0=3.2$cm$^{-3}$ and $r_0=91.5$pc.  The initial temperature of the entire box is set to 100K and the simulation contains only hydrogen.  The boundary conditions of the three walls touching the radiation source are reflective while the other three are outflow.  The original test uses a fixed 128$^3$ Cartesian mesh however, since VSLA attempts to keep the light crossing time of all cells the same regardless of resolution, we have modified the original test slightly to better explore the properties of VSLA.  Rather than a fixed uniform mesh, we initialise a fixed grid as listed in Table~\ref{reftable} which uses a base grid of 64$^3$ cells with three additional levels of refinement at specified radii.

\begin{table}
\centering
\begin{tabular}{@{}lcc@{}}
\hline
Level & N$_{\text{cells}}$ & Radii \\
\hline
6 & $64^3$   & $r>0.25$kpc \\
7 & $128^3$ & $0.25\geq r>0.125$kpc \\
8 & $256^3$ & $0.125\geq r>0.0625$kpc \\
9 & $512^3$ & $0.0625\geq r$kpc \\
\hline
\end{tabular}
\caption{Refinement scheme for Iliev 2009 test 6.}
\label{reftable}
\end{table}

We provide three versions of this simulation: 1) $c_{\text{sim}}=0.01c$, 2) $c_{\text{sim}}=0.08c$, and 3) the VSLA implementation where $c_{\text{sim}}=0.01c$ on Level 9 and is increased by a factor of two at each lower level up to $c_{\text{sim}}=0.08c$ on Level 6.  In this test, the velocity of the I-front is much slower than the speed of light and we therefore do not expect any differences between the three different runs.  This test is simply meant to demonstrate that in the regime where the I-front velocity is slow relative to the speed of light, out VSLA method reproduces what is expected.  In the left panel of Figure~\ref{i6}, we show the position and velocity of the front as a function of time for the first $\sim$15Myr of the simulation.  As expected, we see very good agreement between all three simulations despite the change in the speed of light.  In the bottom left panel of Figure~\ref{i6}, we see that the velocity of the I-front is much shorter than the values of the speed of light used in any of the three experiments.  

\subsubsection{Iliev 2009 test 6 Reversed}
In our next test, we change the location of the radiation source such that it begins emitting in much lower density gas while irradiating a much higher density region, much like one galaxy shining onto another.  The simulation is very similar to the setup in the previous test in that we use the same density profile for the gas, initial temperature, box size, and source luminosity, but we move the radiation source from $(x,y,z)=(0,0,0)$ to $(x,y,z)=(0.8,0.8,0.8)$.  The refinement scheme is exactly the same as in the previous test and is given in Table~\ref{reftable}.  Because the density profile is convex with respect to the source location, we measure the position of the I-front with respect to the opposite corner and subtract the distance from the diagonal.  We then use the change in distance to get a velocity.

Similar to the previous test, we run three versions of the simulation: 1) $c_{\text{sim}}=0.01c$, 2) $c_{\text{sim}}=0.08c$, and 3) the VSLA implementation where $c_{\text{sim}}=0.01c$ on Level 9 and is increased by a factor of two at each lower level up to $c_{\text{sim}}=0.08c$ on Level 6.  The situation here is very different from the previous example because the radiation source begins emitting in a much lower density region.  In this case, we expect the I-front to move considerably faster, much as it would in the low-density IGM, and thus the particular value we choose for $c_{\text{sim}}$ will affect the propagation of the I-front.

In the right panel of Figure~\ref{i6}, we see that the VSLA run agrees well with the $c_{\text{sim}}=0.08c$ simulation in both the position of the I-front and its velocity while the $c_{\text{sim}}=0.01c$ run lags behind.  This is expected since the VSLA has $c_{\text{sim}}=0.08c$ at Level 6 where much of the initial propagation takes place.  When the I-front reaches the higher density region of the box, it slows down considerably to a value much less than the reduced speed of light which allows the I-front in run with $c_{\text{sim}}=0.01c$ to catch up with the other two runs.  

What this test reveals is that using the VSLA technique to set the speed of light to always be slightly faster than the expected velocity of the I-front will save significantly on computational cost while also reproducing the correct behaviour of the I-front expansion.  Comparing the timings of this run, we find that the runs with $c_{\text{sim}}=0.01c$, $c_{\text{sim}}=0.08$, and VSLA took  4319.16, 34425.47, and 14479.48 seconds, respectively.  As expected, the run with $c_{\text{sim}}=0.01c$ is nearly exactly 8 times faster than the run with $c_{\text{sim}}=0.08c$.  Abandoning the multi-stepping cost an extra factor of 3.35 over the run with $c_{\text{sim}}=0.01c$; however the VSLA run is still 2.38 times faster than the run with $c_{\text{sim}}=0.08c$ despite the fact that the evolution of the I-front is indistinguishable between the two runs.  

\begin{figure}
\centerline{\includegraphics[scale=0.4]{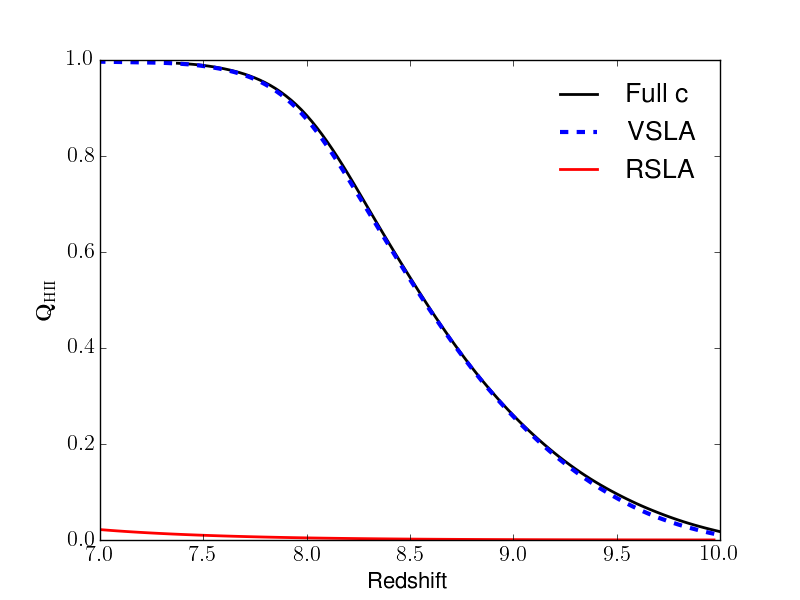}}
\caption{Volume filling factor of HII as a function of redshift in a $230^3$Mpc box using ``Full c", VSLA, and RSLA.  As this simulation is extremely low resolution, only massive haloes form form with extremely massive and luminous star particles.  This extreme choice is meant to mimic a quasar-dominated reionization scenario and in this case VSLA works extremely well.}
\label{lowrescosmo}
\end{figure}

\subsubsection{Low Resolution Cosmological Simulation}
Our final test aims to see how the VSLA algorithm performs when used in a low resolution cosmological simulation.  In order to do this, we set up a cosmological box of size $230^3$Mpc with a base grid on Level 8 with $256^3$ dark matter particles and allow for refinements in the grid up to Level 13.  This gives a maximum physical resolution of $\sim3.5$kpc at $z=7$.  We run three simulations: a ``Full c" run where $c_{\text{sim}}=c$ at all levels, a VSLA run where we set $c_{\text{sim}}=c$ on Level 8 and divide it by a factor of two up to Level 13 where $c_{\text{sim}}=0.03125c$, and a final RSLA run where $c_{\text{sim}}=0.03125c$ on all levels.  We set the star formation and escape fraction of radiation from star particles to be the same for all three simulations and run the simulation to $z=7$.  The crossing time of light across the box at this redshift is $\sim3$Gyr for the RSLA run while it is $\sim94$Myr for the ``Full c" run.  The former is much longer than the age of the Universe at this redshift while the latter is much less than the age of the Universe.  In a box with only a few strong and clustered sources, as we have in this simulation, one would expect that the RSLA simulations lags far behind the ``Full c" simulation.  

In Figure~\ref{lowrescosmo} we plot the volume-weighted HII filling factor as a function of redshift.  We can see that the VSLA simulation tracks the ``Full c" run beautifully and there is near perfect agreement.  Both of these simulations have the entire box completely ionized at $z\sim7.5$.  By contrast, the RSLA simulation is only a few percent ionized at these redshifts. 

We should emphasise that the setup of this simulation is not completely realistic and designed to exploit the potential of VSLA compared to RSLA in terms of agreement with the ``Full c" run.  The star particles that form and ionise the box are strongly clustered around a few different locations and only a small fraction of cells are refined to the maximum level.  What potentially matters for this is how the sources are distributed throughout the box.  If  there are many weaker sources equally distributed around the volume, than he RSLA run will not lag very far behind the ``Full c" run in terms of redshift of reionization.  This might correspond to a case where mini-haloes are the dominant sources of photons during reionization.  The case we have just simulated here where there are a few very strong sources around the box exacerbates the problem of using the RSLA because the time until overlap of the Stromgren spheres is much longer.  We know that the RSLA lags behind until the Stromgren radius is reached.  In this type of simulation, the Stromgren radii are very large for these sources which makes the lag in the RSLA run worse.  This might correspond to a physical scenario where quasars are the dominant sources of photons during reionization and it is well established that for these very bright sources, the ionization fronts never reach their Stromgren radius \citep{Shapiro1987}.

Comparing the timing between the two runs, the RSLA run took 2283s to reach $z=7$ while it took 32,852s and 42,271s for the VSLA and ``Full c" runs, respectively, on 64 cores.  In this case, the VSLA run takes $\sim13.7$ times longer than the RSLA while the ``Full c" run takes $\sim17.7$ times longer.  We can ask why there is such a difference in the timings?  In theory, the ``Full c" run should be 32 times faster than the RSLA run.  However, there is a time period before the first star particle forms when the on-the-fly RT is turned off where the simulations take the same amount of time.  Since we have only run the simulations for a very short period of time, this eats into how much speed-up one can achieve just by using the RSLA.  Running the simulation for longer should lead to a better speed-up for the RSLA.  Furthermore, we have not made an attempt to optimise the number of cores for the computation.  Looking towards the VSLA, this factor is much higher than one would naively expect given our previous tests.  We can expect that abandoning multi-stepping in time costs roughly a factor of $\sim3.5$ in runtime as observed in the previous section.  We still have to account for another factor of 4 in runtime.  This is due to the fact that star particles form before the maximum level of resolution is reached.  If this occurs, the standard version of VSLA will be a factor of two slower than RSLA for every lower resolution level above the maximum level that star particles form.  This is because we have set the speed of light to divide by a factor of two on each subsequent level.  In this specific very low resolution simulation, star particles start forming on Levels 10 and 11 before the simulation has refined down to Level 13.  This can slow down VSLA considerably.  One can mitigate this effect in a few different ways.  One might enforce that star particles can form only when the maximum level of the simulation is reached thereby preventing the slow down.  This solution is potentially useful when running simulations which refine many levels and in particular zoom simulations.  Alternatively, one could change the algorithm slightly so that the highest resolved cells always have the same speed of light as the RSLA run and as more levels are introduced into the simulation, the speed of light increases at the lower resolution levels.  As an example, in this specific case, one could start the VSLA run with $c_{\text{sim}}=0.03125c$ on Level 8.  When the simulation refines to Level 9, we would set $c_{\text{sim}}=0.03125c$ on Level 9 and $c_{\text{sim}}=0.0625c$ on Level 8.  Likewise, when the simulation refines to Level 10, we would set $c_{\text{sim}}=0.03125c$ on Level 10, $c_{\text{sim}}=0.0625c$ on Level 9, and $c_{\text{sim}}=0.0125c$ on Level 8.  If one allows about 7 level of refinement, by the time the simulation reaches the maximum level, we will be back in the original form of the VSLA algorithm.  If one sets the simulation to always resolve the same physical scale (i.e. releasing a new level when the scale factor increases by a factor of two), then this latter algorithm should perform very similarly to the original VSLA and also have the added bonus of a significant speed up compared to the original algorithm.  All in all, the way to take advantage of the VSLA is to configure it to the problem of interest and there is no one-size-fits-all method.


\bsp	
\label{lastpage}
\end{document}